\documentclass[article]{aastex}
\usepackage{epstopdf} 
\usepackage{rotating}
\usepackage{epsfig}  
\usepackage{graphics}
\usepackage{graphicx}
\usepackage{listings}

\begin{document}


\title{Abundance Derivations for the Secondary Stars in Cataclysmic Variables
from Near-Infrared Spectroscopy}

\author{Thomas E. Harrison$^{\rm 1,2}$}

\affil{Department of Astronomy, New Mexico State University, Box 30001, MSC 
4500, Las Cruces, NM 88003-8001}

\email{tharriso@nmsu.edu}

\begin{abstract}
We derive metallicities for 41 cataclysmic variables (CVs) from near-infrared
spectroscopy. We use synthetic spectra that cover the 0.8 $\mu$m $\leq \lambda 
\leq$ 2.5 $\mu$m bandpass to ascertain the value of [Fe/H] for CVs with 
K-type donors, while also deriving abundances for other elements. Using 
calibrations for determining [Fe/H] from the $K$-band spectra of M-dwarfs, we 
derive more precise values for T$_{\rm eff}$ for the secondaries in the 
shortest period CVs, and examine whether they have carbon deficits. In general,
the donor stars in CVs have sub-solar metallicities. We confirm carbon deficits
for a large number of systems. CVs with orbital periods $>$ 5 hr are most 
likely to have unusual abundances. We identify four CVs with CO emission.
We use phase-resolved spectra to ascertain the mass and radius of the donor in 
U Gem. The secondary star in U Gem appears to have a lower {\it apparent} 
gravity than a main sequence star of its spectral type. Applying this result
to other CVs, we find that the later-than-expected spectral types observed for 
many CV donors is mostly an affect of inclination. All of the magnetic CVs, 
except the low accretion rate polar MQ Dra, have donors with subsolar 
metallicities. We find that two systems with unusual spectra, EI Psc and QZ 
Ser, that have large excesses of sodium, and extreme deficits of carbon. 
Synthetic spectra that have a reduced abundance of hydrogen are best able to 
explain the spectra of these two objects.

\end{abstract}

\noindent
{\it Key words:} infrared: stars --- stars: cataclysmic variables --- stars: 
abundances --- stars: individual (RX And, AE Aqr, VY Aqr, UU Aql, TT Ari, SS 
Aur, SY Cnc, V436 Cen, WW Cet, WX Cet, TT Crt, EM Cyg, EY Cyg, SS Cyg, EX Dra, 
MQ Dra, YY/DO Dra, U Gem, AH Her, AM Her, EX Hya, VW Hyi, RZ Leo, ST LMi, CW 
Mon, V426 Oph, V2051 Oph, CN Ori, CZ Ori, V1309 Ori, IP Peg, LS Peg, RU Peg, 
GK Per, KT Per, EI Psc, VV Pup, V893 Sco, MR Ser, QZ Ser, AR UMa, CH UMa, TW 
Vir) 

\begin{flushleft}

$^{\rm 1}$ Visiting Observer, W. M. Keck Observatory, which is operated 
as a scientific partnership among the California Institute of Technology, the 
University of California, and the National Aeronautics and Space 
Administration.\\

$^{\rm 2}$ Visiting Astronomer at the Infrared Telescope Facility, which is 
operated by the University of Hawaii under contract from NASA.\\

\end{flushleft}
\clearpage

\section{Introduction}

Cataclysmic variables (CVs) result when the cool, low mass secondary in
a binary contacts its Roche lobe and transfers material to the white dwarf
primary. To arrive at this stage, the standard evolutionary paradigm 
involves having the more massive component in the binary
evolve off the main sequence and engulf its companion in a short-lived
common envelope (CE) phase (e.g., Iben \& Livio 1993). During the
CE phase, the outer atmosphere of the primary star is removed by interaction
with the secondary star, and this interaction leads to either a stellar
merger, or a more compact binary (Politano \& Weiler 2007). Angular momentum
loss, such as through ``magnetic braking'' (Verbunt \& Zwaan 1981), eventually
leads to a semi-contact binary and a CV is born. In most of the early population
synthesis calculations (cf., Howell et al. 2001, and references therein)
the vast majority of donor stars in CVs were expected to be unevolved. If
true, the donor stars should generally appear to be similar to their main
sequence counterparts, except for the effects introduced by the mass
loss process. Thus, while they will have a slightly different mass-radius 
relationship due to the secondary being out of thermal equilibrium, their 
photospheres should not show any abundance peculiarities.

The standard evolutionary paradigm though is slowly changing to reflect the 
fact that many long period CVs appear to have donors with unusual masses and/or 
radii. This suggests that the secondaries in some subset of CVs have undergone 
significant evolution prior to becoming interacting binaries. Podsiadlowski et 
al. (2003) constructed a population synthesis for CVs that allowed for evolved 
secondaries with initial masses of up to 1.4 M$_{\sun}$. Inclusion of these 
objects allowed them
to explain the wide dispersion in the observed temperatures of donor stars in 
CVs with P$_{\rm orb}$ $>$ 5 hr. This result confirmed the conclusions of
Baraffe \& Kolb (2000), that long period CVs are dominated by systems
containing evolved secondaries. Recently, Goliasch \& Nelson (2015) have
produced a new population synthesis calculation that includes detailed
nuclear evolution. They confirm the results of Podsiadlowski et al., showing
that there is a large range of masses and radii for donors in CVs with 
P$_{\rm orb}$ $>$ 5 hr when nuclear evolved secondary stars are considered. 
Two other interesting results from
this study are that CVs with evolved secondaries do not have a period 
gap\footnote{The observed dearth of CVs with 2 hr $\leq$ P$_{\rm orb}$ $\leq$ 
3 hr (see Kolb et al. 1998).}, and they can evolve to much shorter periods 
than predicted by models where evolved donors are not considered.  
Podsiadlowski et al. suggest that the secondaries that have undergone
pre-CV evolution should show evidence for nuclear processing, in particular, 
products produced via the CNO cycle.

Two alternative paths that might allow the donor stars in CVs to acquire 
peculiar abundance patterns are for them to accrete such material during the 
original common envelope phase, or during the briefer common envelope phase 
that results from a classical nova eruption. Marks \& Sarna (1998) have 
investigated a large number of scenarios for the composition of the surface
layers of CV donor stars that have accreted both types of material, and find
that significant changes are possible. A common theme amongst these models
are large deficits of carbon. In a series of papers we have presented near-IR 
$K$-band spectra that demonstrated that the secondaries of many CVs appeared 
to have significant carbon deficits (Harrison et al. 2004a, 2005ab, 2009). 
Recently, Harrison \& Hamilton (2015, ``H\&H'') used synthetic spectra to 
quantify that SS Cyg, RU Peg, and GK Per had carbon abundances that were 10 
to 30\% of the solar value. In contrast, however, they found that the 
metallicities ([Fe/H]) of the donor stars were solar, or only slightly 
subsolar.

The study by H\&H was limited to the $K$-band and, as they describe, there are 
not many strong atomic absorption lines in that bandpass to derive more precise
values of [Fe/H], nor enough spectral lines from any species other than carbon 
to allow derivation of their abundances. To remedy this issue, we have 
developed the capability to generate synthetic spectra that cover the $IJHK$ 
bandpasses. This allows us to investigate the abundances of other important 
elements for the subset of CVs we have observed using cross-dispersed 
spectrographs.  From modeling these data we find a wide range of values for 
[Fe/H], and for [C/Fe]. In addition, however, we find a number of CVs that 
appear to have unusual abundances of aluminum, magnesium, and/or sodium. Due 
to the lack of an extensive line list (and other uncertainties discussed below)
our abundance analysis has been limited to CVs whose secondaries have spectral 
types earlier than M0. Recently, relationships have been developed to extract 
metallicities for M dwarfs. Using those relationships it is possible to examine
the cooler secondary stars found in short period CVs for abundance anomalies.

We describe the observations in the next section, the generation of 
synthetic spectra in Section 3, the results on CVs with K-dwarf secondaries
in Section 4, the process and data needed to generate metallicities for
M dwarfs in Section 5, discuss our results in Section 6, and present our
conclusions in Section 7.

\section{Observations}

The majority of the new data presented below was obtained with SPEX on
the IRTF. SPEX was used in cross-dispersed mode generating spectra spanning 
the wavelength range from 0.8 to 2.45 $\mu$m. In the $K$-band, the dispersion 
is 5.4 \AA/pix. Nearby A0 V stars were used for telluric correction, and these 
data were telluric-corrected and flux- and wavelength-calibrated using the 
SPEXTOOL package (Cushing et al. 2004). SPEXTOOL attempts to fit, and
remove the H I lines in the telluric A star. While this process can
work quite well, many times there remain residuals at the positions of the
strongest H I lines. Thus, the line strengths and profiles of strong H I
absorption or emission features in these data should not be taken too
seriously. A log of all of our observations for previously unpublished data
is presented in Table \ref{obslog}, and it lists the UT date, start and stop 
times, the length of the exposure times used for each object, the instrument 
used, the orbital period, the orbital phase at the mid-point of the 
observational sequence (except for U Gem where full orbits were covered), and
the outburst state for each of the CVs.

We also present previously unpublished spectra obtained with
NIRSPEC\footnote{ http://www2.keck.hawaii.edu/inst/nirspec/nirspec.html} on 
Keck.  We used NIRSPEC in low-resolution, single order mode with a 0.38" slit. 
The grating tilt was set so as to cover the wavelength region 2.04 $\mu$m 
$\leq$ $\lambda$ $\leq$ 2.46 $\mu$m, with a dispersion of 4.27 \AA/pixel. We 
employed the two-nod script, and used 4 minute exposure times for all of 
the program CVs. To correct for telluric absorption, we observed bright A0V 
stars located close to the program objects so as to minimize their relative 
differences in air mass. These data were reduced using the IDL routine 
REDSPEC\footnote{https://www2.keck.hawaii.edu/inst/nirspec-old/redspec/index.html}, specially developed for NIRSPEC.  

Finally, we used TripleSpec\footnote{http://www.apo.nmsu.edu/arc35m/Instruments/TRIPLESPEC/} 
(Wilson et al. 2004) on the Apache Point Observatory (APO) 3.5 m telescope
to obtain phase-resolved near-IR spectra of U Gem over two orbital periods,
and for 50\% orbital period coverage of QZ Ser. TripleSpec is similar to 
SPEX in that it provides cross-dispersed spectra covering the near-IR, from 
0.95 $\mu$m to 2.46 $\mu$m, at a resolution of R $\sim$ 3500. The 1.1" slit 
was used, and all exposures were four minutes in length. These data were 
reduced using ``Triplespectool,'' a version of SPEXTOOL modified by Michael 
Cushing for use at APO. Nearby A0V stars were used to provide telluric 
correction.

In the following we will also be analyzing the spectra of CVs observed in 
just the $K$-band with the IRTF using SPEX, at Keck with NIRSPEC, or
at the VLT with ISAAC. All of these data have been previously published in 
Harrison et al. (2004a, 2005ab, 2007a, 2009), and Hamilton et al. (2011). 
Unless otherwise noted, all of the spectra have been Doppler corrected using 
published radial velocity data and orbital ephemerides.

\section{Synthetic Spectra for the $IJHK$ Bands}

In H\&H we described the construction of a Python-based program (named 
``$kmoog$'') that used MOOG (Sneden 1973) to generate a large grid of synthetic
$K$-band spectra, and to autonomously search through this grid to derive the 
best fitting result for each program object. As described there, the focus 
of that effort was in modeling the CO features, and the capability to 
investigate the abundances of other elements was not 
implemented. With the desire to expand our coverage to shorter 
wavelengths, we decided to abandon further development of $kmoog$ in favor of 
using ``SPECTRUM'' by R. Gray\footnote{http://www.appstate.edu/~grayro/spectrum/spectrum.html}. The main driver for this decision was the existence of a line list 
covering the visual and near-IR\footnote{http://www.appstate.edu/grayro/spectrum/specftp.html}. As discussed in H\&H, the process of constructing and validating
a $K$-band line list was quite onerous, and one we wished to avoid if possible.

As in H\&H, we fixed the micro-turbulence to 2 km s$^{\rm -1}$ for all
synthetic spectra, a value that appears to best represent normal stars (Husser 
et al. 2013, Smith et al. 2013). Likewise, all of our synthetic spectra were 
run with log$g$ = 4.5. Cesetti et al. (2013) have examined the use of moderate 
resolution data from the IRTF Spectral Library to derive diagnostics for cool 
stars. They found very few features that allow for accurate quantification of 
log$g$ for higher gravity stars. Given the large rotation velocities of CV 
donors, the limited spectral resolution, and the presence of numerous emission 
features, it is not possible to constrain the surface gravity from the data 
discussed herein.

Unlike $kmoog$, SPECTRUM is not natively designed to use a multiple processor
machine that greatly aids in speeding-up the process of generating large grids
of models.  To allow us to quickly construct large grids of synthetic spectra 
using SPECTRUM, we used the GNU shell tool ``{\it parallel}'' with our 32 core
Linux workstation. We found that we could generate models as quickly 
with SPECTRUM $+$ $parallel$, as with $kmoog$.

The other main difference between the use of SPECTRUM vs. $kmoog$ is that 
the latter uses MARCS\footnote{http://marcs.astro.uu.se/} atmosphere models 
when computing synthetic spectra, while SPECTRUM was designed for the input of 
Kurucz\footnote{http://kurucz.harvard.edu/grids.html} models. Thus, before
we began our program we first had to insure that the results generated
with SPECTRUM matched those generated by $kmoog$ (see H\&H for validation of
$kmoog$). We ran a small grid of models with identical parameters using
both $kmoog$ and SPECTRUM. In addition, we also translated the MARCS models
used for this test into the format used by Kurucz to determine if the input
atmosphere model resulted in significant differences. The spectra produced
by the three processes were identical as far as we could tell.

As in H\&H, the next step was to validate the line list using template 
K dwarf spectra.
For this process we used the same set of dwarfs as in H\&H (the data
are from the IRTF Spectral Library, Cushing et al. 2005). We found that
in general, the ``luke.nir'' (covering 6800\AA ~ $\leq$ $\lambda$ $\leq$ 1 
$\mu$m) and ``luke.ir'' (covering 1 $\mu$m $\leq$ $\lambda$ $\leq$ 4 $\mu$m)
line lists were quite reliable. We did adjust the oscillator strengths for 
some of the strongest Mg I lines, but otherwise few changes were necessary.

In Fig. 1 we present the spectra in the $IJHK$ bandpasses for the K dwarf
templates with the best fitting models calculated using SPECTRUM. We identify 
the strongest atomic and molecular absorption features in each panel to show 
the redundancy of spectral features for each of the elements, and as a guide
for when we discuss features in the spectra of the program CVs. Except for
the CN bandhead near 1.1 $\mu$m, the models reproduce all of the strongest
spectral features. The spectral regions near 1.1 $\mu$m and 2.0 $\mu$m
are afflicted with strong telluric features that compromise the observations.

As we did for the $K$-band in H\&H, to determine the best fitting models
we constructed heat maps from a $\chi^{2}$ analysis. Those for the $J$-band, 
covering the spectral types from K0V to K5V, are shown in Fig. 2. We tabulate 
the best fitting value for T$_{\rm eff}$ and [Fe/H] for each of the template 
stars from a $\chi^{2}$ analysis of the synthetic spectral grid for each of 
the bandpasses in Table 2. In this table we also list the averages from those 
fits, and their nominal values (see Table 2 in H\&H). In H\&H, the typical 
error bars in the determination of
[Fe/H] and T$_{\rm eff}$ from the SPEX data using just the
$K$-band were $\pm$ 0.25 and $\pm$ 250 K, respectively. As shown in Table 2, 
the errors in both quantities, when averaging over the four bandpasses, are 
all smaller, and easily within the combined error bars of our models and
the averages of the published values. We must note that the temperature
grid of the Kurucz atmospheres steps in intervals 250 K, while the abundance 
grid at each temperature was [Fe/H] = $+$0.5, $+$0.25, 0.0, $-$0.1, 
$-$0.2, $-$0.3, $-$0.5, and $-$1.00 (we also interpolated between the $-$0.5 
and $-$1.00 models to create atmospheres with [Fe/H] = $-$0.75). Thus, there 
is an uneven quantization in the model values of [Fe/H]. 

The average of the 
differences in temperature between our models and the nominal values is only 
$+$22 K, suggesting our temperature derivations using these low resolution 
spectra are reliable: $\sigma$ = $\pm$ 110 K. For the 
values of [Fe/H], however, the average of the differences is $-$0.11, 
suggesting that we tend to derive slightly lower metallicities than the means 
of the published values. Thus, while the formal error on our derivation of 
[Fe/H] for the templates is only $\sigma$ = $\pm$ 0.04, the offset we find, 
and the actual error bars on the nominal values, suggests that the true 
standard deviation on our metallicities is at least a factor of two larger 
than this. 

In the following subsection will use this same technique for all of the CVs
with K-type secondaries. $\chi^{2}$ minimization will be used to derive 
T$_{\rm eff}$, [Fe/H], and [C/Fe]. Note that the $\chi^{2}$ minimization
process was only run on select regions of the spectra in each of the
photometric bands. As described in H\&H, we avoid regions with strong emission 
features intrinsic to the CV, and regions where strong telluric features 
compromise the calibration. In the $K$-band,
the analysis was first conducted on the spectral region from 2.19 $\mu$m
to 2.29 $\mu$m to arrive at values for T$_{\rm eff}$ and [Fe/H]. Then
the region from 2.19 to 2.36 $\mu$m was used to find the best fitting
value of [C/Fe]. If we were to include the CO region into the initial
$\chi^{2}$ analysis for the $K$-band, the results are greatly biased by the 
strength of the CO features (which might have a non-solar abundance). In the 
$J$-band the analysis region was constrained to 1.16 $\leq$ $\lambda$ $\leq$ 
1.26 $\mu$m. 

In the $I$- and $H$-bands, the available spectral regions for analysis are
compromised by H I and/or Ca II emission lines. There were no CVs for which 
a $\chi^{2}$ analysis was possible in the $I$-band. While there are gaps 
between the Paschen series in the $I$-band, no strong absorption 
features are visible in those gaps. For most of the CVs, the $\chi^{2}$ 
minimization calculation for the $H$-band was run over the wavelength 
interval 1.55 $\leq$ $\lambda$ $\leq$ 1.67 $\mu$m. There were CVs, however, 
where the Brackett series made this impossible (e.g., GK Per). In all
cases, we visually inspected the reasonableness of the fit of the final model 
to the data in the $I$- and $H$-bands that was derived from the $J$- and 
$K$-band analysis.

 For several of the CVs it is clear that 
the abundances of magnesium, sodium, and/or aluminum appeared anomalous. 
In those cases we ran models with the best-fit values for T$_{\rm eff}$ and
[Fe/H], but with small changes in the abundances of the element that seemed
peculiar. We then derived the value for the deficit/enhancement of that 
element from visual inspection. The final results for all of the objects are 
listed in Table \ref{results}. In this table, and in the remainder of the
paper, CV subclasses, orbital periods and inclinations are from the online 
version of the Ritter \& Kolb (2003) catalog (``RKCat''), unless otherwise
noted. In Table \ref{results} we list the object name, the CV subtype, 
P$_{\rm orb}$, T$_{\rm eff}$ of the donor derived from the modeling, its 
metallicity ([Fe/H]), its carbon abundance ([C/Fe]), listing any other 
abundance anomalies in the final column. In several cases, we set the abundance
to a prescribed value before modeling, when this occurs, the value in the 
table is followed by an ``='' sign.

\subsection{The Derived Abundances for CVs with K-type Secondaries Using
Cross-Dispersed SPEX Data}

Using the methods described above, we now present the results from modeling the
program objects. We want to stress again that the presence of H I emission 
throughout the range of these spectra severely limits the $\chi^{2}$ analysis. 
Thus, the precision of the derived metallicities for the CVs are almost
certainly poorer than those derived from the K-type templates. The following 
subsections on the individual CVs are ordered by orbital period. As shown
in Knigge et al. (2011), CVs with orbital periods above 5 hrs are 
expected to have masses of M$_{\rm 2}$ $\sim$ 0.6 M$_{\sun}$ for their
donor stars. The spectral type of a main sequence star with such a mass is 
M0V. Thus, we will start with the assumption that all CVs with P$_{\rm orb}$ 
$>$ 5 hr will probably have K type donors (though our analysis may lead to a
different answer).

\subsubsection{GK Persei}

GK Per is a long period (P$_{\rm orb}$ = 48.1 hr) CV that had a classical
nova eruption in 1901 (see Harrison et al. 2013). The secondary star
in this system is clearly a subgiant. H\&H modeled the $K$-band 
spectrum of GK Per observed with both Keck NIRSPEC and IRTF SPEX, finding 
T$_{\rm eff}$ = 5100 
K, [Fe/H] = $-$0.125, and [C/Fe] = $-$1.0. The cross-dispersed SPEX data are 
shown in Fig. \ref{gkper}. As detailed in Table 3, we find T$_{\rm eff}$ = 
5000 K, [Fe/H] = $-$0.3, and [C/Fe] = $-$0.5. As discussed in H\&H, the lower
S/N of the GK Per data set meant more uncertain results than for their
other two CVs. With the wider spectral coverage, we are better able
to constrain the metallicity and carbon abundance, leading to slightly
revised results (but within the error bars of both results). In addition
to a strong deficit of carbon, we also find that all of the Mg I lines 
(see Fig. 1 for identifications) are weaker than they should be for this 
spectral type. We conclude that the magnesium abundance in the subgiant donor 
is 50\% of the solar value.

\subsubsection{EY Cygni}

EY Cyg is a low inclination dwarf nova of long orbital period, P$_{\rm orb}$
= 11.02 hr (Echevarria et al. 2007). Echevarria et al. found a spectral type
for the secondary of K0V, in agreement with that derived by Kraft (1962).
G\"{a}nsicke et al. (2003) found an anomalous N V/C IV emission line ratio
that suggested a deficit of carbon. Analysis of {\it FUSE} and {\it HST}
spectroscopy by Sion et al. (2004) found a carbon abundance that was 20\%
solar, and a silicon abundance that was 10\% of solar. From the SPEX data
(Fig. \ref{eycyg}) we derive T$_{\rm eff}$ = 5250 K, consistent with a 
spectral type of K0V, and [Fe/H] = 0.0. We confirm the reported carbon 
deficit, finding [C/Fe] = $-$0.5. While the Si I lines in the $J$-band do 
appear to be weaker than expected, the S/N of these data are too poor to 
determine an absolute silicon abundance.

\subsubsection{AE Aquarii}

AE Aqr is an intermediate polar (IP), a system with a highly magnetic
white dwarf that is spinning much more rapidly than the orbital period:
P$_{\rm orb}$ = 9.88 hr, P$_{\rm spin}$ = 0.55 min. Harrison et al.
(2007a) have already published the $IJHK$ spectra from SPEX, and noted
the near-absence of the CO features in the $K$-band. Those data, and the best 
fitting model, are plotted in Fig. \ref{aeaqr}. We derive a temperature of 
T$_{\rm eff}$ = 4750 K ($\sim$ K4), [Fe/H] = 0, and [C/Fe] = $-$1.0. The 
derived temperature is in agreement with the phase-dependent spectral type 
range of K0 ($\phi$ = 0.4, 0.9) to K4 ($\phi$ = 0.65), found by Echevarria et 
al.  (2008). Some of this spectral type range might be due to starspots, as 
Doppler tomography found that $\sim$ 18\% of the northern hemisphere of
the secondary star is covered with spots (Watson et al. 2006). Except
for the deficit of carbon, no other spectroscopic anomalies are evident.

\subsubsection{RU Pegasi}

RU Peg, P$_{\rm orb}$ = 8.99 hr, was modeled by H\&H, and they found
T$_{\rm eff}$ = 4700 K, [Fe/H] = 0, and [C/Fe] = $-$0.75. The SPEX data
are presented in Fig. \ref{rupeg}. The best fitting model has T$_{\rm eff}$ = 
5000 K, [Fe/H] = $-$0.3, and [C/Fe] = $-$0.4. While the global metallicity
in our new result is substantially lower, the absolute carbon deficit
has not changed. Obviously, the values of [Fe/H] and [C/Fe] are intertwined,
and both models suggest a large, total deficit of carbon.
It is not as clear why we now find a hotter star with a lower metallicity 
(though again, within the combined error bars of both results).
As discussed in H\&H (see their Fig. 18), with a limited data set, hotter
models with super-solar abundances are nearly equivalent with cooler models
with sub-solar abundances. Here we have the opposite trend. The increased
wavelength coverage makes the new determination more reliable.

In the following we will use the newly constrained
properties of the secondary stars to estimate the distances to CVs using
spectroscopic parallaxes. When possible, we compare these spectroscopic 
distances to those derived from astrometric parallaxes. Harrison et al. (2004b)
used the Fine Guidance Sensors (FGS) on the $HST$ to measured a parallax for 
RU Peg that leads to a distance of 282 $\pm$ 20 pc. The 2MASS data for RU Peg,
obtained during quiescence, is: ($J - H$) = 0.44, ($H - K$) = 0.16, and
$K$ = 10.46. These colors are consistent with the K4 temperature we derive 
from the spectra. The parallax, however, gives M$_{\rm K}$ = 3.2, whereas a main
sequence K4V has M$_{\rm K}$ = 4.48. The secondary star would have to
be a G4V to explain the observed M$_{\rm K}$! The donor in
RU Peg is three times more luminous than a main sequence star of the same
spectral type.

\subsubsection{SS Cygni}

H\&H modeled the $K$-band data for SS Cyg and found T$_{\rm eff}$ = 4700 K,
[Fe/H] = $-$0.25, and [C/Fe] = $-$0.50. They noted that the Mg I line at
2.28 $\mu$m was weaker than expected, but could not derive an abundance
for magnesium. Modeling the $IJHK$ spectra, Fig. \ref{sscyg}, we find
very similar results: T$_{\rm eff}$ = 4750 K, [Fe/H] = $-$0.3, and [C/Fe] = 
$-$0.40. With the more expansive wavelength coverage, we conclusively find 
that there is a deficit of magnesium, [Mg/Fe] = $-$0.3.

We can use the temperature we found above, and published values that 
characterize the secondary star, to generate an updated spectroscopic parallax.
Bitner et al. (2007) conducted a radial velocity study of SS Cyg and found
that the secondary star was  10 to 50\% larger than a main
sequence dwarf, which implies an absolute magnitude that is brighter
by 0.21 to 0.88 mag. In their paper they assumed a K4.5V star. Here, the
temperature is closer to a K3.5V. For such an object M$_{\rm K}$ $\simeq$
4.40, with their offsets in absolute magnitude, the limits become 3.52 $\leq$ 
M$_{\rm K_{\rm SS}}$ $\leq$ 4.19. With $\langle K \rangle$ = 9.45 (Harrison
et al. 2007b), the spectroscopic distance to SS Cyg is 102 $\leq d \leq$ 153 pc.
Unfortunately, this range includes all of the published astrometric distances 
for this object. The mean, 128 pc, is very similar to the distance that
resulted from the revised FGS analysis\footnote{$\pi$ = 7.41 $\pm$ 0.20 mas. 
Note that this value for the parallax differs from that quoted in the 
abstract of that paper due to further analysis of the data set.} of Harrison 
\& McArthur (2016).  As will be discussed below, evidence suggests that the 
spectral types derived from our spectroscopic observations are inclination 
angle-dependent. For an object with the binary inclination angle of SS Cyg,
$i$ $\sim $ 45$^{\circ}$ (Harrison et al. 2007b), the true spectral type of 
the secondary is likely to be one full subclass earlier, $\sim$ K2.5V. 
Therefore, with 
M$_{\rm K}$ $\simeq$ 4.2, the resulting distance estimate, assuming an 
unbloated main sequence donor, is 112 pc; in agreement with the radio parallax
of Miller-Jones et al. (2013). Given the result for RU Peg, combined with the 
fact that the secondary star in SS Cyg has a subsolar metallicity, and 
deficits of both carbon and magnesium, suggests that the assumption of a main 
sequence luminosity for the donor is suspect.

\subsubsection{RX Andromedae}

RX And has an orbital period of 5.03 hr, and Schreiber et al. (2002)
suggest it has characteristics of both Z Cam dwarf novae (DNe), and VY Scl 
stars. The Z Cam 
subclass of DNe have ``standstills,'' where the system brightens to a level 
well above minimum, but stays a magnitude or more below maximum. VY Scl stars, 
on the other hand, frequently show very low states, where the mass transfer
drops to low levels, and the system fades by a magnitude or more below
its typical minimum light level. VY Scl stars do not show dwarf novae
outbursts. RX And clearly has dwarf novae outbursts, but also extended
standstills and very low states. The shorter orbital period of RX And 
suggests that its donor
star may have a spectral type of M0, or later. We will discuss deriving the
metallicities for M stars in \S 4. The line lists used for modeling
the spectra of K dwarfs with SPECTRUM are not expected to be reliable for M 
dwarfs. However, our main reluctance in modeling M dwarf data using SPECTRUM 
(or MOOG) is the lack of the incorporation of water vapor absorption. 

At M0V, the water vapor absorption features are not very strong, and our 
continuum division process can easily remove them. Thus, we ran a grid of 
models that included T$_{\rm eff}$ = 3750 K to attempt to constrain the
nature of the secondary star in RX And. To demonstrate how well the line list
works for this lower temperature, in Fig. 7 we also plot
our model spectra on top that of the IRTF spectral template HD 19305.
There are no published values for the metallicity of HD 19305, however, using
the techniques described below (\S 4), we derived [Fe/H] = 0.0,
and used this value in generating model spectra. The resulting synthetic 
spectrum does a reasonable job of matching HD 19305 in the $I$, $J$, and 
$K$-bands, though nearly all of the lines from Ca I are poorly fit by the 
model spectrum in these bandpasses. The model $H$-band spectrum does a very 
poor job at reproducing the observations. This is presumably due to 
ignoring molecules such as H$_{\rm 2}$O.

We found that the best fit for RX And was achieved with the T$_{\rm eff}$ = 
3750 K model, and [Fe/H] = $-$0.3, but the donor star has several abundance 
anomalies. While the carbon deficit was mild, [C/Fe/] = $-$0.1, the abundances
of both Mg and Al were significantly diminished (e.g., Mg I at 1.173 and 1.771
$\mu$m; the Al I doublets 1.314 and 2.11 $\mu$m). The donor star in RX And 
could have a normal carbon abundance if the effective temperature was slightly
higher. In agreement with the results of Sepinsky et al. (2002) who found a 
solar abundance for carbon. Sepinsky et al. also found a normal abundance for 
silicon, and the Si I lines in the $J$-band for RX And appear to be at their 
proper strength. The Mg I and Al I lines were reproduced for HD 19305 by the 
synthetic spectrum. The strengths of both of these species decline with 
decreasing temperature, but this would require a much cooler donor star than
M0 to explain their strength here.

Using the techniques described in \S4, the metallicity analysis procedure
suggests that the donor in RX And has a spectral type of M2, and [Fe/H] = 
$+$0.07.  This is a later spectral type than would be expected at this orbital
period. As shown in Table \ref{obslog}, the secondary was close to
superior conjunction, thus any irradiation would have made it appear hotter
than it would on average. Obviously, synthetic spectra using models with 
cooler effective temperatures are necessary to quantify the properties of the 
donor in RX And. A cooler star would have weaker  Mg I and Al I lines,
so their deficiencies can thus be explained.

It is also important to note the strong Ca II triplet emission in the 
$I$-band. These three lines are narrower than the nearby H I Paschen series, 
suggesting they have an origin on the secondary star. We will see further 
evidence of Ca II emission below. Such lines could be used to generate a 
radial velocity curve for the donor objects in systems where the
absorption features cannot be easily seen.

\subsection{Abundance Analysis for CVs with Expected K Dwarf Donors Using
$K$-band Data Only}

Over the years, we have collected $K$-band spectra for a large number of
CVs. Most of these data were presented in Harrison et al. (2004a, 2005ab).
The spectral types and carbon abundances were qualitatively examined in
those papers. Here we have generated models for those objects. The results
for these objects are listed in Table \ref{results}; we briefly discuss each 
object below. As noted earlier, having only a single bandpass of moderate 
resolution data leads to poorer constraints on both T$_{\rm eff}$ and
[Fe/H]. However, the results for some objects indicate very low values for the
metallicities, and argue for deeper, multi-bandpass infrared spectroscopy.
Some of the data analyzed below have low S/N. Where the $\chi^{\rm 2}$ 
analysis is ambiguous, we have chosen the {\it least} unusual model as
the final result. 

\subsubsection{SY Cancri}

SY Cnc is a long period dwarf nova of the Z Cam subclass. The $K$-band
spectrum was presented in Harrison et al. (2004a), where a spectral type
of G2 was derived. Connon Smith et al. (2005) obtained a radial velocity
curve for this object and find a low mass for the secondary star (M$_{\rm 2}$
= 0.36 M$_{\sun}$), and conclude that its spectral type is near M0. While
the $K$-band data are of low quality, the strength of the blue Ca I features,
Na I, Ca I, and CO are most consistent with T$_{\rm eff}$ = 5500 K, [Fe/H] =
0.0, and [C/Fe] = 0.0. The temperature we derive corresponds to a spectral 
type of G8V, and is more consistent with the type of donor expected at this 
orbital period (Knigge 2006). Note that SY Cnc was in outburst 
at the time of this spectrum, and irradiation might result in finding a hotter
temperature for the 
secondary star. The visibility of the donor in $K$-band spectroscopy 
during outburst suggests that an excellent radial velocity curve and 
$v$sin$i$ rotation rate could be obtained for this bright object ($K$ = 10.9) 
using NIR spectroscopy. Such data are needed to better determine the masses 
of both components in this system.

\subsubsection{CH Ursae Majoris}

Harrison et al. (2004a) presented the $K$-band spectrum of CH UMa and derived
a spectral type of K7 $\pm$ 2, consistent with the M0 type found by Friend et 
al.  (1990). This is cooler than expected for this orbital period 
(P$_{\rm orb}$ = 8.232 hr). UV spectroscopy reveals that CH UMa has a large 
N V/C IV ratio, suggesting a deficit of carbon (G\"{a}nsicke et al. 2003). 
Sion et al. (2007) found that their best fit white dwarf model for the $FUSE$
spectrum of CH UMa indicated very low abundances for both carbon and silicon. 
The $\chi^{\rm 2}$ analysis indicates that the best fit temperature is near
T$_{\rm eff}$ = 4750 K. Such a value would be closer to that expected
for this orbital period. While the donor star appears to have a near-solar
metallicity, the CO features in the $K$-band spectrum are exceptionally weak, 
and our analysis finds [C/Fe] = $-$1.0. The Si I lines in the $K$-band 
spectrum are not prominent, even at solar metallicities for these 
temperatures, but they do not appear to be unusual in the spectrum of CH UMa.

\subsubsection{V1309 Orionis}

In Harrison et al. (2005a) we presented the $K$-band spectra of five polars
with M-type companions, and found that they all appeared to have normal 
carbon abundances (see below). Since that time, we obtained a $K$-band spectrum
of the very long period polar V1309 Ori (P$_{\rm orb}$ = 7.98 hr) using
NIRSPEC on Keck (Howell et al. 2010). We have generated a grid of synthetic 
model spectra and find that, assuming a solar metallicity, the temperature
is between 4750 K $\leq$ T$_{\rm eff}$ $\leq$ 5000 K. Hotter temperatures
do not have the strong NaI and Ca I features seen here, while cooler 
temperatures would require reduced metallicities to reproduce the strength
of these lines. Our result for T$_{\rm eff}$ is similar to the late K-type 
derived by Howell et al.  (2010), but is in contrast to Garnavich et al. 
(1994) who suggested an M0V. 

Given the poor quality and limited spectral range
of the data for V1309 Ori, we cannot fully exclude cooler donor stars. CO is 
clearly not in emission in this object, nor are their any signs of cyclotron 
emission (see Harrison \& Campbell 2015). We conclude that carbon 
is highly deficient no matter the spectral type. For T$_{\rm eff}$ = 4750 K, 
we derive  [C/Fe] = $-$1.0. Mg I at 2.28 $\mu$m is also very weak. The 
profile of the Mg I line at 2.1 $\mu$m is corrupted by He I emission. To 
reproduce the strength of the 2.28 $\mu$m line implies that [Mg/Fe] $\sim$ 
$-$0.50. Szkody \& Silber (1996) found that the N V/C IV ratio was large,
suggesting a carbon deficit. Thus, there is at least one polar whose secondary star does not 
appear to be ``normal.'' It is interesting that this deviation occurs in 
the longest period polar.

\subsubsection{EM Cygni}

EM Cyg is an eclipsing Z Cam dwarf novae that has an orbital period of 6.982 
hr. Thus the donor star should have a mid-K spectral type. Harrison et al. 
(2004a) had a difficult time classifying its $K$-band spectrum, finding that 
some features suggested a K0V, while others were more consistent with a K5V. 
This ambiguity can be understood as the best fitting family of models from our 
analysis have both sub-solar metallicities, and modest deficits of carbon. 
Godon et al. (2009) have modeled $FUSE$ data for EM Cyg and found that 
nitrogen, sulphur and silicon are highly overabundant, while carbon
{\it appeared} to have a solar abundance. Unfortunately, our spectrum of
EM Cyg is too noisy to examine the strength of the Si I lines, though
absorption features are present at the locations of each of the strongest
such lines.

\subsubsection{V426 Ophiuchi}

V426 Oph is a bright, Z Cam-type dwarf nova that was observed by Harrison
et al. (2004a). They found an early K spectral type, consistent with the
K3 type derived by Hessman (1988). Evidence is mounting that V426 Oph
is actually an IP (Ramsay et al. 2008, and references therein). The favored
synthetic model spectrum has T$_{\rm eff}$ = 4750 K, [Fe/H] = 0.0, and with
[C/Fe] = $-$0.5, producing an excellent fit to the SPEX data.

\subsubsection{TT Crateris}

$K$-band spectra for TT Crt have been presented in Harrison et al. (2004a)
where a spectral type of K5V was found. Szkody et al. (1992) and Thorstensen 
et al. (2004a) found that the donor star in TT Crt (P$_{\rm orb}$ = 6.44 hr) is 
easily seen in visual spectroscopy, and appears to be a mid to late K-type 
star. Our best fitting model agrees with these three estimates, having T = 
4500 K, and [Fe/H] = 0.0. The CO features are weaker than they should be at 
this spectral type, and we derive a carbon abundance of [C/Fe] = $-$0.50.

\subsubsection{AH Herculis}

AH Her is a DNe with a period of 6.19 hr, and would be expected to have a 
mid/late-K type secondary. The $K$-band data for AH Her was obtained with SPEX
and, as described in Harrison et al. (2004a), this DNe was going into outburst 
at the time of these observations. The data 
quality is poor, but the two best fitting families of models have either 
T$_{\rm eff}$ = 5500 K, [Fe/H] = 0, and [C/Fe] = 0, or T$_{\rm eff}$ = 
4750 K, [Fe/H] = $-$0.75, and [C/Fe] = $-$0.4. The hotter model matches most 
of the spectrum except for the Ca I lines near 1.98 $\mu$m. Given the orbital 
period, the lower temperature for the donor would be expected, and indicates 
a low metallicity secondary with a significant carbon deficit. Our result may 
have been compromised by irradiation and/or dilution, though the system was 
near inferior conjunction.

\subsubsection{CZ Orionis}

CZ Ori is a DNe with a period of 5.25 hr. Ringwald et al. (1994) estimated 
that the secondary star has a spectral type of M2.5 $\pm$ 1 from the presence 
of TiO absorption. While the $K$-band spectrum of CZ Ori is poor (see Fig.
\ref{czori}), it is clear 
that the CO features have an unusual profile, suggesting CO emission is 
present. Using the technique described in H\&H, we added CO emission spectra 
to the models generated by SPECTRUM and find that a donor star with T = 4250 K,
[Fe/H] = $-$1.0 and [C/Fe] = 0.0, combined with a CO emission spectrum 
broadened to a velocity of 800 km s$^{\rm -1}$, reproduces the observations. 
This is only the second CV shown to have CO emission features, the other being
the ultra-short period system WZ Sge (Howell et al. 2004, Harrison 2016); 
several additional examples will be discussed below. A temperature of 4250 K 
corresponds to a spectral type near K7. Hotter stars fit nearly equally as 
well, and 
have larger metallicities (e.g., [Fe/H] = $-$0.75 for T$_{\rm eff}$ = 4750 K).
Cooler stars require much lower metallicities to fit these data, and also 
need to have carbon deficits. If the metallicity was truly as low as 
indicated here, the TiO features would probably have been too weak to have 
extracted a spectral type from optical data. Higher quality, multi-bandpass 
infrared spectroscopy of this object is needed.

\subsubsection{EX Draconis}

EX Dra has P$_{\rm orb}$ = 5.04 hr, and Baptista et al. (2000) found
M$_{\rm 2}$ = 0.54 M$_{\sun}$. This mass for the secondary would suggest
a mid-M type star, while the expected spectral type for this period would be 
near the K/M boundary. Harrison et al. (2004a) derived a spectral type of
K7 from the SPEX data. They noted that the CO features and the Mg I lines were 
weaker than expected at this spectral type. Our analysis finds a best fitting
model of T$_{\rm eff}$ = 4000 K, [Fe/H] = $-$0.5, [C/Fe] = $-$0.2,
and [Mg/Fe] = 0.0. Except for the carbon deficit, there are no obvious
abundance anomalies.

\section{Deriving Metallicities for CVs with M-type Donor Stars}

In our infrared spectroscopic surveys of shorter period CVs, we presented
evidence that the carbon abundances in many of the secondaries with M spectral
types appeared deficient. The main issue with such analysis was that the
template stars to which we were comparing the CV spectra had known
spectral types, but unknown metallicities. Thus, if the spectral type of the 
CV donor was somewhat uncertain, comparing it to the spectra of several 
template stars might lead to erroneous conclusions due to the possibility
that the template stars might have significantly non-solar values for [Fe/H].
A proper qualitative analysis requires us to be able to compare stars of
known properties to those that we seek to characterize.

While models for the structure of the atmospheres of M dwarfs have improved 
(e.g., Baraffe et al. 2015), it is still difficult to generate synthetic
spectra that are adequate for abundance measures due to incomplete line
lists, and the presence of water vapor features throughout the NIR (however,
see \"{O}nehag et al. 2012, and Bean et al. 2006). Thus, alternative methods 
have been sought that calibrate various spectral features for M dwarfs with 
known metallicities determined using their association with hotter stars, for 
which such derivations are simpler (see Rojas-Ayala et al. 2010, Mann et al. 
2013, 2014, and Newton et al. 2014). Mann et al. (2013, 2014) have performed a 
rigorous analysis and developed relationships to quantify the value of [Fe/H] 
for dwarfs spanning the spectral type range K5 to M9. These relationships use 
the equivalent widths of the Na I doublet at 2.20 $\mu$m, and the Ca I triplet
at 2.26 $\mu$m, along with an ``H2O index'' (from Covey et al. 2010)
to derive [Fe/H] from $K$-band spectra. The H2O index gives
the temperature dependence, and thus an estimate of the spectral type.

In Mann et al. (2013) the relationship for determining [Fe/H] using these 
three features for earlier spectral types ($<$ M4) is given, while in Mann 
et al. (2014), a different formulation is found that covers the range M4 to M9.
We can now derive [Fe/H] for the various template M dwarfs that can
used for comparison to the CV spectra, as well as attempt to derive [Fe/H]
directly for the donor stars. Note that due to the weakness of the CO
features in some of the CVs, the value of the H2O index could
be erroneous (since its reddest continuum bandpass is at 2.37 $\mu $m). In
these cases, we can use the calibration by Newton et al. for [Fe/H] that only 
uses the Na I doublet. That relationship is only valid for spectral types
from M2 to M6. The 1$\sigma$ error bars on [Fe/H] for the Newton et al. (2014)
relation is $\pm$ 0.12 dex, while Mann et al. (2013, 2014) claim an error
of $\pm$ 0.07 dex for their calibration.

There is one vitally important caveat that could dramatically alter our 
metallicity estimates using these techniques, and that is the dependence of the
derived value of [Fe/H] on the abundances of carbon and oxygen. Veyette
et al. (2016) have used PHOENIX models of cool stars to attempt to probe
how the derived values of [Fe/H] using the above techniques are dependent on 
the C/O abundance ratio.
Less carbon in the atmosphere means more H$_{\rm 2}$O, leading to
changes in the continuum of the cooler M dwarfs. As an example, for an object
with [C/Fe] = $-$0.2 and [O/Fe] = 0.0, the derived abundance for a solar
metallicity star, using the Mann et al. (2013) relations with T$_{\rm eff}$ = 
3500 K, would be [Fe/H] = $-$0.15. For T$_{\rm eff}$ = 3000 K, it would 
be [Fe/H] = $-$0.4. Thus, if the carbon abundance is low, we could dramatically
underestimate the true metallicity of the secondary star. We will return
to this subject below.

With the ability to constrain [Fe/H] for the templates and donor stars, we
can fully investigate whether the lines of any other elements are at unusual
strengths in the CV secondaries. Like the preceding section, we first begin with
the analysis of cross-dispersed spectra for the individual program objects
ordered by orbital period. We then examine the objects with only $K$-band
data. Unless otherwise indicated, we have used visual inspection to
determine the best fitting spectral type and to note any abundance anomalies.
The results are listed in Table \ref{results}. The value of T$_{\rm eff}$
listed in this table uses the calibration in Rajpurohit et al. (2013) to turn 
estimated spectra types into temperatures. That calibration assumes solar
metallicity.

\subsection{Cross-Dispersed Spectroscopy}

\subsubsection{SS Aurigae}

SS Aur is a typical DNe with an orbital period of 4.39 hr, and Harrison
et al. (2005b) found a spectral type of M4 for the donor star. Using the
Mann et al. relationship gives [Fe/H] = 0.03, but the H2O index 
suggests a spectral type near M1. Thus, the value for [Fe/H] might be 
influenced by a noisy continuum at the red end of the $K$-band. Using the 
Newton et al. calibration, we derive [Fe/H] = $-$0.05. The donor star in
SS Aur appears to have a solar metallicity. In Fig. \ref{ssaur} we present
the SPEX data for SS Aur, and compare it to Gl 388, for which the 
metallicity calculation gives [Fe/H] = $+$0.18, and indicates a spectral type
of M3.4. Except for the $I$-band, where the continuum of SS Aur is poorly
defined, the data presented in this figure for both objects have been 
continuum divided. The CO region of the spectrum for SS Aur has low S/N, but 
suggests that the features are similar in strength to that of the M3V 
template. 

The normality of the CO features is confirmed with the $K$-band spectrum of SS
Aur obtained with NIRSPEC (Fig. \ref{ssaurkeck}). Except for the slightly bluer 
continuum of SS Aur, the M3V template is an excellent match for SS Aur.
The metallicity analysis using the NIRSPEC data gives [Fe/H] = $-$0.13,
and a spectral type of M2.7. It is clear that SS Aur was in a higher state of 
activity when observed with NIRSPEC than with SPEX, as the He I line at 2.06 
$\mu$m is much stronger, and the donor star absorption features at 2.1 $\mu$m 
are corrupted by He I emission in the NIRSPEC data. The AAVSO database 
indicates that SS Aur went into outburst six days after the Keck observations. 
For the SPEX observations, SS Aur had just returned to quiescence from an 
outburst that had its maximum three weeks prior.

The Ca I lines near 1.98 $\mu$m appear to be very weak in the SPEX data
for SS Aur, but the telluric correction is almost certainly responsible for 
this deviation. The Mg I features at 1.5 and 1.7 $\mu$m also appear to be 
weaker in SS Aur than the template, but the depth of those lines is altered
by emission lines of He I. Potassium, aluminum, and manganese appear to have 
solar abundances.  Note that the Ca II triplet in the $I$-band is in
emission, and superposed on the TiO bandhead at 8432 \AA. We conclude
that the donor star in SS Aur is completely consistent with a near-solar 
metallicity dwarf of spectral type M3.

Given the normality of the donor star in SS Aur, we can estimate a distance
to this system. With $K$ = 12.00, and M$_{K_{\rm M3}}$ = 6.5, $d$
 = 126 pc. The $HST$ parallax gave $d$ = 167 $\pm$ 9 pc. The spectroscopic 
parallax is not in close agreement with the astrometric parallax. For that
to occur, requires that the secondary in SS Aur to actually have the 
luminosity of an M2V (the spectroscopic distance is then $d$ = 155 pc). We 
will encounter a similar issue for U Gem. 

\subsubsection{U Geminorum}

U Gem has an orbital period that is very similar to SS Aur, but
we will find that it has a very different secondary star. U Gem was observed
using TripleSpec at APO over a time interval of 8.3 hrs, with a 30 minute
break two thirds of the way through to obtain telluric standard data, as
well as observe the M-type star GJ 402. Thus, our data set spanned almost two
complete orbital periods. The Triplespectool routine uses night sky emission
lines on the exposed frames for wavelength calibration. We determined a 
wavelength solution for each pair of spectra, resulting in a very precise 
calibration of the data. 

The TripleSpec data for U Gem are presented in Fig. \ref{ugemspec}. As
first noted by Harrison et al. (2000), the CO features are extremely weak.
The metallicity analysis leads to [Fe/H]  =  $+$0.05, and a spectral type
of M3.9. Given the weak CO features, we use the Newton et al. relationship to
find [Fe/H]  =  $+$0.20. In Fig. \ref{ugemspec} we also plot the data
for GJ 402 obtained on the same night, with the same instrument. Published
spectral types for this star range from M3V to M5V. The metallicity analysis
gives a spectral type of M3.8, and [Fe/H]  =  $+$0.27. Thus, it should
provide an excellent template for U Gem. As shown in Fig. \ref{ugemspec}, the 
match between U Gem and GJ 402 in the wavelength interval 0.94 $\leq \lambda 
\leq$ 1.34 $\mu$m is very good. The only issue in this bandpass is the poor 
S/N of the spectrum of GJ 402 through the telluric feature at 1.112 $\mu$m. The 
FeH feature at 0.99 $\mu$m shows that they both have similar metallicities. The
Mg I, K I, and Al I features all overlap. 

The two spectra do not match as well in the $H$-band, with the middle sections
of the continuum of U Gem well above that of GJ 402. Naively, one would 
assume this deviation is due to H I emission from the Brackett series, since 
the Doppler correction for the secondary's motion makes the H I lines appear
very broad (FWHM $\sim$ 1400 km s$^{\rm -1}$); but this explanation is 
inadequate. First, the influence of the two strongest lines of the Bracket 
series in the $H$-band, Br10 ($\lambda$1.74 $\mu$m) and Br11 ($\lambda$1.68 
$\mu$m), on the spectrum of U Gem can just be discerned, as they fill-in some 
of the absorption features of the M dwarf. They would have to be extremely 
broad (and much stronger) to create an excess in the continuum between these 
two wavelengths. This would also require broader H I emission lines in the
$H$-band than seen in either the $J$- or $K$-bands. Secondly, adding in a 
strong Brackett continuum at $\lambda <$ 1.6 $\mu$m should raise the blue end 
of the $H$-band above that of an M4V, and this is not seen. The excess in the 
$H$-band is not due to H I emission.

Suspiciously, the excess in the $H$-band is centered near 1.65 $\mu$m, the 
location of the peak of the H$^{-}$ opacity minimum. As discussed in Wing 
\& J\o rgensen (2003), and displayed in Fig. \ref{giantdwarf}, lower gravity 
giants have larger H$^{-}$ opacity ``bumps'' in the $H$-band, as compared to 
higher gravity dwarfs. In Fig.
\ref{ugemHcomp}, we plot the $H$-band spectra of U Gem, GJ 402, and that of
HD4408, an M4III. The excess due to H$^{-}$ is obvious in the red giant
spectrum. The strong and broad absorption due to water vapor seen in the
dwarfs at the red end of the $H$-band, is weaker in the red giant. The
water vapor features in red giants are much weaker than those of dwarfs with 
the same effective temperature. Thus, the most 
obvious explanation for the peculiarity of the $H$-band spectrum is that the 
secondary of U Gem has an {\it apparent} gravity that is much lower than that 
of GJ 402.

The Na I doublet in the $K$-band was easily seen 
in the individual spectra, and we have used IRAF to individually measure the 
positions of each of the lines of the doublet to allow us to construct the 
radial velocity curve shown in Fig. \ref{ugemrv}. $\chi^{\rm 2}$ analysis of 
sinewaves with various amplitudes fit to these data finds a semi-amplitude of 
K$_{\rm 2}$ = 310 $\pm$ 10 km s$^{\rm -1}$. This result agrees with the value 
of 309 $\pm$ 3 km s$^{\rm -1}$ found by Friend et al. (1990), and K$_{\rm 2}$ 
= 302 km s$^{\rm -1}$ found by Naylor et al. (2005). 

Naylor et al. attempted to determine the $v$sin$i$ rotation rate of the 
secondary, 
and depending on the template they used, found values from 96 km s$^{\rm -1}$
(with an M5.5V), to 142 km s$^{\rm -1}$ (with an M3V). Using the M4V template 
GJ 402 for comparison, and working in the $J$-band (5$^{\rm th}$ order, a 
dispersion of $\Delta \lambda$ = 1.7 \AA), we find a best fit value for 
$v$sin$i$ = 155 $\pm$ 20 km s$^{\rm -1}$. This large value may result from the 
fact that GJ 402 has a slightly higher metallicity than U Gem. The 
spectral features for a higher metallicity object are going to be slightly 
deeper, and thus would require a larger velocity convolution to best match the
observed spectrum (see Shahbaz 1998). To test this, we used SPECTRUM to 
generate synthetic
$J$-band spectra using T$_{\rm eff}$ = 3,200 K, log$g$ = 4.5, and MARCS models 
with [Fe/H] = 0.0 and $-$0.50. Given the uncertainties in modeling M dwarf
atmospheres we limited our analysis to 1.24 $\leq \lambda \leq$ 1.26 $\mu$m
(essentially the two K I lines). A model generated using the parameters of
GJ 402 reproduced these two lines.  The best fit ($\chi^{\rm 2}$) value 
for the rotation velocity of U Gem using the solar metallicity model has 
$v$sin$i$ = 145 km s$^{\rm -1}$, while the sub-solar metallicity model gives
$v$sin$i$ = 120 km s$^{\rm -1}$.

It is obviously critical to have the appropriate metallicity template when 
using this technique to estimate $v$sin$i$. Thus we note that, except 
for Gl 51 and Gl 65A, the late-type templates used by Naylor et al. for their 
$v$sin$i$ derivation have sub-solar metallicities. Gl 51 and 
Gl 65A are slightly cooler (T$_{\rm eff}$ = 3030 K, $\simeq$ M5V) than the 
donor in U Gem. The closest temperature match is GJ 213, but it has the lowest 
metallicity of the sample ([Fe/H] = $-$0.22, Mann et al. 2015). Convolving
the spectrum of GJ 463 gave their highest $v$sin$i$ estimate, but there are 
no measurements of [Fe/H] for this star (though it has a similar temperature 
to U Gem, T$_{\rm eff}$ = 3300 K, Houdebine \& Mullan 2015). Given the 
uncertainties in this process, our result for $v$sin$i$ is certainly 
consistent with the conclusions of Naylor et al. This confirms their result 
that U Gem has a massive white dwarf, M$_{\rm 1}$ $\simeq$ 1.2 M$_{\sun}$.
The mass of the secondary that they derive, M$_{\rm 2}$ = 0.44 M$_{\sun}$, is 
similar to those of main sequence M2/M3 dwarfs.

Given the parameters of the solution found by Naylor et al. (2005), the surface
gravity of the secondary in U Gem would be log$g$ = 4.8. A value similar to 
those of normal M dwarfs (cf., Fernandez et al. 2009). Our larger values of 
K$_{\rm 2}$ and $v$sin$i$ result in M$_{\rm 2}$ = 0.42 M$_{\sun}$, and an 
equatorial radius of R$_{\rm 2}$ = 0.58 R$_{\sun}$, leading to
log$g$ = 4.5. This would imply a gravity at the equator that is one half that 
of a typical M dwarf. Applying the relations in Warner (1995) to U Gem, the 
equatorial radius is 1.64 times larger than the volume radius ``R$_{\rm L}$'' 
of the Roche lobe (the radius of the equivalent sphere). Thus, using our 
result, the surface gravity of the equivalent sphere is then log$g$ = 5.0. 
Since U Gem has a large inclination ($i$ = 69$^{\circ}$), the equatorial 
regions of the secondary will dominate the spectrum. Since these regions are 
further away from the center of the star, they will be cooler, leading to a 
cooler observed spectral type. This is, of course, the effect of ``gravity 
darkening'' found for rotating stars, first quantified by von Zeipel (1924). We 
explore whether we can reconcile the results for U Gem with the expected levels
of gravity darkening in \S6.2.

 This inference would resolve the paradox 
from the $HST$ parallax measurement, where the luminosity of the secondary star
was equivalent to an M2V, but the photometry and spectroscopy indicated an M4V 
(Harrison et al. 2000). It also explains the ``too cool for comfort'' finding 
of Friend et al (1990), where the observed spectral type was later than 
expected if the donor was to obey the main sequence mass-radius relationship.
A more robust value of $v$sin$i$ for U Gem is required to confirm this result.

To attempt to quantify the carbon deficit, we have continuum divided
the $K$-band spectra of both U Gem and GJ 402 to remove the water vapor
features. In the process, we also remove much of the broader absorption
due to CO, and thus these features will also be weakened by this
process, resulting in an artificial deficit in carbon. We have used {\it kmoog} 
to generate model spectra to determine a {\it relative value} for the carbon 
deficit between U Gem and GJ 402. As noted above, the line lists and atmosphere
models for M dwarfs are poorly known, so the following process has considerable 
uncertainty. {\it kmoog} was run with T$_{\rm eff}$ = 3200 K, [Fe/H] = 0, and 
$g$ = 4.5, using a MARCS atmosphere model, over a wide range of carbon 
abundances (0.0 $\leq$ [C/Fe] $\leq$ $-$2.5). The best fit results are shown 
in Fig. \ref{ugemKcomp}. The Na I doublet and Ca I triplet in the synthetic 
spectra do a reasonable job fitting the observations. We find that we need to 
artificially have a carbon abundance of [C/Fe] = $-$0.5 to match the features 
in GJ 402, while U Gem has to have [C/Fe] = $-$1.5. Thus, we estimate that the 
C abundance for the donor star in U Gem is $\sim$ 10\% of solar, in agreement 
with the value derived from UV spectroscopy of the white dwarf primary
(Long \& Gilliland 1999).

One additional insight from this modeling process are the strengths of the 
$^{\rm 13}$CO features in U Gem. Harrison et al. (2005b) commented on the
ease at which the $^{\rm 13}$CO$_{\rm (2,0)}$ ($\lambda$2.345 $\mu$m) feature 
was seen in their SPEX data for U Gem, given the weakness of the $^{\rm 12}$CO 
features. In Fig. \ref{ugemKcomp}, we identify the locations of the 
$^{\rm 12}$CO bandheads with solid lines, and the $^{\rm 13}$CO bandheads with 
dashed lines. At the normal isotopic ratio of $^{\rm 12}$CO/$^{\rm 13}$CO = 89
used in the GJ 402 model, only the $^{\rm 13}$CO$_{\rm (2,0)}$ bandhead at 2.345
$\mu$m is visible. In U Gem, this feature is of similar strength to the
$^{\rm 12}$CO$_{\rm (2,0)}$ bandhead at 2.294 $\mu$m! The 
$^{\rm 13}$CO$_{\rm (3,1)}$ bandhead at 2.374 $\mu$m falls at the position
of a strong telluric feature that is always difficult to remove. The
next $^{\rm 13}$CO bandhead at 2.403 $\mu$m ($^{\rm 13}$CO$_{\rm (4,2)}$),
however, falls in a cleaner region, and is clearly present in U Gem.
In our synthetic spectrum for U Gem we used an isotopic ratio of 
$^{\rm 12}$CO/$^{\rm 13}$CO = 1. Such a ratio is beyond the maximum ratio (4)
that can be achieved by the CNO cycle. With the uncertainties inherent to 
modeling M dwarf spectra, a more precise value will require better 
synthetic data. We conclude that $^{\rm 13}$C is highly 
overabundant in the donor star of U Gem.

Given the caveat noted earlier, assuming the results by Veyette et al. (2016)
are correct, the actual value for [Fe/H] for U Gem is much higher than derived 
from the metallicity relationships due to the lack of carbon in its spectrum.
If we assume the O abundance is normal in U Gem (leading to C/O = 0.055), then
we have underestimated [Fe/H] by $\sim$ $-$0.3. If this offset is true,
U Gem has a very high, super-solar metallicity of [Fe/H] $\simeq$ 0.35,
bringing it into closer agreement with GJ 402.

\subsubsection{WW Ceti}

The long-term AAVSO light curve of WW Cet is unusual, showing frequent 
DNe outbursts, periods of intermediate brightness that suggest Z Cam-like 
behavior, but also low states like VY Scl. Ringwald et al. (1996) find that 
the behavior of WW Cet is not consistent with a Z Cam classification, nor are 
its low states like the VY Scl stars. The orbital period
of WW Cet, P$_{\rm orb}$ = 4.22 hr, is just a bit shorter than that of SS Aur 
and U Gem. Thus, the odd behavior of WW Cet is puzzling. The SPEX data
for WW Cet are presented in Fig. \ref{wwcet}. The metallicity relationship
gives a spectral type of $\sim$M0V, and [Fe/H] = $-$0.17, suggesting that noise
at the red end of the $K$-band might be causing issues with the H$_{\rm 2}$O
index. However, input of the same H$_{\rm 2}$O index into the metallicity
relationship as that for U Gem leads to an identical value of [Fe/H]. Using 
only the Na I equivalent width we derive [Fe/H] = $-$0.34. It appears certain 
that the secondary in WW Cet has a subsolar metallicity.

The IRTF Spectral Library fortunately has spectra for a number of sub-solar
metallicity M dwarfs in the spectral range M2V to M4V. We compared the
spectra of each of these to WW Cet, and find that the best fit results
from the M2.5V dwarf Gl 381. The continuum of WW Cet is slightly bluer
than this star, with the excess increasing to the blue, suggesting the
presence of emission from the white dwarf $+$ accretion disk. The metallicity 
relationship gives [Fe/H] = $-$0.06 for Gl 381, while Gaidos et al. (2014) 
find [Fe/H] = $-$0.07. All of the atomic absorption features in the template
are stronger than their counterparts in WW Cet. This would be expected
given the lower metallicity computed for WW Cet. Even accounting for
the difference in metallicity, the CO features in WW Cet are weaker than
the template (note the excess at the red end of the $K$-band). Sion et al. 
(2006) found that the carbon abundance of the white dwarf was 10\% solar, and 
silicon was 50\% solar. Clearly, the deficit of carbon is not as extreme in 
WW Cet as it is in U Gem. Building on the experience gained above, we estimate 
that the carbon abundance of the donor in WW Cet is [C/Fe] $\geq$ $-$0.3.

\subsubsection{LS Pegasi}

LS Peg, P$_{\rm orb}$ = 4.19 hr, had previously been classified as an SW Sex 
star (Taylor et al. 1999). Recently, however, it has 
been re-classified as an IP by Ramsay et al. (2008). They believe that the 
lack of an X-ray beat period, a phenomenon common to IPs, is due to the fact 
that the magnetic and spin axis of the white dwarf are aligned. The
cross-dispersed spectrum of LS Peg is shown in Fig. \ref{lspeg}. Except,
perhaps, for the Na I doublet at 0.82 $\mu$m, no absorption features from
the secondary are evident. The spectrum is dominated by H I and He I emission.
In addition, there are strong CO emission features. He II at
2.18 $\mu$m appears to be present, and a small peak on the red wing of
H I Br11 might be [Fe II]. It is also clear that Ca II is in
emission, as two of the members of the high order H I Paschen series (Pa13
and Pa15) are elevated above their expected levels. Of the eight IPs for
which we have obtained NIR spectra, none of them have had CO in emission 
(Harrison et al. 2007a, 2010). It is now apparent, however, that Ca II 
emission was present in the spectrum of V603 Aql (Fig. 4 in Harrison et al.
2007a). The same set of unidentified emission lines in the spectrum of LS Peg
(see Fig. \ref{lspeg}), were also present in the spectrum of V603 Aql. 

\subsubsection{KT Persei}

KT Per is a Z Cam-type dwarf nova with a period of 3.90 hr (Thorstensen \&
Ringwald 1997). They derived a spectral type of M3 for the secondary star. The 
IRTF data for KT Per are presented in Fig. \ref{ktper}. 
The shorter orbital period, as compared to U Gem and SS Aur, would lead to the 
expectation of a mid-M spectral type for the donor star. The metallicity 
analysis gives [Fe/H] = $-$0.53, and a spectral type of M1. The Na I relation 
gives [Fe/H] = $-$0.96, though if the spectral type of the secondary was truly 
an M1V, this could possibly be erroneous (see Newton et al. 2014). 
Unfortunately, the IRTF Spectral Library does not have a template with similar 
properties. The closest match is Gl 806 for which the metallicity relation 
gives [Fe/H] = $-$0.08, and a spectral type of M2.2. Mann et al. (2015) lists 
Gl 806 as an M2.0 with [Fe/H] = $-$0.15. It is clear that the continuum of Gl 
806 closely matches that of KT Per in the $K$-band. At shorter wavelengths,
there is an increasing contribution from the white dwarf $+$ accretion
disk. It is obvious that {\it all} of the spectral absorption features in the
spectrum of KT Per are much weaker than those in Gl 806. This includes the 
water vapor features, confirming the very low metallicity estimate.
Given this result, the CO features then appear to be normal. The Ca II 
triplet lines in the $I$-band are in absorption for the donor star.

Given the low level of contamination in the $K$-band, it is possible to estimate
a distance to KT Per if we assume a normal M2V. The 2MASS catalog has
$K$ = 12.63. To insure that this measurement reflects true quiescence,
we consulted the $NEOWISE$ single exposure source 
catalog\footnote{http://irsa.ipac.caltech.edu/cgi-bin/Gator/nph-scan?projshort=WISE\&mission=irsa}. There are 83 measurements for KT Per in this database, and 
they show that the faintest it gets is W1(3.6 $\mu$m) = 12.25. For 
a normal M2V, ($K$ $-$ W1) = 0.15 (Pecaut \& Mamajek 2013). KT Per is slightly 
redder than expected, but clearly the 2MASS value must be close to quiescence. 
Assuming M$_{K}$ = 6.05, gives $d$ = 207 pc, just within the error bars on the 
distance derived from its parallax: $d$ = 180$^{+36}_{-28}$ pc (Thorstensen et 
al.  2008). Thorstensen et al. note that there is a common proper motion
companion to KT Per which has a mid-K spectral type. Using the data in the
2MASS catalog for this object, the NIR colors are consistent with a K5V, and 
the resulting distance to this companion is 299 pc. There is either
significant contamination to the $K$-band luminosity of KT Per, or else
the secondary is 0.8 mag more luminous than a main sequence M2V, having the
absolute magnitude of an M1V. The 2MASS colors suggest the contamination
is low, and this is supported by the ease at which Thorstensen \& Ringwald 
detected the donor in optical spectra. Unfortunately the orbital inclination 
for KT Per is unconstrained. If we assume an offset of one spectral subtype due
to inclination effects, the spectral type and distance can be reconciled
if $i$ $\sim$ 40$^{\circ}$.

\subsubsection{TT Arietis}

Our final short period CV with cross-dispersed SPEX data is TT Ari, 
P$_{\rm orb}$ = 3.30 hr. The data are presented in Fig. \ref{ttari},
and show no absorption features from a late-type secondary star. However,
the Ca II triplet is in emission, and presumably originates on the donor.
If we assume that the secondary in TT Ari has a spectral type of M4,
we can add a hot (T$_{\rm eff}$ = 10,000 K) blackbody continuum source until 
the spectral slope in the $K$-band is reproduced. From this iterative process,
we find that the blackbody source has 3.5$\times$ the luminosity of the 
secondary in the $K$-band. Given that TT Ari has $K_{\rm 2MASS}$ = 10.88, and 
an M4V has M$_{K}$ = 7.43, the distance to TT Ari is d $\sim$ 104 pc. 

\subsection{Metallicity Analysis for CVs with Expected M Dwarf Donors Using 
$K$-band Data Only}

\subsubsection{TW Virginis}

TW Vir has a period very similar to SS Aur and U Gem. 
The SPEX data were presented in Harrison et al. (2005a). TW Vir has a blue
continuum, with unusually strong Na I doublet and Ca I triplet features.
Running the metallicity analysis for this 
spectrum finds a spectral type of M0V, and [Fe/H] = $-$0.3. Clearly, this
result is compromised by the significant contamination of the secondary by the 
white dwarf and accretion disk. Subtracting a blackbody with T$_{\rm eff}$ = 
10,000 K that has 67\% of the $K$-band flux results in a continuum that
is consistent with an M1V, while subtraction of a hot blackbody that supplies
77\% of the $K$-band flux results in a spectrum whose continuum matches that 
of an M4V. The metallicity analysis on the subtracted spectra yield [Fe/H] = 
0.56, and 1.2, respectively. The M1V dwarf HD 42581 in the IRTF spectral 
library has [Fe/H] = 0.2, while the M4.5V dwarf Gl 268AB from the same source 
has [Fe/H] = 0.26. It is clear that both Na I doublets, and the Ca I triplet in 
the subtracted TW Vir spectra are much stronger than those in either
HD42581 or Gl 268AB. 

Obviously, such high values of [Fe/H] for TW Vir are not reasonable, being
as high, or higher, than seen anywhere in the Galaxy. Using the inferences of 
Veyette et al. (2016), to bring the metallicity of TW Vir in line with normal
stars would require an oxygen deficiency. If [O/Fe] = $-$0.2, the derived
abundance for TW Vir would be reduced by $\Delta$[Fe/H] = $-$0.6. It is
not clear how to produce such a deficit without it being pre-existing.
As shown by Veyette et al., there are few such objects in the solar
neighborhood. The result that we will derive for QZ Ser in the next
section may provide the necessary insight for explaining this object. Clearly,
however, new higher S/N data spanning the entire NIR will be essential for
further progress.

\subsubsection{CW Monocerotis}

CW Mon is a typical DNe with an orbital period of 4.238 hr, and exhibits
grazing eclipses (Kato et al. 2003). CW Mon has a significant X-ray 
luminosity (Verbunt et al. 1997), suggesting that it might be an IP.
Kato et al. found a possible white dwarf spin period that would aid in
this classification, but such a periodicity only appears to show up during
outbursts. CW Mon was observed with NIRSPEC, and the spectrum is shown in
Fig. \ref{cwmon}. The metallicity analysis indicates a spectral type of
M2.3, and [Fe/H] = $-$0.23. The IRTF template Gl 806 has a spectral type
of M2.2V, and [Fe/H] = $-$0.15, and proves to be an excellent match for
CW Mon. There are no abundance anomalies, though the continuum of Gl 806
appears to be slightly bluer than CW Mon, indicating a slightly later
spectral type for the donor. The IRTF spectral template star Gl 381 (M2.9, 
[Fe/H] = $-$0.06) provides a much better match to the continuum, but not to 
the depth of the Na I and Ca I absorption features.

Verbunt et al. list a distance for CW Mon of 290 pc, though the origin
of this ``adopted distance'' is not obvious. In the 2MASS catalog,
CW Mon has ($J - H$) = 0.53, ($H - K$) = 0.32, and $K$ = 13.02, roughly
consistent with an M2/3. The AAVSO database shows that CW Mon was at
minimum light during the epoch of the 2MASS observations. If the
secondary star is a normal M2V, the distance to CW Mon is d = 247 pc. If
the ``U Gem effect'' is universal, the large inclination of CW Mon, $i$
= 65$^{\circ}$ (Szkody \& Mateo 1986), would suggest a true spectral type 
closer to M1V, giving d = 311 pc, closer to the Verbunt et al. estimate.
As noted by Kato et al., at that distance, the visual maxima of CW Mon are 
nearly one magnitude less luminous than expected.

\subsubsection{YY/DO Draconis}

YY/DO Dra (see Patterson \& Eisenman 1987 for comments on its designation)
is an IP with an orbital period of 3.969 hr, and a spin period of 264.6 s
(Patterson \& Szkody 1993). Like a small subset of its brethren (see
Szkody et al. 2002), it shows infrequent dwarf novae outbursts of short 
duration. Metallicity analysis on these data find a spectral type of M1.8, 
and [Fe/H] = $-$0.20. The actual continuum of YY/DO Dra is better matched by 
the M2.5V template Gl 381 (M2.9, [Fe/H] = $-$0.06). The NIRSPEC data of YY/DO
Dra are shown in Fig. \ref{yydra}, along with the spectrum of Gl 381. Clearly, 
the absorption lines in the template are stronger than those seen in YY/DO Dra, 
consistent with their difference in metallicity. It is also obvious, 
however, that the CO features in YY/DO Dra are weaker than those of the 
template. We estimate that [C/Fe] $\leq$ $-$0.3.

As noted in the discussion of U Gem, weaker CO features should lead to
an underestimation of the metallicity if the conclusions of Veyette et al.
(2016) are correct (and we assign the entire deficit to carbon). In the case 
of YY/DO Dra, it clearly has weak CO, but the derived value of [Fe/H] and
the spectral comparison are consistent with it being lower metallicity than 
that of the template. If [C/Fe] = $-$0.2, the results of Veyette et al.
would lead to the derived metallicity being underestimated by $\Delta$[Fe/H] 
= $-$0.15. This offset would imply that both objects have the same 
metallicities, which is clearly not true. 

\subsubsection{UU Aquilae}

The spectrum of UU Aql was discussed in Harrison et al. (2005a) where a
spectral type of M2 was found to match the continuum, but not the absorption
features. Especially noteworthy was the absence of CO absorption. The 
metallicity analysis suggests a spectral type of K8, and [Fe/H] = $-$0.34. The 
NaI-only relation of Newton et al. gives [Fe/H] = $-$0.98. The expectation at 
the orbital period of UU Aql is for a mid-M type 
donor.  We compare the spectrum of UU Aql to the M2V template Gl 806 in Fig. 
\ref{uuaql}, and it is clear that CO is in emission in this system. This 
explains the discrepant spectral type estimation from the H2O index. The 
weakness of the blue Ca I lines, and the Na I doublet in UU Aql, vis-\'{a}-vis 
Gl 806, suggests that it has a much lower metallicity. If we use the H2O index 
of Gl 806 for UU Aql, the Mann et al. (2013) relation gives [Fe/H] = $-$0.72.
Even with the uncertainty induced by CO emission, the secondary
star of UU Aql appears to have an extremely low metallicity.

In the 2MASS catalog, UU Aql has ($J - H$) = 0.69, ($H - K$) = 0.24,
and $K$ = 13.33, consistent with an M2V. There are no AAVSO data for the
epoch of these observations, but they occurred halfway between two maxima, 
suggesting that UU Aql was near quiescence. If we assume a normal M2V donor, 
this
suggests a distance of d = 286 pc. Sion et al. (2007) found that from modeling
$FUSE$ observations, they could not settle on a distance for UU Aql,
with both d = 150 pc, and d = 300 pc producing nearly identical results. A 
model of theirs with the silicon and carbon abundances at 10\% the solar value,
 however,
gave a distance of 314 pc. Perhaps dramatically lowering the global abundances 
in their white dwarf/accretion disk models might lead to more robust fits. It 
is important to note that Woolf \& West (2012) show that low metallicity M 
dwarfs are less luminous than more metal-rich stars. For [Fe/H] = $-$0.5, they 
found $\Delta$M$_{V}$ = $+$0.53. If that offset is applied here, the distance 
to UU Aql is reduced to 224 pc.

\subsubsection{CN Orionis}

The orbital period of CN Ori is about 30 s shorter than UU Aql, so 
one would expect them to appear similar. The NIRSPEC spectrum for CN Ori is 
shown in Fig. \ref{cnori}, where we compare it to Gl 806. The continua of
the two sources match quite well. The metallicity analysis gives the spectral
type as M1.1, and [Fe/H] = $-$0.49. While more subtle than seen in UU Aql, CO 
emission is clearly present in the spectrum of CN Ori, so the spectral type 
calculated from the H2O index will be corrupted. If we again assume the H2O 
index of Gl 806, we find [Fe/H] = $-$0.71 for CN Ori. Essentially identical to 
UU Aql. H\&H modeled several CO emission scenarios and showed that when 
the velocity width of the emission is low, the underlying CO absorption 
features of the secondary star appear much narrower, with small blue emission 
wings. The red end of the model spectrum at the bottom of their Fig. 21 looks 
nearly identical to that of CN Ori (see also Fig. \ref{czori}).
As we found for UU Aql, the main absorption features in the spectrum
of CN Ori are much, much weaker than their counterparts in Gl 806. This
supports the very low metallicity calculation. 

Verbunt et al. (1997) lists a distance of 295 pc. The 2MASS colors of CN 
Ori are ($J - H$) = 0.50, ($H - K$) =  0.20, and $K$ = 13.10. The AAVSO 
light curve data shows that the 2MASS observations fell in-between two 
outbursts. The ($J - H$) color is slightly bluer than an M2V by 0.1 mag,
while ($H - K$) matches such an object exactly. The estimated distance
is 257 pc. Again, due to the low metallicity of the secondary star, this
may be an overestimate.

\subsubsection{IP Pegasi}

IP Peg is an eclipsing CV with an orbital period of P$_{\rm orb}$ = 3.797 hr.
Ribeiro et al. (2007) modeled $JHK$ light curves of IP Peg in quiescence and 
found $i$ = 84.5, T$_{\rm 2}$ = 3100 $\pm$ 500 K, and a distance of 115 pc.
The NIRSPEC data for IP Peg is shown in Fig. \ref{ippeg}, where it is
compared to the M0V template HD 19305. Metallicity analysis on the NIRSPEC
data finds a spectral type of M0.9, and [Fe/H] = 0.18. The same analysis
on HD 19305 derives a spectral type of M0.7, and [Fe/H] = 0.02. Thus, HD 19305
should provide a reasonable comparison, and this is confirmed in Fig. 
\ref{ippeg}. Except for the Al I doublet at 2.1 $\mu$m, the main atomic
absorption features in IP Peg are slightly stronger than the template,
consistent with the metallicity difference between the two objects. The Al I
features might be compromised by He I emission, though the always much stronger
He I emission line at 2.06 $\mu$m is quite weak in IP Peg. The CO features
in IP Peg appear to be marginally weaker than the template, though the
telluric correction at the red end of the NIRSPEC spectrum is not of the
highest quality.

As shown in Table 1, the secondary of IP Peg was at inferior conjunction at
the time of the NIRSPEC observations, and the AAVSO light curve database
shows that the system was quiescent (though it would go into outburst 10 days 
later). Thus, it is somewhat surprising that we obtain a much earlier spectral
type than found by Ribeiro et al. ($\sim$ M3.5). 2MASS data was obtained
at $\phi$ = 0.35, and with ($J - K$) = 0.88, suggests an M2V. With 
$K_{\rm 2MASS}$ = 11.72, and assuming the secondary is a normal M1V, the 
distance to IP Peg is 172 pc.

\subsubsection{AM Herculis}

AM Herculis is the proto-type for the polars, and has an orbital period of 
3.09 hr. As shown in Campbell et al. (2008), AM Her has strong cyclotron 
emission in the $K$-band during much of its orbit. The SPEX data for AM Her 
presented in Harrison et al. (2005a) were obtained during a polar high state, 
and near inferior conjunction, when the cyclotron emission is at its strongest. 
Campbell et al. observed AM Her over a full orbit during a low state using the 
low resolution ``prism'' mode of SPEX. 
During that time they found a spectral type of M5V. The metallicity analysis 
on the higher resolution SPEX data in Harrison et al. (2005a) finds a spectral 
type of M1.2, and [Fe/H]= $-$0.25. To attempt to determine whether this 
metallicity is at all reliable, we have continuum-divided the moderate
resolution SPEX data for AM Her to compare with the spectral templates. The M4V
template HIP9291 (H2O index gives M4.7) has [Fe/H] = 0.15, while Gl 213 (M4.6)
has [Fe/H] = $-$0.31. The resulting match between the spectra of Gl 213 and AM 
Her is excellent, except that the CO features in AM Her are weaker than those 
in the template.  Running the metallicity routine on the low resolution SPEX 
data from Campbell et al. (2008) at phase 0.5, {\it when the cyclotron emission
is near its minimum}, we derive a spectral type of M5.0, and [Fe/H]= $-$0.44.
It is possible that SPEX prism mode spectroscopy is of insufficient 
resolution for these routines to work reliably, however, the results using 
such data are consistent with the those derived using the other techniques. We 
conclude that secondary star in AM Her has a low metallicity, and probably a 
carbon deficit. Higher spectral resolution observations during a low state, 
and at phases of low cyclotron emission, are needed to confirm this result.

Thorstensen (2003) obtained a parallax for AM Her that leads to a distance of
79$^{\rm +8}_{-6}$ pc. An M5V at this distance would have $K$ = 13.00,
while an M4V would have $K$ = 11.9. Campbell et al. (2008) present a $JHK$
light curves for AM Her, and from their ellipsoidal model, $\langle K \rangle$
= 11.45, and $i$ = 50$^{\circ}$. The donor in AM Her has the luminosity
of an M3.5V, 1.5 subtypes earlier than that derived from spectroscopy.

\subsubsection{The Polars from Harrison et al. 2005b}

We lump together the results for the polars discussed in Harrison et al. 
(2005b) to examine their similarities. There are four short period polars that 
were observed using NIRSPEC on Keck: AR UMa, MR Ser, ST LMi, and VV Pup. They 
also observed the longer period low accretion rate polar MQ Dra 
(= SDSS1553+5516). None of the donor stars in these systems appeared to
be unusual in any way, with normal carbon abundances for their spectral types.
As shown in Table \ref{results}, the secondaries in all four short period 
polars have subsolar metallicities, while MQ Dra has a super-solar 
metallicity. 

Harrison \& Campbell (2015) discussed the field strengths for the four short
period polars, and examined/modeled $WISE$ mid-IR light curve data for these
systems. AR UMa has a large field strength, thus it has no cyclotron 
emission in the NIR, and its $K$-band spectrum will be uncorrupted from 
non-thermal emission sources. The spectral type from the H2O index, M4.5, agrees
with the classification in Harrison et al. (2005b). It is likely that there 
is weak cyclotron emission from the $n$ = 2 harmonic in the $K$-band for MR 
Ser. In modeling its infrared light curve data, Harrison \& Campbell (2015) 
used a donor with T = 2400 K ($\sim$ M8). This contrasts with the M4.5 type 
derived from the metallicity analysis. The NIRSPEC and SPEX data for MR Ser
(see Fig. 2 in Harrison et al. 2005b) show no evidence for the short or long
wavelength turnovers due to water vapor that would be expected from a late
type M dwarf. Harrison \& Campbell (2015) present a phase-resolved set of SPEX
prism spectra for MR Ser, but their data did not cover a full orbit, and at no
phase does the cyclotron emission in this system drop to very low levels like
that in AM Her.

ST LMi can have strong cyclotron emission in
the $K$-band during part of its orbit; fortunately the NIRSPEC observations 
missed those phases. The spectral type derived from the metallicity analysis, 
M6.1, agrees with the M6V classification from Campbell et al. (2008). Harrison 
\& Campbell find that the field strengths of the two poles in VV Pup are B =
57 and 33 MG, thus there are no cyclotron harmonics in the $K$-band.
Harrison et al. (2005b) derived a spectral type of M7, in agreement with that 
from the H2O index. Infrared light curves of MQ Dra are presented in Szkody
et al. (2008) and show that there is no evidence for cyclotron emission
in the $K$-band. Thus, the metallicity analysis for this object should be 
robust.

Thorstensen et al. (2008) have published parallaxes for four of the five
polars discussed here. For VV Pup, they found $d$ = 124$^{\rm +17}_{\rm -14}$ 
pc, AR UMa has $d$ = 86$^{\rm +9}_{\rm -8}$ pc, MR Ser has $d$ = 
126$^{\rm +14}_{\rm -12}$ pc, and MQ Dra has $d$ = 162$^{\rm +27}_{\rm -21}$.
From Harrison \& Campbell (2015), $\langle K_{\rm AR~UMa} \rangle$ = 13.2,
and $\langle K_{\rm MR~Ser} \rangle$ = 14.0. From Szkody et al.  (2008),
$\langle K_{\rm MQ~Dra} \rangle$ = 13.7, while Szkody \& Capps (1980) found
 $\langle K_{\rm VV~Pup} \rangle$ = 15.1. The absolute $K$ magnitudes of AR 
UMa and MR Ser (M$_{K}$ = 8.5) are identical to that of an M5V, while VV Pup
is a full magnitude less luminous, being more consistent with an M6V. MQ 
Dra has the most luminous secondary of these four objects, M$_{K}$ = 7.6, 
suggesting the donor star has a luminosity slightly fainter than that of an 
M4V main sequence star. The classifications derived from their luminosities 
are in good agreement with those found from the spectral analysis. The donor 
stars in these five systems are very similar to normal main sequence stars.

\subsubsection{RZ Leonis}

The NIRSPEC data for RZ Leo have been presented in Howell et al. (2010), and
Hamilton et al. (2011). RZ Leo is a short period DNe in the
WZ Sge family. This subclass of CVs have infrequent, but large amplitude
outbursts. Both Howell et al. and Hamilton et al. classify the secondary
spectral type as M4. We have used measurements of the wavelength of the Na I 
doublet to Doppler correct the NIRSPEC data (see Fig. \ref{rzleo}). The 
metallicity analysis on the resulting spectrum gives a spectral type of M3.5, 
and [Fe/H] = $-$0.44. However, measurement of the Na I doublet and its local
continuum is probably compromised by the broad H I Br$\gamma$ emission. In
addition, the Ca I triplet has a much broader profile than it should. Comparison
to the various late-type templates finds that Gl 268AB (M4.8, [Fe/H] = 0.26) 
provides an excellent match to the continuum and to both Na I doublets of
RZ Leo. This suggests that the donor has a solar metallicity. While the CO 
features in RZ Leo appear to be only slightly weaker than the template, the 
first overtone of $^{\rm 13}$CO at 2.345 $\mu$m is very strong. The data are 
noisy, but suggest that the $^{\rm 13}$CO$_{\rm (4,2)}$ feature at 2.403 
$\mu$m is also present. 

It is rare for such a short period system to have such a prominent secondary.
We suspect that there is contamination from a hot source that makes the
continuum appear bluer than expected. If so, the spectral type of the
secondary is probably later than M5, and that would imply a substantial
carbon deficit. For example, if we subtract-off a 10,000 K blackbody that
supplies 30\% of the $K$-band flux (Fig. \ref{rzleo}), the continuum of RZ Leo 
is very similar to that of GJ 1111 (M6.8, [Fe/H] = 
$-$0.17). The donor in RZ Leo would then have a substantial subsolar abundance, 
and a significant carbon deficit (note, however, that $^{\rm 13}$C would not 
be strongly enhanced in this scenario). New multi-bandpass near-infrared 
spectroscopy of this system would be extremely valuable.

\subsubsection{V893 Scorpius}

V893 Sco is an eclipsing SU UMa system. Hamilton et al. (2011) presented
a $K$-band spectrum obtained using ISAAC on the VLT. They derived a spectral 
type of M6V. The spectrum has low S/N, but boxcar smoothing by 21 pixels,
and then running the metallicity analysis, leads to a spectral type of M5.2,
and [Fe/H] = $-$0.82. There are no templates in the library with such
a low metallicity. The closest, Gl 399 (M5.0, [Fe/H] = $-$0.58), is not
a very good match for the spectrum of V893 Sco. While there is a significant
CO/H$_{\rm 2}$O break in the spectrum of V893 Sco, the absorption from Na I
and CO are extremely weak. Hamilton et al. (2011) examined the luminosity
of this CV, and found that it had an observed $K$-band magnitude that was 2.5 
mags brighter than an M6V at the distance implied by the parallax. The
$K$-band spectrum of V893 Sco suffers from substantial contamination. 
Multi-bandpass data will be needed to confirm the low metallicity of this 
system.

\subsubsection{VW Hydri}

The $K$-band spectrum from ISAAC on the VLT was presented in Hamilton et al.
(2011). VW Hyi is an SU UMa dwarf novae for which Hamilton et al. estimated 
a secondary star mass of 0.11 M$_{\sun}$, and derived a spectral type 
of M4. The metallicity analysis run for this spectrum finds a spectral type of 
M2.9, and [Fe/H] = $-$0.32. The Newton et al. relation gives [Fe/H] = $-$0.34.
There are a number of IRTF templates with similar properties for comparison.
We find the best match occurs with
HD95735 (M3.0, [Fe/H] = $-$0.33), though Gl 213 (M4.6, [Fe/H] = $-$0.31) 
provides a very similar fit. Thus, the low metallicity is confirmed, and
the carbon abundance appears to be normal.

\subsubsection{EX Hydrae}

Hamilton et al. (2011) presented $K$-band spectra for EX Hya, which is a short 
period IP. They derived a spectral type of M5V. The metallicity analysis on the
NIRSPEC spectrum gives a spectral type of M2.0, and [Fe/H] = $-$0.55. However, 
M2V templates, such as Gl 806, have bluer continua than EX Hya. The spectral 
template Gl 273 (M4.2, [Fe/H] = $-$0.13) provides a better match to the 
continuum, though all of the absorption features in the spectrum of EX Hya are 
much weaker than in this object. The low metallicity template Gl 299 (M5.0, 
[Fe/H] = $-$0.58) provides a good match to the strength of the Na I doublet and 
Ca I triplet absorption features, though its continuum is too red, and the CO 
features are weaker in EX Hya. Assuming no contamination in the $K$-band,
we conclude that the donor in EX Hya has spectral type near M4, a low 
metallicity, and [C/Fe] $\lesssim$ $-$0.30.

Beuermann et al. (2003) obtained a parallax for EX Hya using the FGS on the
$HST$, giving a distance of 64 pc. The 2MASS catalog has ($J - H$) = 0.32,
($H - K$) = 0.08, and $K$ = 11.69. These colors are much bluer than a mid-M
type star, confirming the conclusion in Hamilton et al. (2011), that there
is some contamination of the $K$-band flux. With the astrometric distance,
and the 2MASS $K$-band magnitude, M$_{K}$ = 7.76. This is 0.3 mag fainter than 
an M4V. This suggests that the true spectral type of the donor in EX Hya is 
closer to M5, and contamination is responsible for the resulting continuum.  
This would imply an even larger carbon deficit.

\subsubsection{VY Aquarii}

VY Aqr is an SU UMa-type dwarf nova with an orbital period of 1.51 hr.
Harrison et al. (2009) present the $K$-band spectrum of this source obtained
with NIRSPEC. They showed that the continuum of the observed spectrum was
most consistent with an M0V. The metallicity analysis gives a similar answer:
K9.6V (and [Fe/H] = $+$0.37). Harrison et al. found that analysis of archival
VLT data for VY Aqr suggested a spectral type closer to M6, though the
CO features were much too weak for {\it any} M dwarf. The problem with such
a classification is that subtraction of a hot blackbody contaminating source 
does not lead to a realistic spectrum. To merely flatten the spectrum so that 
it superficially resembles an M dwarf requires us to subtract off a hot source
that supplies 95\% of the $K$-band flux. The result is a spectrum with 
ludicrously strong Na I and Ca I absorption features: [Fe/H] = 13.9! 
{\it Though the data are noisy, it is clear that VY Aqr does not have CO in 
emission.} In \S5, we will examine synthetic spectra for stars with hydrogen 
deficits. Those types of models appear to be required to explain unusual 
objects like VY Aqr.

Thorstensen (2003) derives an astrometric distance of 97$^{\rm +15}_{\rm -12}$
pc to VY Aqr. The 2MASS data for VY Aqr are ($J - H$) = 0.42, ($H - K$) =
0.27, and $K$ = 14.59. There are no AAVSO measurements for this period, but
using the {\it NeoWISE} database, the faintest VY Aqr gets is $W1$ = 14.4.
This suggests VY Aqr was in quiescence at the time of the 2MASS observations,
giving M$_{K}$ = 9.5. Assigning 100\% of the $K$-band luminosity
to the donor star, implies a spectral type of M6. While this agrees with
the spectral type found by Harrison et al. (2009), the $K$-band spectra
remain inconsistent with such a classification.

\subsubsection{V436 Centauri}

The SU UMa system V436 Cen has a period that is nearly identical to VY 
Aqr. The $K$-band spectrum has been presented by Hamilton et al. (2011). 
While the S/N of those data are not high, the secondary is quite prominent
in the VLT spectrum. Hamilton et al. estimated a spectral type of M8V for this 
object. Our analysis derives a spectral type of M6.9, and [Fe/H] = $-$0.31. 
Fortunately, the IRTF spectral library has several late-type templates
for comparison. LP944-20 is listed as having a spectral type of M9, but
the metallicity program gives M7.7, and [Fe/H] = $-$0.60. The same analysis
for Gl 644c gives M7.4, [Fe/H] = $-$0.32. The match of the latter to V436 Cen
is much better than the former, and suggests that the metallicity of the
donor in V436 Cen is low, while the C abundance is normal.

\subsubsection{V2051 Ophiuchi}

The orbital period of V2051 Oph, an eclipsing SU UMa system, is 6.5 s shorter 
than V436 Cen. Hamilton et al. derive a spectral type of M7 for the donor
from VLT data. The metallicity analysis estimates a spectral type of M9.2,
and [Fe/H] = $-$0.53. There are no templates in the library with this late of 
spectral type, and this low of an abundance, to use for comparison. Analysis 
for Gl 752B gives M8.6, and [Fe/H] = $-$0.11. Gl 752B provides an excellent
match to the Na I doublet, and to the CO features. We conclude that the 
noisiness of the continuum near Na I is leading to both a very low value for
the metallicity, and the very late spectral type. Except for VY Aqr, all of 
the donor stars for the SU UMa CVs with periods near 1.5 hr appear to have 
subsolar metallicities. 

\subsubsection{WX Ceti}

WX Cet is an ultra-short period CV that appears to be an intermediary
between the normal SU UMa DNe, and the WZ Sge stars. Sterken et al. (2007)
show that WX Cet has normal, short duration SU UMa type eruptions, but also
rare superoutbursts of large amplitude. They find that the time between 
short eruptions is about $\sim$200 d, while superoutbursts occur every 
$\sim$880 d. These recurrence times are four to five times longer than seen in 
the typical SU UMa system. WX Cet is extremely faint in the near-IR. It does not
appear in the 2MASS Point Source Catalog, though it is present in the $J$-band 
image. Aperture photometry gives $J_{\rm 2MASS}$ = 17.2. The spectrum presented
in Fig. \ref{wxcet}, has been smoothed by 30\AA. Besides the strong H I and 
He I emission lines, absorption from the first overtone of $^{\rm 12}$CO is 
clearly present. The metallicity regimen assigns a spectral type of M4.9, 
with [Fe/H] = $-$0.57.  It is likely that the continuum is contaminated by 
white dwarf $+$ accretion disk emission. If 25\% of the $K$-band flux is 
supplied by this source, the estimated spectral type is M6.5, and the 
metallicity is [Fe/H] = $-$0.25. If the donor provides only 50\% of the 
$K$-band luminosity, the spectral type is M9.4, and [Fe/H] = 0.35. The
IRTF template Gl 752 (M8.6, [Fe/H] = $-$0.11) is a reasonable match to the
spectrum where the secondary supplies 75\% of the $K$-band flux. We conclude
that the donor in WX Cet is a late M-type star, and has a near-solar
abundance with normal levels of carbon.

\section{Other Interesting Objects}

There are two objects for which we have obtained spectra for that appear to
be extremely unusual CVs: QZ Ser and EI Psc. Both appear to have K-type
secondary stars, while having short orbital periods. Here we analyze the 
spectra of these two objects.

\subsection{QZ Serpentis}

Thorstensen et al. (2002a) recount the discovery of QZ Ser, and used 
time series spectroscopy to derive an orbital period of 1.996 hr. Unexpectedly,
for this orbital period, the secondary appeared to have a spectral type
of K4 $\pm$ 2. They also believed that the sodium abundance was probably
enhanced. We observed QZ Ser with TripleSpec, and the data are presented
in Fig. \ref{qzser}. As found by Thorstensen et al., the continuum of
this spectrum is matched by that of a K5V. Beyond that, the spectrum is
highly unusual: the only emission lines are from He I, there is no evidence 
for CO absorption, and the Na I lines are extremely strong
across all bandpasses. Especially prominent is the Na I doublet at 1.638
$\mu$m, a feature that is usually lost amongst numerous absorption features
found in this spectral region. We generated synthetic spectra over
a temperature range of 4000 $\leq$ T$_{\rm eff}$ $\leq$ 5000 K, log$g$ = 4.5,
and $-$1.00 $\leq$ [Fe/H] $\leq$ 0.5. None of these models matched the
depth of the Na I, Ca I, or the Mg I lines. To fit these common absorption 
features required huge overabundances of these three elements. After
adjusting their abundances, the resulting models fit the strong lines better, 
but they did not reproduce the depth of the Fe I lines, even in the metal-rich 
models.

As an alternative, taking the suggestion of Thorstensen et al. that the
donor star might be deficient in hydrogen and have enhanced helium 
abundances, we ran models with {\it solar abundance patterns}, while changing 
the abundances of H and He. We found the best fitting spectrum occurred
when the hydrogen abundance was reduced to 30\% of its normal value, and the
helium abundance was increased by a factor of ten. With these changes,
iron, magnesium and calcium have normal abundances. This is 
not true for carbon, sodium, or aluminum. The blue end of the $H$-band is
sensitive to both the H and C abundances: reduce either one too much,
and the continuum becomes too blue. The final model, plotted in Fig. 
\ref{qzser}, has [C/Fe] = $-$1.7! The sodium abundance needed to explain the 
absorption features is 32$\times$ the solar value. The aluminum abundance also 
has to be higher to fit the data, and we found [Al/Fe] = $+$0.5. None of our 
models fit the Na I doublet at 1.638 $\mu$m. This doublet comes from 
transitions between the 2p$^{\rm 6}$4p$^{\rm 2}$P$^{\rm o}$ and 
2p$^{\rm 6}$6s$^{\rm 2}$S$_{\rm 1/2}$ (we used the oscillator strengths listed 
in Cunha et al. 2015). These lines are more prominent in red giants (ibid.), 
than in dwarfs, and may suggest a lower gravity for this peculiar star. 

Obviously, we are in uncharted territory for abundance analysis. To enable
us to produce realistic synthetic spectra for an object where the absolute 
abundances of H or He are non-solar, we need to start with a stellar atmosphere 
that has the correct abundance patterns. While we believe the best solution to
the difficulty in modeling the spectrum of QZ Ser is a hydrogen deficiency,
in all of our models, the Na overabundance is very large.

How could such an anomaly arise? There are two proton capture cycles beyond the 
CNO cycle that are likely responsible for the peculiar abundances of QZ 
Ser: the NeNa cycle, and the MgAl cycle (see Wallerstein et al. 1997, and
references therein). Prantzos et al. (1991) model the production of sodium in 
2 and 15 M$_{\sun}$ stars. They found that just before He ignition occurs, the 
sodium abundance in the innermost regions of the star are enhanced by a factor 
of five in both models. For the lower mass star, the altered values of Na were
constrained to the inner 0.43 M$_{\sun}$ of the H-exhausted core. Inside
this radius, the carbon abundance would also be very low. If the donor star of 
QZ Ser was once a red giant whose outer atmosphere has been stripped, simple 
stellar evolution models can probably explain the peculiar Na abundances. 
This would require an initial mass for the donor in QZ Ser that is higher than 
typically envisioned under the standard evolutionary paradigm for CVs.

A similar story holds for the MgAl cycle. Proton capture onto Mg can result
in Al. To operate this cycle, however, appears to require more massive
stars ($\sim$5 M$_{\sun}$, Ventura et al. 2013). Smiljanic et al. (2016) have 
surveyed the Na and Al abundances in giants and dwarfs. They only found
large Al enhancements for red giants with M $>$ 3 M$_{\sun}$. It appears 
that it will require an even more massive progenitor to explain the 
overabundance of Al seen in QZ Ser, if it must arise from simple stellar 
evolution.

Alternatively, the peculiar atmosphere of QZ Ser might be acquired during
the common envelope phase of a classical novae (CNe) eruption. The rapid proton 
capture process (the ``rp-process''), during explosive hydrogen burning can
produce a wide variety of nucleosynthetic products. Denissenkov et al. (2014)
have shown that significant production of Na or Al in a CNe eruption is
possible, but almost certainly requires enrichment of material from an 
underlying ONe white dwarf. These same models also produce significant amounts 
of carbon. If the relative abundances of our model are even approximately
correct, the accretion of CNe processed material cannot explain QZ Ser.

Above, we encountered two systems, VY Aqr and TW Vir, that had anomalously 
strong metal absorption lines for their spectral types. Instead of having
super-solar metallicities, it is just as likely that they suffer from
hydrogen deficits. Even an object such as U Gem has some parallels with
QZ Ser. Unfortunately, hydrogen-deficient model atmospheres are not yet
available for stars cooler than T$_{\rm eff}$ $\sim$ 10,000 K (cf.,
Behara \& Jeffery 2006), so further investigation of these CVs is not
currently possible.

\subsection{EI Piscium}

Thorstensen et al. (2002b) present spectra and time series photometry of
EI Psc and find an ultra-short period of 64 minutes, well below the
period minimum for CVs with donor stars of normal composition (see Howell et 
al. 2001, and references therein). Thorstensen et al. believe that the He 
abundance is enhanced.  Harrison et al. (2009) presented a $K$-band
spectrum for EI Psc, obtained with NIRSPEC. As discussed in Harrison et al.,
and Thorstensen et al. (2002b), the spectral type appears to be close to
a K5V. The issue with this spectral classification, however, is that the
Ca I triplet is much weaker in EI Psc than the Na I doublet (see Fig. 
\ref{eipsc}). For K dwarfs, these two features have a similar strength, and
it is not until the mid-M dwarfs ($\sim$ M5V) where the ratio of the 
equivalent widths for the Na I and Ca I features becomes as large
as seen here (2.2). Like QZ Ser, there is no evidence for CO absorption
features in this spectrum. Unlike QZ Ser, there is H I emission. If we
assume the secondary has a temperature of 4500 K, the best fit model to 
the spectrum, has [Fe/H] = $-$0.2, and [C/Fe] = $-$1.7. Like QZ Ser, 
Na I appears to be enhanced, and we find [Na/Fe] = $+$0.9 provides a good
match to the two doublets in the $K$-band. 

While we did not need to invoke an H deficit to explain the NIRSPEC data, it 
is likely that the short orbital period of this system requires it to be 
hydrogen deficient. Thorstensen et al. show that if the donor had an initial 
mass of 1.2 M$_{\sun}$, and mass transfer began shortly after the end of H 
burning, the secondary star could evolve to short periods while maintaining a 
high temperature. This model predicts a depletion of C, and enhanced levels of 
N in the photosphere. Using our results for QZ Ser as a guide, if the donor in 
EI Psc has an H deficiency of 30\% like the Thorstensen et al. model, the Na I 
enhancement would be much smaller, and the abundances of the other metals
would have to be substantially subsolar. 

\section{Results}

We have used near-IR spectra to derive metallicities and effective temperatures
for forty one cataclysmic variables spanning most of the major subclasses. We 
have extended the program described in H\&H to enable us to generate synthetic 
$IJHK$ spectra, allowing us to model cross-dispersed spectroscopy of CVs
we have obtained with SPEX on the IRTF, or TripleSpec at APO. We then 
constructed grids of models to compare to both cross-dispersed and $K$-band 
spectra for analysis of those CVs with K type secondary stars. We have used the
recent metallicity calibrations that employ $K$-band spectra to derive
temperatures and values of [Fe/H] for CVs with M-type donor stars. These
calibrations were applied to template spectra enabling the direct comparison
of similar objects, bolstering the determinations of T$_{\rm eff}$, [Fe/H], and
allowing for estimates of [C/Fe]. As found in H\&H, and implied in the data from
Harrison et al. (2004a, 2005a,b, 2009), carbon deficits were found for numerous
(21/36) objects. The trend, however, is that these deficits were more likely to
occur in long period CVs (P$_{\rm orb}$ $>$ 4 hr), then in short period
systems. We discuss this, and the other main results in the subsections that 
follow.

\subsection{Carbon Deficits and CO Emission}

H\&H investigated whether it was possible for CO emission to create apparent 
carbon deficits. They found that the presence of CO emission, presumably 
from the accretion disk, overlaying the absorption features of the secondary 
star creates a spectrum with unique CO absorption feature profiles: blue 
emission wings, and dramatically narrower bandheads. Using those model results,
it is very simple to identify CO emission even in noisy data. We found
four CVs with such emission: UU Aql, LS Peg, CN Ori, and CZ Ori. With the 
inclusion of WZ Sge, this brings the number of CVs that exhibit CO emission to 
five.  {\it None} of the other CVs discussed above show evidence for CO 
emission. 
Thus, any derived carbon deficits are real, though the quality of the spectra 
often limited our ability to derive precise values of [C/Fe]. The ability to 
compare templates with known metallicities to CVs with M-type secondaries makes 
the identification of carbon deficits in the shorter period CVs more robust 
than previous efforts that lacked this information.

One relevant question is why do these five CVs show CO emission? As 
described in H\&H, CO emission is believed to originate in the outer, cooler 
parts of the accretion disk. We believe that there are three possible 
explanations. The first is that CO emission is a transient phenomenon. The 
second is that the mass transfer rate in these five systems is just right to 
create the environment amenable to CO emission. Or third, the structure 
of their accretion disks is somehow altered by composition, leading to 
conditions that fosters CO emission. 

Only a larger spectroscopic survey of CVs can answer the first possibility, 
though the CO emission in WZ Sge was present for at least a full year (see 
Harrison 2016). The low S/N, low resolution $K$-band spectrum of WZ Sge 
presented by Dhillon et al. (2000), is highly suggestive of CO emission being 
present in WZ Sge as far back as 1997. This would rule it out as an artifact 
of the 2001 eruption. Given its long inter-outburst intervals and low 
accretion rate, one would expect the disk structure in WZ Sge to be relatively 
stable. Thus its CO emission might be persistent. We have no second epoch
observations for the other sources with CO emission. It is interesting
to note that the observations of CN Ori and UU Aql were obtained just a
few days after they had returned to minimum light following outbursts.

We have a handful of other objects that have observed more than once. 
In H\&H, GK Per, SS Cyg, and RU Peg were observed twice in one week with
no change in the CO features. RU Peg and SS Cyg were also previously observed 
by Harrison et al. (2004a), and Dhillon et al. (2002) observed SS Cyg. 
There is no evidence for CO emission in those data. We have three 
epochs of observation for SS Aur, including Harrison et al. (2005a), and it 
has never shown the signature of CO emission. It will take additional epochs 
of observation to ascertain whether CO emission is transitory.

If we assume CO emission is tied to accretion rate, the second explanation does 
not seem especially viable, as LS Peg clearly has a very large accretion rate, 
while WZ Sge has one of the lowest. The third explanation might be a
possibility, given that the donors in CZ Ori, UU Aql and CN Ori all have large 
subsolar values of [Fe/H].  Wehrse \& Shaviv (1995) explore the role that metal 
opacities have on the structure of accretion disks.

The group of objects with the most extreme values for their carbon deficits
([C/Fe] $\leq$ $-$1.0) are a heterogeneous set: AE Aqr, CH UMa, V1309 Ori, 
U Gem, QZ Ser, VY Aqr, and EI Psc. Essentially one object from each of
the subclasses in the survey. AE Aqr, U Gem, V1309 Ori and CH UMa were 
previously shown to have deficits of carbon from UV studies. Hamilton et al. 
(2011) have discussed the direct correlation between weak CO features in 
$K$-band spectra, and UV-derived carbon deficits. It is tempting to suggest 
that like QZ Ser, the objects with the most extreme carbon deficits are 
hydrogen deficient. In contrast to Harrison et al. (2005b), we now have 
evidence that polars are not always ``normal,'' as both V1309 Ori and AM Her
have carbon deficits.

\subsection{Effective Temperature vs. Orbital Period}

Knigge et al. (2011) published a compilation of the distribution of CV donor
spectral types vs.  P$_{\rm orb}$. They show that a reasonable fit to the 
distribution results from both standard evolutionary models, as well as models 
with different angular momentum loss prescriptions. Though the distribution of 
spectral types in their Fig. 15 could reasonably be summarized by saying that 
the spectral type of a CV donor star is M4 $\pm$ 2 for P$_{\rm orb}$ $<$ 6 hr. 
The origin of the spectral 
types plotted in their figure come from a diverse set of methodologies (see 
Knigge 2006 for references). Given that the majority of the donor stars in the 
current survey have subsolar metallicities, most of the spectral types in that 
compilation are, at best, suspect. Some of them have been shown to be wrong 
(such as the L0 spectral type for VY Aqr). Our version of this diagram (less 
GK Per) is presented as Fig. \ref{teffvsporb}. We 
code the various subclasses using different symbols. We use T$_{\rm eff}$
in this diagram as that is quantity we derive from our synthetic spectra 
modeling. We use the effective temperature scale from Rajpurohit et al. (2013) 
to convert M spectral types found above to T$_{\rm eff}$.

While there is a general trend of shorter period CVs to have cooler donor
stars, the spread at any P$_{\rm orb}$ or T$_{\rm eff}$ can be significant.
For example, there are six CVs with T$_{\rm eff}$ = 4750 K that have
orbital periods spanning the range six to ten hours. In contrast, there
are ten CVs with P$_{\rm orb}$ $\sim$ 4 hr, that span the range 2800 K
$\leq$ T$_{\rm eff}$ $\leq$ 4600 K. A similar grouping is found below
 P$_{\rm orb}$ $<$ 2 hr, where 2500 K $\leq$ T$_{\rm eff}$ $\leq$ 3050 K.
Orbital period is a poor predictor for the {\it observed} donor star 
temperature.

Because main sequence stars have a well defined mass-radius relationship, 
their mean density is therefore defined. Mann et al. (2015) present the 
relationship for the mass range 0.1 M$_{\sun}$ $\leq$ M $\leq$ 0.7 M$_{\sun}$. 
Fortunately, metallicity plays no significant role. For late-type dwarfs, Mann 
et al. also found that there is a fairly tight correlation between radius and
temperature. Thus, for hydrogen burning main sequence stars, there is an easily
defined relationship between mean density and temperature. Given that the mean 
density of a Roche-lobe filling star is closely tied to orbital period 
($\langle \rho \rangle$ =
107P$_{\rm orb}^{\rm -2}$ gm cm$^{\rm -3}$, equation 2.3b in Warner 1995), 
and the mean density of the Roche lobe filling star is $\langle \rho \rangle$ =
3M/4$\pi$R$_{\rm L}^{\rm 3}$, there are only a small number of variables 
involved that can act to create a range in observed temperature at any 
particular orbital period. The equivalent spherical radius of 
the Roche-lobe filling donor star, R$_{\rm L}$, is a function of the mass
ratio, and dependent on the size of the semi-major axis of the orbit. For
P$_{\rm orb}$ = 4.3 hr and M$_{\rm 1}$ = 1 M$_{\sun}$, R$_{\rm 2}$ = 0.5
and 0.4 R$_{\sun}$, at $q$ = 0.5 and 0.25, respectively. For main sequence 
stars, these two donor stars would differ by $\sim$ 350 K. Thus, a large 
difference in the mass ratio can have a significant influence on the observed 
temperature of the donor star at a fixed orbital period.

Returning to the objects with periods near 4.3 hr, the mean density of the
donor at this period is 6.1 gm cm$^{\rm -3}$. As shown in Fig. \ref{teffvsporb},
a main sequence star with this density has T$_{\rm eff}$ $\sim$ 3900 K, hotter 
than that observed for any of the CV donors at this period. That
the secondary stars of CVs are cooler than the equivalent main sequence
stars has been discussed repeatedly (e.g., Echevarria 1983, Friend et al.
1990, Beuermann et al. 1998, Baraffe \& Kolb 2000). For
U Gem, the secondary has a mass of 0.42 M$_{\sun}$, and a mean density of
13.3 gm cm$^{\rm -3}$. A main sequence star of this mass and density
should have a temperature of 3500 K ($\sim$ M2V). Yet the best fitting spectral 
template suggests T$_{\rm eff}$ = 3050 K. As discussed earlier, the $HST$
parallax indicated that the donor in U Gem has the luminosity of an M2V. 
Thus, it is only the {\it observed} spectral type that appears incongruent.

As noted above, the large inclination of U Gem ($i$ = 69$^{\circ}$) means
that the observed luminosity is dominated by the equatorial regions of the 
donor, and the enlargement in this direction from the Roche geometry means a 
lower gravity, and a cooler appearance. In the case of U Gem, the observed 
spectral type is two subtypes later than expected. SS Aur, with an orbital 
period very similar to U Gem, assuming all things being equal, should have 
approximately the same type donor star. The observed spectral type was found 
to be identical to an M3. As in the case of U Gem, the $HST$ parallax suggested
that the luminosity of SS Aur was equivalent to that of an M2V. The observed
spectral type would then be one subtype too late. The RKCat lists the 
inclination of SS Aur as $i$ = 38$^{\circ}$. It appears that inclination is 
playing a significant role in what we observe (we address contamination 
issues below).

The other non-magnetic systems near this orbital period are RX And 
(P$_{\rm orb}$ = 5.03 hr, T$_{\rm eff}$ = 
3500 K = M1.5V, $i$ = 51$^{\circ}$), TW Vir (P$_{\rm orb}$ = 4.38 hr, 
T$_{\rm eff}$ = 3600 K = M1V, $i$ = 43$^{\circ}$), CW Mon (P$_{\rm orb}$ = 4.24 
hr, T$_{\rm eff}$ = 3300 = M2.5V, $i$ = 65$^{\circ}$), and WW Cet (P$_{\rm orb}$
= 4.22 hr, T$_{\rm eff}$ = 3300 = M2.5V, $i$ = 54$^{\circ}$). If we blindly 
apply the offsets found from U Gem and SS Aur to these other systems, we
would predict that RX And should have a secondary with a spectral type that
is slightly earlier than M0V, TW Vir should have an M0V donor, 
CW Mon should have an M0.5V, and WW Cet should have an M1V. Secondary stars 
with these spectral types would have temperatures between $\sim$ 3600 K and 
3900 K, {\it in good agreement with the predictions from the relationship for 
main sequence stars}. KT Per, which has an orbital period of 3.9 hr, was found 
to have an inferred secondary star spectral type that agrees with the main 
sequence mass-radius relationship at its orbital period (though it has a very 
low metallicity). The donor in AM Her was found to have a spectral type of 
M5, while the parallax suggests the luminosity of an M3.5V. The inclination of
AM Her is $i$ = 50$^{\circ}$. 

If inclination is the source of the later-than-expected spectral types in CVs,
than we are postulating that the gravity darkening of the secondary stars is 
responsible for this result. The gravity darkening can be expressed as 
T$_{\rm eff}$ $\propto$ $g^{\beta}$, where $\beta$ is the gravity darkening 
exponent. For radiative stars $\beta$ = 0.25, while for convective stars 
$\beta$ = 0.08. This assumes that the stars are in radiative equilibrium.  
If the donor in U Gem is truly an M2V (T$_{\rm eff}$ = 3400 K), yet it appears 
to be an M4V (T$_{\rm eff}$ = 3050 K), and the derived values for the polar and
equatorial gravities noted above are assumed, we derive that $\beta$ = 0.09. 
A highly unusual limb darkening law does not appear to be necessary to explain 
the observations. However, the inclination of U Gem is not 90$^{\circ}$, but 
69$^{\circ}$. This would imply that the actual equatorial temperature is cooler 
than we have found, and the limb darkening exponent would have to be larger 
than we have derived. Fortunately, Espinosa Lara \& Rieutord (2012) have 
theoretically investigated the gravity darkening for the Roche-lobe filling 
component in semi-contact binaries. While not calculating a model with the 
exact properties of U Gem, we can interpolate the values of the gravity 
darkening exponent from their Fig. 4, giving $\beta \simeq$ 0.22. For U Gem, 
this would imply an equatorial temperature of T$_{\rm eff}$ = 2640 K.

In a similar context, Shahbaz (2003, see also Shahbaz 1998, and Bitner \&
Robinson 2006) have constructed the mean spectrum of the 
Roche-lobe filling secondary in the black hole X-ray transient system GRO 
J1655-40 (V1033 Sco) in an attempt to measure its mass ratio. For modeling this
system, the Roche surface is broken into 2048 surface elements. For each 
element the surface gravity is determined, and an effective temperature 
calculated using the appropriate gravity darkening formulation. The most 
appropriate NEXTGEN model spectrum is then assigned to that element. The 
spectra for all of the elements are then summed together to create the 
effective spectrum. For GRO J1655-40, $i$ = 70$^{\circ}$, and the secondary 
star has an apparent spectral type of mid-F. 
For their models they used both of the two standard limb darkening coefficients
($\beta$ = 0.08, and $\beta$ = 0.25). For the smaller exponent, the
temperature across the stellar surface, assuming $\langle$ T$_{\rm eff}$ 
$\rangle$ = 6336 K (F6.5V), varied from 6490 K (F5.5V) at the poles, to 5120 K 
(K1V) at the equator. As expected, for $\beta$ = 0.25, the temperature range 
was larger: 3230 K (M3V) to 6780 K (F3V). In the end, their best fitting model 
had $\langle$ T$_{\rm eff}$ $\rangle$ = 6600 K with $\beta$ = 0.08. 

We conclude that for the shorter period systems, inclination appears to be
the main source of the ``cooler than expected'' spectral types. Any remaining 
scatter can then be attributed to chemical peculiarities in the individual 
objects, or the various systems could have dramatically different binary mass 
ratios. With analogies to QZ Ser, the odd abundance patterns in U Gem and TW 
Vir suggest hydrogen deficiencies, and thus one would not expect them to 
exactly mimic the properties of normal main sequence stars. 

The other main peculiarity in the P$_{\rm orb}$ $-$ T$_{\rm eff}$ diagram is 
the existence of a set of six CVs (AE Aqr, CH UMa, V1309 Ori, V426 Oph, SS Cyg,
and AH Her), with T$_{\rm eff}$ = 4750 K, but with a large spread in orbital 
period. In every case, we find that these objects have significant
carbon deficiencies. The three longer period CVs of these six appear to
have solar values of [Fe/H] (ignoring V1309 Ori where we set [Fe/H] = $+$0.0). 
SS Cyg and AH Her have subsolar metallicities. The most reasonable explanation
for this result is that the donor star underwent nuclear evolution prior
to the system formally becoming a CV, and these objects have unusual
radii and/or masses. RU Peg is an example of a donor in a long period CV that 
is more than a magnitude more luminous than it would be if it had the radius 
of a main sequence star. 

As first shown in Podsiadlowski et al. (2003), and re-addressed
by Goliasch \& Nelson (2015), the inclusion of evolved donors in CV population 
studies leads to a considerable spread in T$_{\rm eff}$ at orbital periods 
$>$ 5 hr.  This effect arises due to some donors being zero age main sequence 
stars at the time they become semi-contact binaries, while others have 
undergone significant nuclear processing. Our results, combined with the 
growing numbers of donors in long period CVs that have unusual physical
properties (e.g., Connon Smith et al. 2005, Thoroughgood et 
al. 2005, Bitner et al. 2007, Echevarria et al. 2007, Neustroev \& Zharikov 
2008, Rodr\'{i}guez-Gil et al. 2009), confirm the conclusions of Baraffe \& Kolb
(2000) that CVs with evolved secondaries dominate the 
population. Many CV donor stars must have had initial masses that were larger
than 1 M$_{\sun}$, and our measurements of magnesium deficiencies for several 
long period CVs strengthens this argument. 

\subsection{Contamination Effects}

In the preceding we have mostly ignored the possibility of contamination
influencing the determination of donor star spectral types. Here we wish to 
briefly explore how a hot source, such as the white dwarf, or hot spot on the 
accretion disk, might alter the spectral type derivation. Qualitatively, the
addition of a hot blackbody source to the spectrum of a late-type star
should result in the derivation of an earlier spectral type than reality.
This argument can be mostly blunted by noting that the depth of various
atomic and molecular absorption features are dependent on the temperature
of the underlying donor star. Thus, the match of the continua {\it and}
absorption features of two objects with similar metallicities confirms they 
have similar spectral types.

Here, however, we have the possibility of non-solar metallicities as well
as non-standard abundance patterns, and thus cannot blindly assume that a match 
of the continuum and absorption features in two objects signals concordance 
of temperature. To test this, we simply have added blackbodies with
various fractions of the $K$-band flux to the spectrum of a M-type template
star (note that for the analysis of CVs with K-type secondaries, we have
continuum divided them, removing any subtle contamination effects). If
the contamination is at a low level, $\sim$ 25\%, we find that a donor with
a true spectral type of M5V could be confused with an M3V. Larger levels
of contamination, however, will not result in further confusion, as the 
depression of the continuum due to water vapor at both ends of the $K$-band 
spectrum of an M5V remains imprinted on the continuum of the contaminated 
spectrum. Thus, the spectrum of a highly contaminated M5V cannot be confused 
with templates that have spectral a type of M2V or earlier.

It is possible for those objects with very low values of [Fe/H] to have weak
water vapor features. Their continua would be easier to reproduce with
larger contamination fractions. The weakening/dilution of the spectral features
that comes with a larger contamination would imply an even lower
metallicity. This suggests that several of the shortest period systems with
the lowest metallicities in Table 3, such EX Hya and V893 Sco, are affected
by this process. But we have already identified that these two objects
suffer significant contamination. Thus, the process of deriving T$_{\rm eff}$
from NIR spectra is surprisingly robust.

The other argument that we are not finding earlier spectral types than reality
comes from analysis of those objects with astrometric distances. In nearly
every case, the luminosity of the CV system in the $K$-band implies an
earlier spectral type for the donor than found here. There is no doubt
that the $K$-band spectra of U Gem and SS Aur suffer little contamination:
Harrison et al. (2000) found that the contamination of the $K$-band 
luminosities of both objects was less than 5\%. Yet we find 
that in both cases, the observed spectral type is too late to explain their 
observed $K$-band luminosities. In SS Aur, we need an M2V to explain the 
$K$-band luminosity, while we derive a donor spectral type of M3V. An M2V is a 
full magnitude more luminous than an M3V, and the contaminating source would
have to supply more than 70\% of the observed flux. In U Gem the situation
is even worse, with the contaminating source having to supply over 90\%
of the observed luminosity. While contamination is present at some level in 
every CV, it certainly is not seriously affecting our results.

\subsection{The Metallicities of the Donors in Magnetic CVs}

One of the most consistent findings from our study is the sub-solar 
metallicities of the IPs and polars. The only real outlier is MQ Dra, which 
appears to have a slightly super-solar value for [Fe/H]. MQ Dra 
(= SDSS1553+5516) has been classified as a low-accretion rate polar (LARP, 
Schwope et al.  2002). From X-ray observations and the width of the cyclotron 
features seen in optical spectroscopy, the derived plasma temperatures at the 
accretion shock region in LARPs are much cooler than typically found in polars 
($\sim$ 1 keV, vs. $\gtrsim$ 5 keV). The derived accretion rates in LARPs are 
also very low compared with normal polars (10$^{\rm -14}$ M$_{\sun}$ 
yr$^{\rm -1}$, vs.  $>$ 10$^{\rm -13}$ M$_{\sun}$ yr$^{\rm -1}$). In addition, 
no ``high states,'' periods of increased mass transfer rate that frequently 
occur in polars, have yet been observed for a LARP.
Schmidt et al. (2005) suggest that these objects are ``pre-polars,'' in that 
the orbital period has not yet evolved to short enough values for the 
secondaries in these systems to contact their Roche lobes. The cyclotron 
emission then results from mass transfer through a stellar wind from the 
donor star.

The position of MQ Dra in the T$_{\rm eff}$ vs. P$_{\rm orb}$ diagram is very
unusual, having a donor spectral type that is much cooler than the non-magnetic 
CVs at its orbital period. The derived spectral type is nearly as 
late as those seen in the shortest period polars. The derived inclination
from light curve modeling in Szkody et al. (2008) indicates an inclination
near $i$ = 35$^{\circ}$. If we assume it is a normal main sequence star,
and its spectral type is 1 subtype earlier than observed, it would have a 
radius of R$_{\rm 2}$ = 0.16 M$_{\sun}$. To fill its Roche lobe, it would need 
to have a radius that was twice this size. Schmidt et al. (2005) and Schwartz 
et al. (2001) found similar results for other LARPs. 

The other short period polars would be expected to have temperatures near 
T$_{\rm eff}$ = 3100 K (M4V), consistent with our results (not accounting for 
inclination effects). Thus, the suggestion that LARPs are pre-polars appears
to be valid. The super-solar metallicity of MQ Dra could also be 
interpreted as a sign of youth. It would be extremely useful to obtain $K$-band
spectra of all of the other LARPs to explore their metallicities, and to 
confirm/refine the spectral types of their donor stars.

\section{Conclusions}

We have ascertained the metallicity for the donor stars in a large number
of CVs using NIR spectroscopy. For CVs with K-type donor stars, temperatures
and abundances have been derived through the comparison of synthetic spectra
to observations. For CVs with M-type stars, the values for [Fe/H] were
derived from recently developed relationships for field M dwarfs that employ 
$K$-band spectroscopy. We find that the donor stars in CVs generally 
have subsolar metallicities. This is consistent with results for field stars 
in the solar neighborhood, where the local stellar population is shown to be a 
mix of thin and thick disk stars (Nordstr\"{o}m, et al. 2004). An analysis of 
the kinematics of CVs by Ak et al. (2015) derives a similar result. The donors
in longer period CVs appear to have more dramatic abundances anomalies than
do the short period systems. In agreement with predictions from population
synthesis modeling. The presence of Mg deficits in several long period CVs
argues that the donor stars in these systems had large initial masses.

A caveat for the determination of [Fe/H] exists for CVs with M-type donors, as 
variations in the C/O ratio could lead to the derivation of artificially low 
values for [Fe/H] (Veyette et al. 2016). This concern is especially relevant 
given that many CVs have carbon deficits. Our comparison of the high S/N 
spectrum of U Gem to that of GJ 402 showed that the predicted metallicity offset
implied by the observed carbon deficit produced a better agreement between the 
two objects. However, application of a similar offset to the derived [Fe/H] 
value for carbon-deficient YY/DO Dra made it {\it less} consistent with the 
best-fitting template. Due to the extreme carbon deficits seen in some
donors, CVs could provide the data to better establish and/or calibrate the
effect of the C/O ratio on the metallicity estimates for normal field stars. 
This could be easily accomplished with new multi-bandpass NIR spectroscopy of 
those CVs in Table 3 with proven carbon deficits.

It is also now abundantly clear that while the donor stars in CVs may 
superficially resemble their main sequence cousins, many of them have undergone
significant nuclear evolution prior to becoming a CV. The inclusion of such 
objects into recent population studies is a welcome change, and demonstrates 
that stellar evolution processes alone are able to explain many of our results. 
It would be extremely useful to have model atmospheres for cool stars that had
both hydrogen deficiencies and enhanced levels of helium, so as to enable the 
generation of more realistic synthetic spectra. New models for the outbursts of
dwarf novae that employ non-solar abundance patterns and/or hydrogen 
deficiencies should also be considered, so as to investigate whether 
composition has any affect on outburst properties. The same is true for models 
of classical novae eruptions, as the composition of the accreted material 
appears to alter the nature of these events (Shen \& Bildsten 2009).

We close by noting that the total mass of the U Gem system is larger than
the Chandrasekhar limit, and if the mass lost to CNe eruptions is not too
great, the system could become a Type Ia supernova. Note that if the donor
star in U Gem is even modestly hydrogen deficient, one of the main arguments 
against single-degenerate CVs as SNIa progenitors is removed: the lack of
H I emission lines in their post-deflagration spectra. Pakmor et al. (2008) 
have found that models for the amount of matter stripped from a main sequence 
donor star in an SNIa explosion is roughly consistent with the limits implied 
by observations. Hydrogen deficits in CV donors would only aid this result. 

\acknowledgements TEH was partially supported by a grant from the NSF 
(AST-1209451). This publication makes use of data products from the Two Micron 
All Sky Survey, which is a joint project of the University of Massachusetts and 
the Infrared Processing and Analysis Center/California Institute of Technology,
funded by the National Aeronautics and Space Administration and the National 
Science Foundation. We acknowledge with thanks the variable star observations 
from the AAVSO International Database contributed by observers worldwide and 
used in this research. This publication makes use of data products from NEOWISE,
which is a project of the Jet Propulsion Laboratory/California Institute of 
Technology. NEOWISE is funded by the National Aeronautics and Space 
Administration. The author wishes to acknowledge S. Howell and P. Szkody
who aided in obtaining observations of several of the targets included here.

\clearpage
\section{References}
\begin{flushleft}
Ak, T., Bilir, S., \"{O}zd\"[o]nmez, A., Soydugan, F., Soydugan, E., et al.
2015, ApSS, 357, 72\\
Baptista, R., Catal\'{a}n, M. S., \& Costa, L. 2000, MNRAS, 316, 529\\
Baraffe, I., Homeier, D., Allard, F., \& Chabrier, G. 2015, A\&A, 577, 42\\
Baraffe, I., \& Kolb, U. 2000, MNRAS, 318, 354\\
Bean, J., Sneden, C., Hauschildt, P. H., \& Johns-Krull, C. M., 2006, ApJ, 652, 1604\\
Behara, N. T., \& Jeffery, C. S. 2006, BaltAst, 15, 115\\
Beuermann, K., Harrison, T. E., McArthur, B. E., Benedict, G. F., \& G\"{a}nsicke, B. T. 2003, A\&A, 412, 821\\
Beuermann, K., Baraffe, I., Kolb, U., \& Wiechhold, M. 1998, A\&A, 339, 518\\
Bitner, M. A., Robinson, E. L., \& Behr, B. B. 2007, ApJ, 662, 564\\
Bitner, M. A., \& Robinson, E. L. 2006, AJ, 131, 1712\\
Campbell, R. K., Harrison, T. E., \& Kafka, S. 2008, ApJ, 684, 409\\
Cesetti, M., Pizzella, A., Ivanov, V. D., Morelli, L., Corsini, E. M., et al.
2013, A\&A, 549, 129\\
Connon Smith, R., Mehes, O., Vande Putte, D., \& Hawkins, N. A. 2005, MNRAS,
360, 364\\
Covey, K. R., Lada, C. J., Roman-Zuniga, C., Meunch, A. A., Forbrich, J., \&
Ascenso, J. 2010, ApJ, 722, 971\\
Cushing, M.C. Rayner, J.T., \& Vacca, W.D., ApJ, 2005, 623, 1115\\
Cunha, K., Smith, V. V., Johnson, J. A., Bergemann, M., M\'{e}sz\'{a}ros, S., 
et al.  2015, ApJ, 798, L41\\
Cushing, M. C., Vacca, W. D., \& Rayner, J. T. 2004, PASP, 116, 362\\
Denissenkov, P. A., Truran, J. W., Pignatari, M., Trappitsch, Ritter, C. et al.
MNRAS, 442, 2058\\
Dhillon, V. S., Littlefair, S. P., Marsh, T. R., Sarna, M. J., \& Boakes, E. H.
2002, A\&A, 393, 611\\
Dhillon, V. S., Littlefair, S. P., Howell, S. B., Ciardi, D. R., et al. 2000,
MNRAS, 314, 826\\
Echevarria, J., Connon Smith, R., Costero, R., Zharikov, S., \& Michel, R.
2008, MNRAS, 387, 1564\\
Echevarria, J., Michel, R., Costero, R., \& Zharikov, S. 2007, A\&A, 462, 1069\\
Echevarria, J., 1983 RMxAA, 8 , 109\\
Espinosa Lara, F., \& Rieutord, M. 2012, A\&A, 547, 32\\
Fernandez, J. M., Latham, D. W., Torres, G., Everett, M. E., Mandushev, G.,
et al. 2009, ApJ, 701, 764\\
Friend, M. T., Martin, J. S., Smith, R. C., \& Jones, D. H. P. 1990, MNRAS,
246, 637\\
Gaidos, E., Mann, A. W., L\'{e}pine, S., Buccino, A., James, D., et al. 2014,
MNRAS, 443. 2561\\
G\"{a}nsicke, B. T., Szkody, P., de Martino, D., Beuermann, K., Long, K. S.,
et al. (2003), ApJ, 594, 443\\
Garnavich, P. M., Szkody, P., Robb, R. M., Zurek, D. R., \& Hoard, D. W.
1994, ApJ, 435, 141\\
Godon, P., Sion, E. M., Barrett, P. E., \& Linnell, A. P. 2009, ApJ, 699, 1229\\
Goliasch, J., \& Nelson, L. 2015, ApJ, 809, 80\\
Gonzalez, G., G., \& Wallerstein, G. 2000, AJ, 119, 1839\\
Hamilton, R. T., Harrison, T. E., Tappert, C., \& Howell, S. B. 2011, ApJ, 728, 16\\
Harrison, T. E. 2016, ApJ, 816, 4\\
Harrison, T. E., \& McArthur, B. E. 2016, AAS 227, 239.07\\
Harrison, T. E., \& Hamilton, R. T. 2015 (H\&H), AJ, 150, 142\\
Harrison, T. E., \& Campbell, R. K. 2015, ApJSupp, 219, 32\\
Harrison, T. E., Bornak, J., McArthur, B. E., \& G. F. Benedict 2013, ApJ,
767, 7\\
Harrison, T. E., Bornak, J., Rupen, M. P., \& Howell, S. B. 2010, ApJ, 710,
325\\
Harrison, T. E., Bornak, J., Howell, S. B., Mason, E., Szkody, P., et al. 2009,
AJ, 137, 4061\\
Harrison, T. E., Campbell, R. K., Howell, S. B., Cordova. F. A., \& Schwope,
A. D. 2007a, ApJ, 656, 444\\
Harrison, T. E., Howell, S. B., Szkdoy, P., \& Cordova, F. A. 2007b, AJ, 133,
162\\
Harrison, T. E., Howell, S. B., Szkody, P., \& Cordova, F. A. 2005a, ApJ,
632, L123\\
Harrison, T. E., Osborne, H. L., \& Howell, S. B. 2005b, AJ, 129, 2400\\
Harrison, T. E., Osborne, H. L., \& Howell, S. B. 2004a, AJ, 127, 3493\\
Harrison, T. E., Johnson, J. J., McArthur, B. E., Benedict, G. F., Szkody, P.
2004b, AJ, 127, 460\\
Harrison, T. E., McNamara, B. J., Szkody, P., \& Gilliland, R. L. 2000, AJ,
120, 2649\\
Hessman, F. V., 1988, A\&AS, 72, 515\\
Houdebine, E. R., \& Mullan, D. J. 2015, ApJ, 801, 106\\
Howell, S. B., Harrison, T. E., Szkody, P., \& Silvestri, N. M. 2010, AJ, 139, 1771\\
Howell, S. B., Harrison, T. E., \& Szkody, P. 2004, ApJ, 602, L49\\
Howell, S. B., Nelson, L. A., \& Rappaport, S. 2001, ApJ, 550, 897\\
Husser, T. -0., Wende-von Berg, S., Dreizler, S., Homeier, D., Reiners, A.,
et al. 2013, A\&A, 553, 6\\
Iben, I., \& Livio, M. 1993, PASP, 150, 1373\\
Kato, T., Uemura, M., Kiyota, S., Tanabe, K., Koizumi, M. et al. 2003,
PASJ, 55, 489\\
Kolb, U., King, A. R., \& Ritter, H. 1998, MNRAS, 298, 29\\
Knigge, C., Baraffe, I., \& Patterson, J. 2011, ApJSupp, 194, 28\\
Knigge, C. 2006, MNRAS, 373, 484\\
Kraft, R. P. 1962, ApJ, 135, 408\\
Long, K. S., \& Gilliland, R. L. 1999, ApJ, 511, 916\\
Mann, A. W., Feiden, G. A., Gaidos, E., Boyajian, T., \& von Braun, K. 2015,
ApJ, 804, 64\\
Mann, A. W., Deacon, N. R., Gaidos, E., Ansdell, M., Brewer, J. M., et al.
2014, AJ, 147, 160\\
Mann, A. W., Brewer, J. M., Gaidos, E., Lepine, S., \& Hilton, E. J. 2013,
AJ, 145, 52\\
Marks, P. B., \& Sarna, M. J. 1998, MNRAS, 301, 699\\
Miller-Jones, J. C. A., Sivakoff, G. R., Knigge, C., K\"{o}rding, E. G.,
Templeton, M., et al. 2013, Sci, 340, 950\\
Naylor, T., Allan, A., \& Long, K. S. 2005, MNRAS, 361, 1091\\
Newton, E. R., Charbonneau, D., Irwin, J., Berta-Thompson, Z. K., Rojas-Ayala,
B., et al. 2014, AJ, 147, 20\\
\"{O}nehag, A., Heiter, U., Gustafsson, B., Piskunov, N., Plez, B., \& Reiners,
A. 2012, A\&A, 542, 33\\
Neustroev, V. V., \& Zharikov, S. 2008, MNRAS, 386, 1366\\
Nordst\"{o}m, B., Mayor, M., Andersen, J., Holmberg,J., Pont, F., et al. 2004,
A\&A, 418, 989\\
Pakmor, R., R\"{o}pke, F. K., Weiss, A., \& Hilllebrandt, W. 2008, A\&A, 489,
943\\
Parsons, S. G., Marsh, T. R., Copperwheat, C. M., Dhillon, V. S., Littlefair,
S. P., et al. 2010, MNRAS, 402, 2591\\
Patterson, J. 1998, PASP, 110, 1132\\
Patterson, J., \& Szkody, P. 1993, PASP, 105, 1116\\
Patterson, J., \& Eisenman, N. 1987, IBVS 3079\\
Pecaut, M. J., \& Mamajek, E. E. 2013, ApJS, 208, 9\\
Podsiadlowski, Ph., Han, Z., \& Rappaport, S. 2003, MNRAS, 340, 1214\\
Politano, M., \& Weiler, K. P. 2007, ApJ, 665, 663\\
Prantzos, N., Coc, A., \& Thibaud, J. P. 1991, ApJ, 379, 729\\
Rajpurohit, A. S., Rehl\'{e}, Allard, F., Homeier, D., Schultheis, M., et al.
2013, A\&A, 556, 15\\
Ramsay, G., Wheatley, P. J., Norton, A. J., Hakala, P., \& Baskill, D. 2008,
MNRAS, 387, 1157\\
Ribeiro, T., Baptista, R., Harlaftis, E. T., Dhillon, V. S., \& Rutten, R. G.
M. 2007, A\&A, 474, 213\\
Ringwald, F. A., Thorstensen, J. R., Honeycutt, R. K., \& Smith, R. C. 1996,
AJ, 111, 2077\\
Ringwald, F. A., Thorstensen, J. R., \& Hamwey, R. M. 1994, MNRAS, 271, 323\\
Ritter, H., \& Kolb, U. 2003, A\&A, 404, 301\\
Rodr\'{i}guez-Gil, P., Torres, M. A. P., G\"{a}nsicke, B. T., Mu\~{n}oz-Darias,
T., Steeghs, D., et al. 2009, A\&A, 496, 805\\
Rojas-Ayala, B., Covey, K. R., Muirhead, P. S., \& Lloyd, J. P. 2010, ApJL,
720, L113\\
Schmidt, G. D. , Szkody, P., Vanlandingham, K. M., Anderson, S. F., Barentine,
J. C., et al. 2005, ApJ, 630, 1037\\
Schwartz, R., Schwope, A. D., \& Staude, A. 2001, A\&A, 374, 189\\
Schwope, A. D., Brunner, H., Hambaryan, V., Schwarz, R., Staude, A., et al.
2002, in ASP Conf. Vol. 261, The Physics of Cataclysmic Variables and Related 
Objects, ed. B.T. G\"{a}nsicke, K. Beuermann, \& K. Reinsch (San Francisco: 
ASP), 102\\
Schreiber, M. R., G\"{a}nsicke, B. T., \& Mattei, J. A. 2002, A\&A, 384, 6\\
Sepinski, J. F., Sion, E. M., Szkody, P., \& G\"{a}nsicke, B. T. 2002, ApJ,
574, 937\\
Shahbaz, T. 2003, MNRAS, 339, 1031\\
Shahbaz, T. 1998, MNRAS, 298, 153\
Shen, K. J., \& Bildsten, L. 2009, ApJ, 692, 324\\
Sion, E. M., Godon, P., Cheng, F., \& Szkody, P. 2007, AJ, 134, 886\\
Sion, E. M., Cheng, F. H., Godon, P., \& Szkody, P. 2004, RevMex, 20, 194\\
Sion, E. M., Winter, L., Urban, J. A., Tovmassian, G. H., Zharikov, S., et al.
2004, AJ, 128, 1795\\
Smiljanic, R., Romano, D., Bragaglia, A., Donati, P., Magrini, L., et al.
2016, A\&A, 589, 115\\
Smith, V. V., Cunha, K., Shetrone, M. D., Meszaros, S.,Allende Prieto, C., et
al. 2013, 765, 16\\
Sneden, C. 1973, PhD thesis, Univ. Texas\\
Sterken, C., Vogt, N., Schreiber, M. R., Uemura, M., \& Tuvikene, T. 2007,
A\&A, 463, 1053\\
Szkody, P., Linnell, A. P., Campbell, R. K., Plotkin, R. M., Harrison, T. E.,
et al. 2008, ApJ, 683, 967\\
Szkody, P., Nishikida, K., Erb, D., Mukai, K., Hellier, C., et al. 2002, AJ,
123, 413\\
Szkody, P., \& Silber, A. 1996, AJ, 112, 289\\
Szkody, P., Williams, R. E., Margon, B., Howell, S. B., \& Mateo, M. 1992,
ApJ, 387, 357\\
Szkody, P., \& Mateo, M. 1986, ApJ, 301, 286\\
Szkody, P., \& Capps, R. W. 1980, AJ, 85, 882\\
Taylor, C. J., Thorstensen, J. R., \& Patterson, J. 1999, PASP, 111, 184\\
Thoroughgood, T. D., Dhillon, V. S., Steeghs, D., Watson, C. A., Buckley, D. A.
H., et al. 2005, MNRAS, 357, 881\\
Thorstensen, J. R., Lepine, S., \& Shara, M. 2008, AJ, 136, 2107\\
Thorstensen, J. R., Fenton, W. H., \& Taylor, C. J. 2004a, PASP, 116, 300\\
Thorstensen, J. R. 2003, AJ, 126, 3017\\
Thorstensen, J. R., Fenton, W. H., Patterson, J., Kemp, J., Halpern, J., et al.
2002a, PASP, 114, 1117\\
Thorstensen, J. R., Fenton, W. H., Patterson, J. O., Kemp, J., Krajci, T.
et al. 2002b, ApJ, 567, L49\\
Thorstensen, J. R., \& Ringwald, F. A. 1997, PASP, 109, 483\\
Ventura, P., Di Criscienzo, M., Carini, R., \& D'Antona, F. 2013, MNRAS, 431,
3642\\
Verbunt, F., Bunk, W. H., Ritter, H., \& Pfeffermann, E. 1997, A\&A, 327, 602\\
Verbunt, F., \& Zwaan, C. 1981, A\&A, 100, 7\\
Veyette, M. J., Muirhead, P. S., Mann, A. W., \& Allard, F. 2016, 
arXiv160504904\\
von Zeipel, H. 1924, MNRAS, 84, 665\\
Wallerstein, G., Iben, I., Parker, P., Boesgaard, A. M., Hale, G. M. 1997,
RvMP, 69, 995\\
Warner B., 1995, Cataclysmic Variable Stars. Cambridge Univ. Press,
Cambridge, p 33\\
Watson, C. A., Dhillon, V. S., Shahbaz, T. 2006, MNRAS, 368, 637\\
Wehrse, R. \& Shaviv, G.  1995,  in  ASP  Conf.  Ser.  78,  Astrophysical  
Applications  of Powerful New Databases, ed. S. J. Adelman \& W.L. Wiese (San 
Francisco, CA: ASP), p301\\
Wilson, J. C., Henderson, C. P., Herter, T. L., Matthews, K., Skrutskie, M. F.,
et al. 2004, SPIE, 5492, 1295\\
Wing, R. F., \& J\o rgensen, U. G. 2003, JAAVSO, 31, 110\\
Woolf, V. M., \& West, A. W. 2012, MNRAS, 422, 1489\\
\end{flushleft}
\begin{deluxetable}{lccccccc}
\tablecolumns{8}
\tablewidth{0pt}
\centering
\tablecaption{Observation Log}
\tablehead{
\colhead{Object}&\colhead{UT Date}&\colhead{Start/Stop Times}&\colhead{Exp.}&\colhead{Instrument}&\colhead{P$_{\rm orb}$ (hr)}&\colhead{Obs. Phase}&\colhead{State} }
\startdata
GK Per & 2004-08-15& 13:53 to 14:19&240s&SPEX   & 48.1&0.94& Min.\\
EY Cyg & 2004-08-16& 11:05 to 11:57&180s&SPEX   &11.02&0.97& Min.\\
AE Aqr & 2004-08-15& 08:05 to 09:48& 90s&SPEX   & 9.88&0.61& Min.\\
RU Peg & 2004-08-15& 10:25 to 10:59&240s&SPEX   & 8.99&0.85& Min.\\
SS Cyg & 2004-08-15& 08:55 to 09:20&120s&SPEX   & 6.60&0.40& Min.\\
CZ Ori & 2005-02-17& 05:00 to 05:20&180s&NIRSPEC& 5.25&0.66& Nearing min.\\
RX And & 2004-08-15& 13:07 to 13:41&240s&SPEX   & 5.03&0.59& Min.\\
SS Aur & 2005-09-01& 15:02 to 15:26&240s&SPEX   & 4.39&0.12& Min.\\
SS Aur & 2007-03-05& 06:00 to 07:18&120s&NIRSPEC& 4.39&0.99& Min. \\
U Gem  & 2012-02-03& 02:22 to 07:51&240s&TSPEC  & 4.25&\nodata&Min.\\
U Gem  & 2012-02-03& 08:40 to 10:34&240s&TSPEC  & 4.25&\nodata&Min.\\
CW Mon & 2005-02-17& 05:29 to 05:40&180s&NIRSPEC& 4.24&0.28&Min.\\
WW Cet & 2004-08-15& 12:20 to 12:54&240s&SPEX   & 4.22&0.60&Min.\\
LS Peg & 2004-08-15& 11:34 to 12:08&240s&SPEX   & 4.19&0.68&Min.\\
IP Peg & 2003-09-06& 06:39 to 07:00&120s&NIRSPEC& 3.98&0.03&Min.\\
DO Dra & 2005-02-17& 09:50 to 09:59&240s&NIRSPEC& 3.97&0.86&Min.\\
CN Ori & 2005-02-17& 05:44 to 05:49&180s&NIRSPEC& 3.92&0.33&Min.\\
KT Per & 2004-08-16& 14:31 to 14:56&240s&SPEX   & 3.90&0.16&Intermed.\\
TT Ari & 2004-08-15& 14:56 to 15:22&240s&SPEX   & 3.30&0.08&Min.\\
QZ Ser & 2012-02-03& 11:08 to 12:45&240s&TSPEC  & 2.00&0.67&$V$ = 15.6\\
\enddata
\label{obslog}
\end{deluxetable}

\begin{sidewaystable}
\centering
\caption{Temperature and [Fe/H] Values for the Spectral Templates}
\begin{tabular}{lcccccc}
\hline
\hline
Spectral Type  & $I$-band & $J$-band & $H$-band & $K$-band & Modeling Means& Nominal Values\\
\hline
K0V& 5500, $+$0.5& 5500, $+$0.25& 5250, $+$0.25& 5500, +0.0& 5437$\pm$125, +0.25$\pm$0.20 &5320$\pm$114, +0.41$\pm$0.12\\
K1V& 5250, $+$0.0& 5000, $-$0.3& 5000, $-$0.2& 5250, $-$0.3&5125$\pm$125, $-$0.20$\pm$0.12 &5189$\pm$53,  $-$0.06$\pm$0.06\\
K2V& 5000, $-$0.1& 5000, $-$0.1& 4750, $-$0.1& 5250, $+$0.0&5000$\pm$204, $-$0.08$\pm$0.05 &5023$\pm$66,  $+$0.05$\pm$0.10\\
K3V& 4500, $-$0.1& 4500, $-$0.2& 4750, $+$0.0& 4750, $-$0.1&4750$\pm$204, $-$0.10$\pm$0.08 &4837$\pm$138, $+$0.05$\pm$0.10\\
K4V& 4500, $+$0.0& 4500, $-$0.1& 4750, $+$0.0& 4750, $-$0.1&4625$\pm$144, $-$0.05$\pm$0.06 &4689$\pm$174, $+$0.03$\pm$0.18\\
K5V& 4500, $+$0.0& 4250, $-$0.3& 4750, $-$0.5& 4750, $-$0.1&4562$\pm$239, $-$0.23$\pm$0.22 &4615$\pm$29, $-$0.14$\pm$0.08\\
K7V& 4250, $-$0.1& 4250, $-$0.5& 4500, $-$0.1& 4250, $-$0.3&4312$\pm$125, $-$0.25$\pm$0.19 &4110$\pm$102, $-$0.21$\pm$0.10\\
\hline
\end{tabular}
\label{chisq}
\end{sidewaystable}
\begin{deluxetable}{lccclll}
\tablecolumns{8}
\tablewidth{0pt}
\centering
\tablecaption{Observation Log}
\tablehead{
\colhead{Object}&\colhead{Subtype}&\colhead{P$_{\rm orb}$}&\colhead{T$_{\rm eff}$}&\colhead{[Fe/H]}&\colhead{[C/Fe]}&\colhead{Other Abundances} }
\startdata
GK Per   &IP        &47.92     &5000 &$-$0.3 &$-$0.5   &[Mg/Fe] = $-$0.3 \\
EY Cyg   &UG        &11.02     &5250 &$+$0.0 &$-$0.5   &\nodata          \\
AE Aqr   &IP         &9.88     &4750 &$+$0.0 &$-$1.0   &\nodata \\
SY Cnc   &ZC         &9.18     &5500 &$+$0.0 &$+$0.0   &\nodata \\
RU Peg   &SS         &8.99     &5000 &$-$0.3 &$-$0.4   &[Mg/Fe] = $-$0.14\\
CH UMa   &UG         &8.24     &4750 &$+$0.0 &$-$1.0   &\nodata \\
V1309 Ori&AM         &7.98     &4750 &$+$0.0=&$-$1.0   &[Mg/Fe] = $-$0.5  \\
EM Cyg   &ZC         &6.98     &4500 &$-$0.5 &$-$0.2   &\nodata \\
V426 Oph &ZC         &6.85     &4750 &$+$0.0 &$-$0.5   &\nodata \\
SS Cyg   &UG         &6.60     &4750 &$-$0.3 &$-$0.4   &[Mg/Fe] = $-$0.3  \\
TT Crt   &UG         &6.44     &4500 &$+$0.0 &$-$0.5   &\nodata \\
AH Her   &UG/ZC      &6.19     &4750 &$-$0.7 &$-$0.4   &\nodata \\
CZ Ori   &UG         &5.25     &4250 &$-$1.0 &$+$0.0=  &\nodata \\
EX Dra   &UG         &5.04     &4000 &$-$0.5 &$-$0.2   & [Mg/Fe] = $+$0.0 \\
RX And   &ZC?        &5.03     &3500 &$+$0.1 &$+$0.0   &\nodata \\
MQ Dra   &AM         &4.39     &2800 &$+$0.2 &$+$0.0   &\nodata \\
SS Aur   &UG         &4.39     &3100 &$-$0.1 &$+$0.0   &\nodata \\
TW Vir   &UG         &4.38     &3600 &$+$0.6 &\nodata  &\nodata \\
U Gem    &UG         &4.25     &3050 &$+$0.2 &$-$1.0   &\nodata \\
CW Mon   &UG         &4.24     &3300 &$-$0.2 &$+$0.0   &\nodata \\
WW Cet   &UG         &4.22     &3300 &$-$0.3 &$\geq -$0.3&\nodata \\
YY/DO Dra&IP         &3.97     &3200 &$-$0.2 &$\leq -$0.3&\nodata \\
UU Aql   &UG         &3.92     &3375 &$-$0.7 &\nodata  &\nodata \\
CN Ori   &UG         &3.92     &3375 &$-$0.7 &\nodata  &\nodata \\
KT Per   &ZC         &3.90     &3580 &$-$0.5 &$+$0.0   &\nodata \\
IP Peg   &UG         &3.80     &3625 &$+$0.2 &$\leq$0.0&\nodata \\
AM Her   &AM         &3.09     &2900 &$-$0.4 &$\leq$0.0&\nodata \\
QZ Ser   &SU &2.00 &4530 &$+$0.0=&$-$1.7 &[Na/Fe] = $+$1.5, [Al/Fe] = $+$0.5 \\
AR UMa   &AM         &1.93     &2925 &$-$0.5 &$+$0.0   &\nodata \\
ST LMi   &AM         &1.90     &2775 &$-$0.4 &$+$0.0   &\nodata \\
MR Ser   &AM         &1.89     &3000 &$-$0.2 &$+$0.0   &\nodata \\
RZ Leo   &WZ         &1.83     &2925 &$-$0.3 &$\leq$0.0&\nodata \\
V893 Sco &SU         &1.82     &2860 &$-$0.8 &\nodata  &\nodata \\
VW Hyi   &SU         &1.78     &3050 &$-$0.3 &$+$0.0   &\nodata \\
VV Pup   &AM         &1.67     &2690 &$-$0.3 &$+$0.0   &\nodata \\
EX Hya   &IP         &1.64     &3050 &$-$0.6 &$\leq -$0.3&\nodata \\
VY Aqr   &SU         &1.51     &2800 &$+$0.4 &$\lesssim$ $-$1.0&\nodata \\
V436 Cen &SU         &1.50     &2670 &$-$0.3 &$+$0.0   &\nodata \\
V2051 Oph&SU         &1.50     &2500 &$-$0.1 &$+$0.0   &\nodata \\
WX Cet   &SU         &1.40     &2600 &$+$0.0:&\nodata  &\nodata \\
EI Psc   &SU         &1.07     &4530 &$-$0.2 &$-$1.7   &[Na/Fe] = $+$0.9\\
\enddata
\label{results}
\end{deluxetable}
\clearpage
\renewcommand{\thefigure}{1a}
\begin{figure}
\centerline{{\includegraphics[width=12cm]{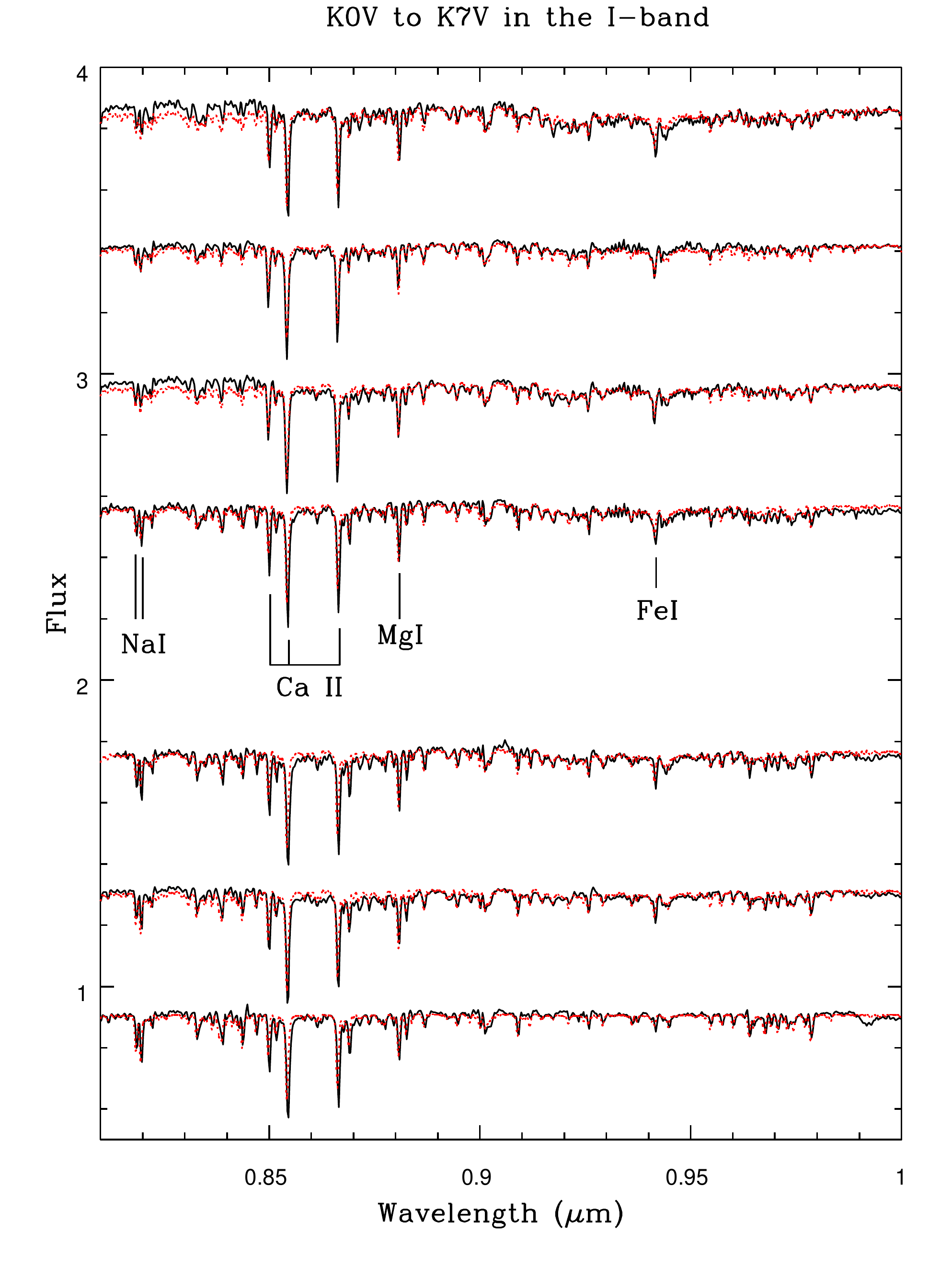}}}
\caption{Spectra of the K dwarf templates, K0V top, K7V bottom (black),
over the wavelength range 0.81 $\mu$m $\leq$ $\lambda$ $\leq$ 1.0 $\mu$m.
The synthetic model spectra generated using SPECTRUM are in red. The strongest 
absorption lines have been identified.}
\label{ibandcomp}
\end{figure}

\renewcommand{\thefigure}{1b}
\begin{figure}[htb]
\centerline{{\includegraphics[width=12cm]{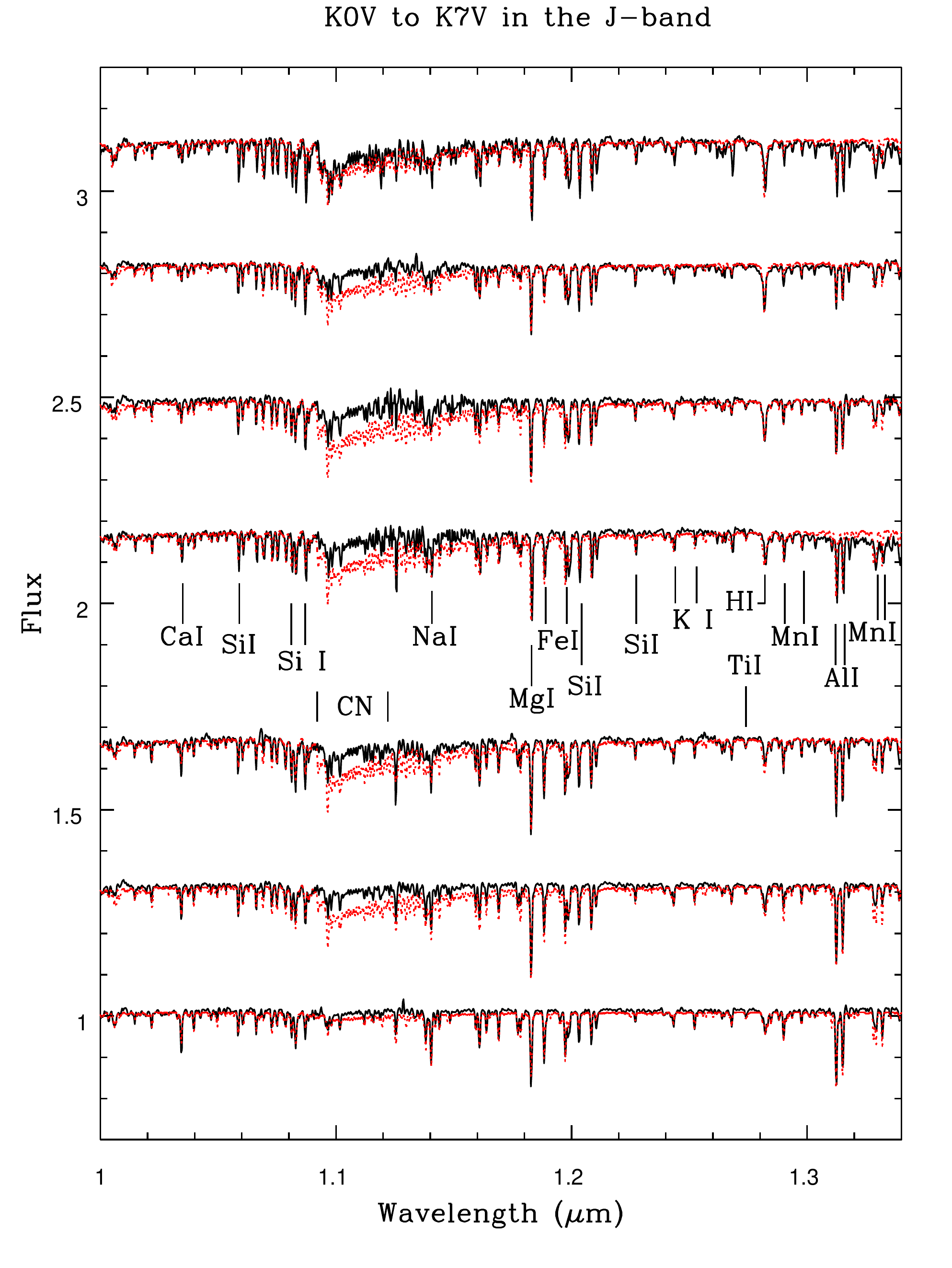}}}
\caption{The same as 1a, but for the wavelength interval 1.0 $\mu$m $\leq$ $\lambda$ $\leq$ 1.34 $\mu$m.}
\label{jbandcomp}
\end{figure}
\renewcommand{\thefigure}{1c}
\begin{figure}[htb]
\centerline{{\includegraphics[width=12cm]{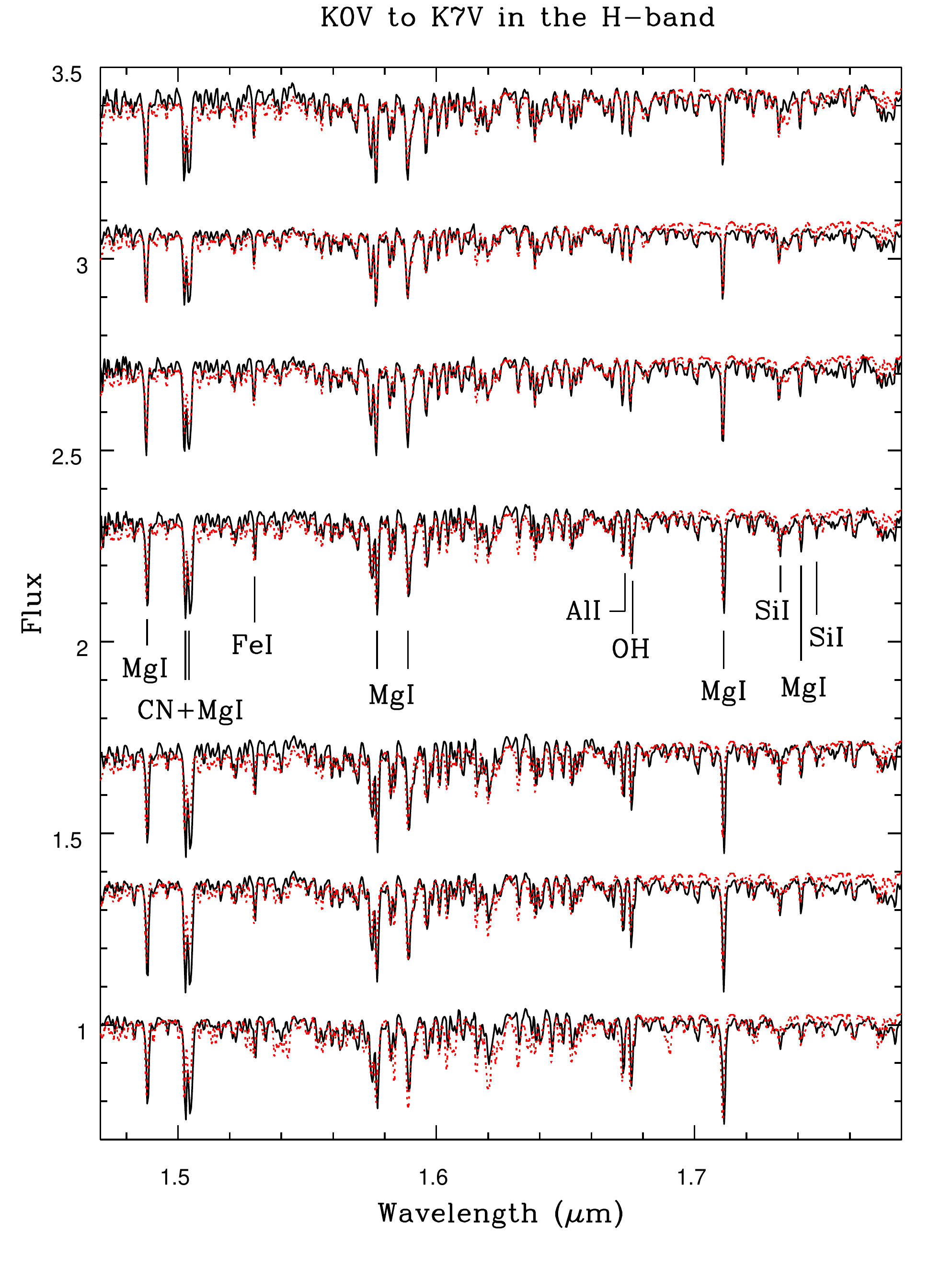}}}
\caption{As in 1a, but for the wavelength interval 1.47 $\mu$m $\leq$ $\lambda$ $\leq$ 1.78 $\mu$m.}
\label{hbandcomp}
\end{figure}
\renewcommand{\thefigure}{1d}
\begin{figure}[htb]
\centerline{{\includegraphics[width=12cm]{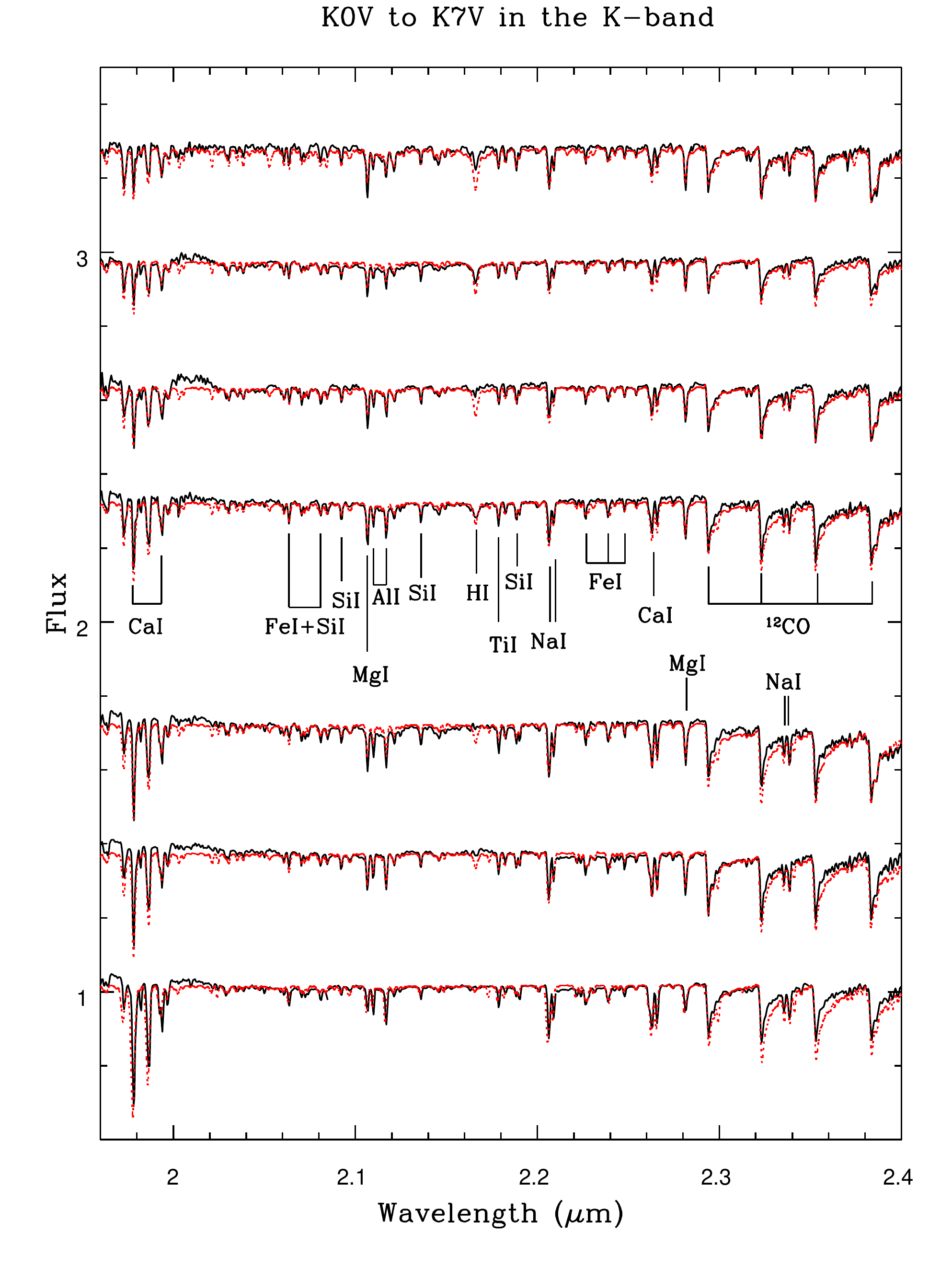}}}
\caption{As in 1a, but for the wavelength interval 1.96 $\mu$m $\leq$ $\lambda$ $\leq$ 2.4 $\mu$m.}
\label{ibandcomp}
\end{figure}

\renewcommand{\thefigure}{2}
\begin{figure}[htb]
\centerline{{\includegraphics[width=12cm]{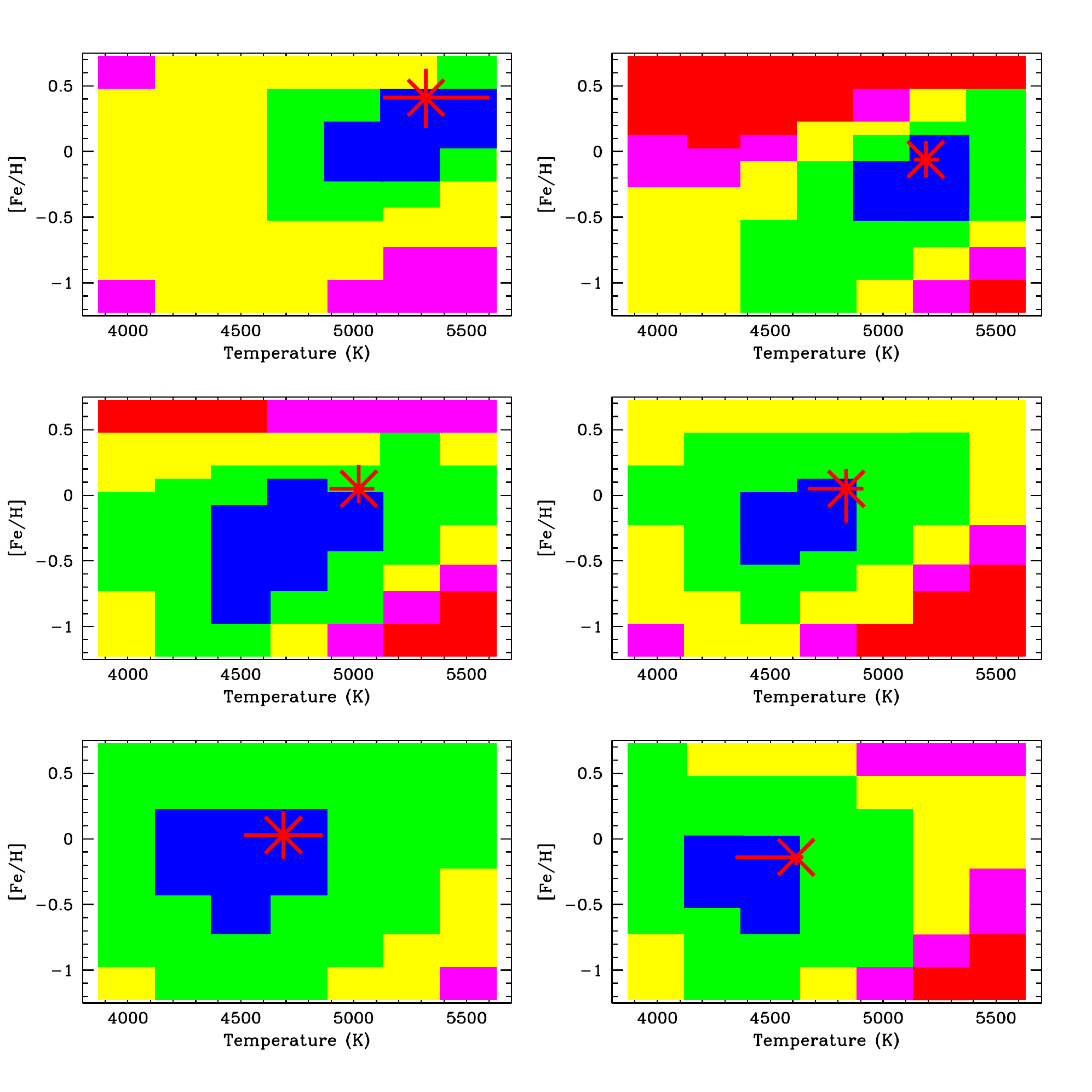}}}
\caption{The heat maps resulting from a $\chi^{2}$ analysis of a grid
of synthetic spectra compared to those of the K dwarf templates (K0V top
left, K5V bottom right, K7V not 
shown) for the $J$-band. The red ``X'' in each panel marks the location of 
the nominal value for T$_{\rm eff}$ and [Fe/H]. The line segments that pass 
through the vertex of the each of the Xs are not error bars, but display the 
full range of published measurements for each star (see Table 2 for their 
formal error bars). In this heat map, blue denotes $\chi^{2}_{\rm red}$ 
$\leq$ 0.7, green is 0.7 $\leq$ $\chi^{2}_{\rm red}$ $<$ 0.9, yellow is 0.9 
$\leq$ $\chi^{2}_{\rm red}$ $<$ 1.5, magenta is 1.5 $\leq$ 
$\chi^{2}_{\rm red}$ $<$ 2.5, and red is $\chi^{2}_{\rm red}$ $\geq$ 2.5.}
\label{jhm}
\end{figure}

\renewcommand{\thefigure}{3}
\begin{figure}[htb]
\centerline{{\includegraphics[width=12cm]{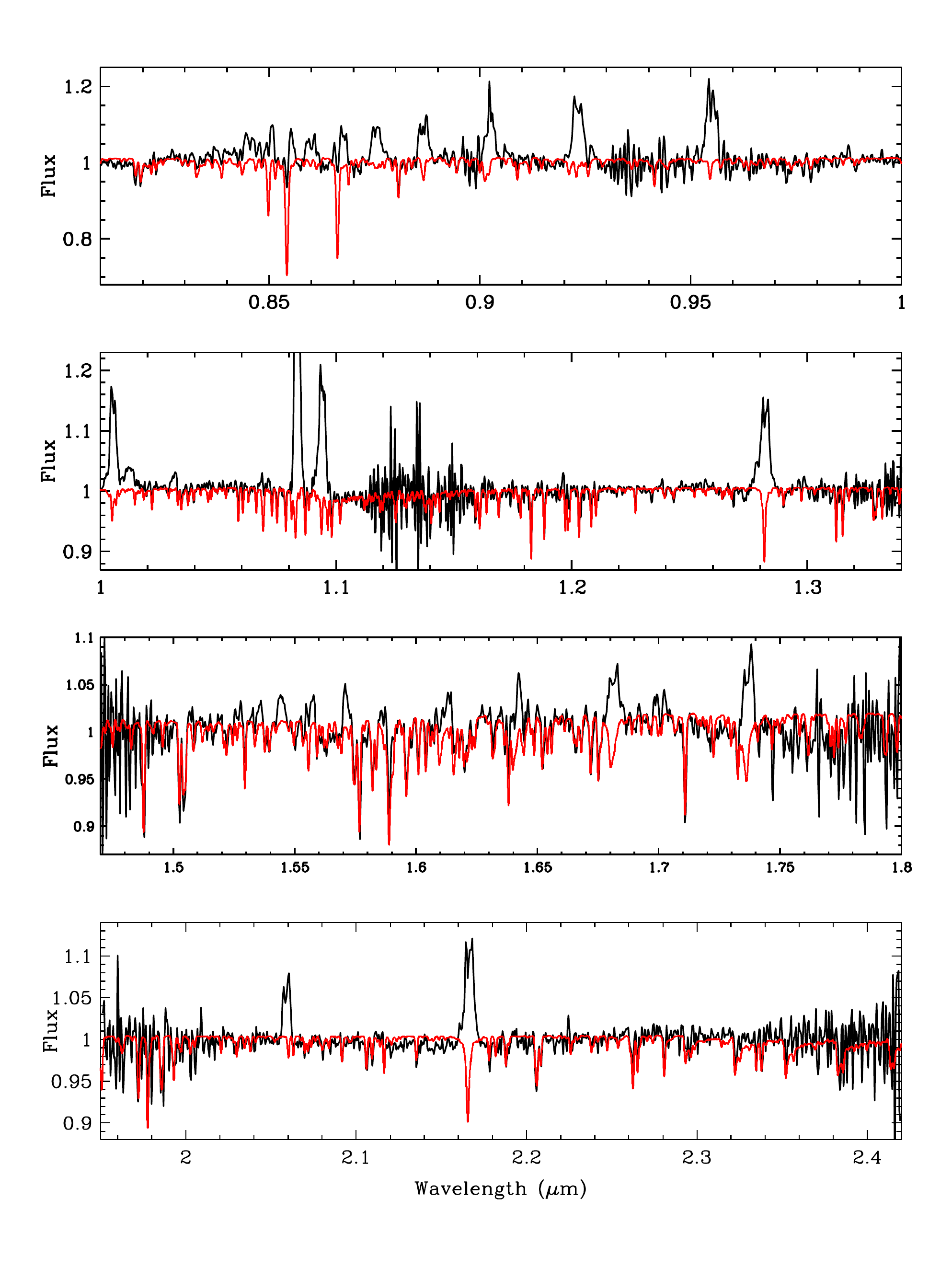}}}
\caption{The spectrum of GK Per (black) and the best fitting model
spectrum (red).}
\label{gkper}
\end{figure}

\renewcommand{\thefigure}{4}
\begin{figure}[htb]
\centerline{{\includegraphics[width=12cm]{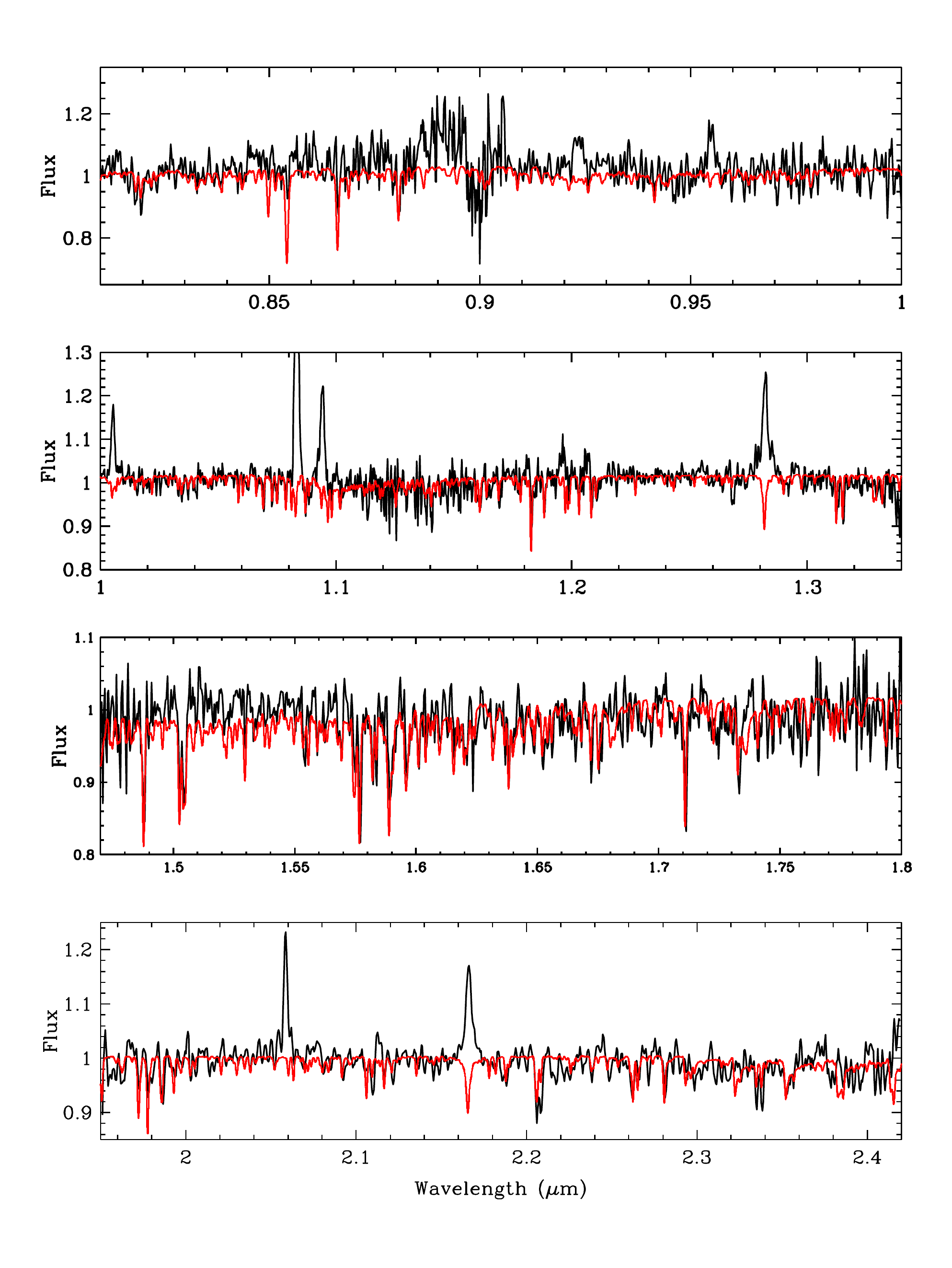}}}
\caption{The spectrum of EY Cyg (black) and the best fitting model
spectrum (red).}
\label{eycyg}
\end{figure}

\renewcommand{\thefigure}{5}
\begin{figure}[htb]
\centerline{{\includegraphics[width=12cm]{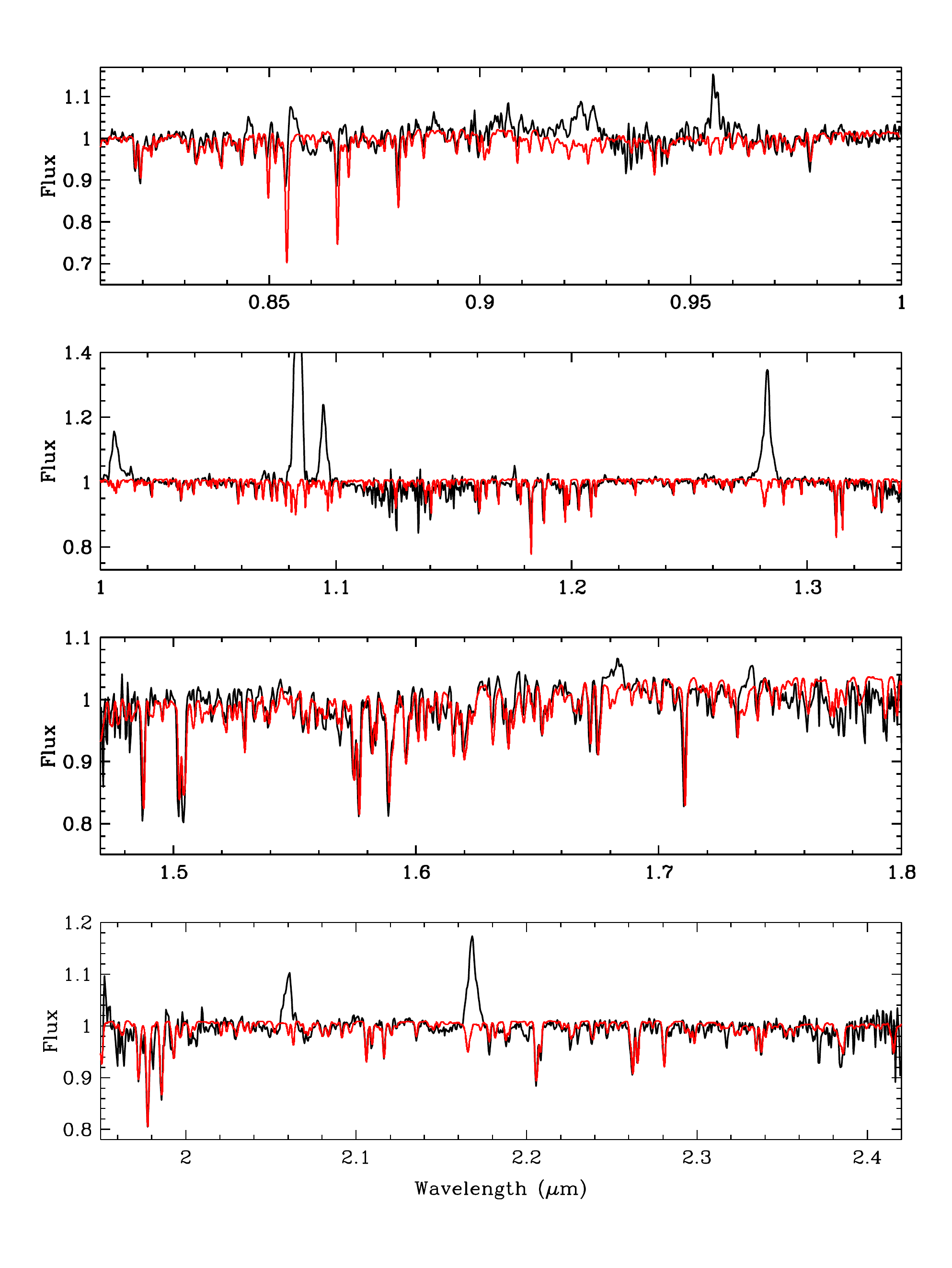}}}
\caption{The spectrum of AE Aqr (black) and the best fitting model
spectrum (red).}
\label{aeaqr}
\end{figure}
\renewcommand{\thefigure}{6}
\begin{figure}[htb]
\centerline{{\includegraphics[width=12cm]{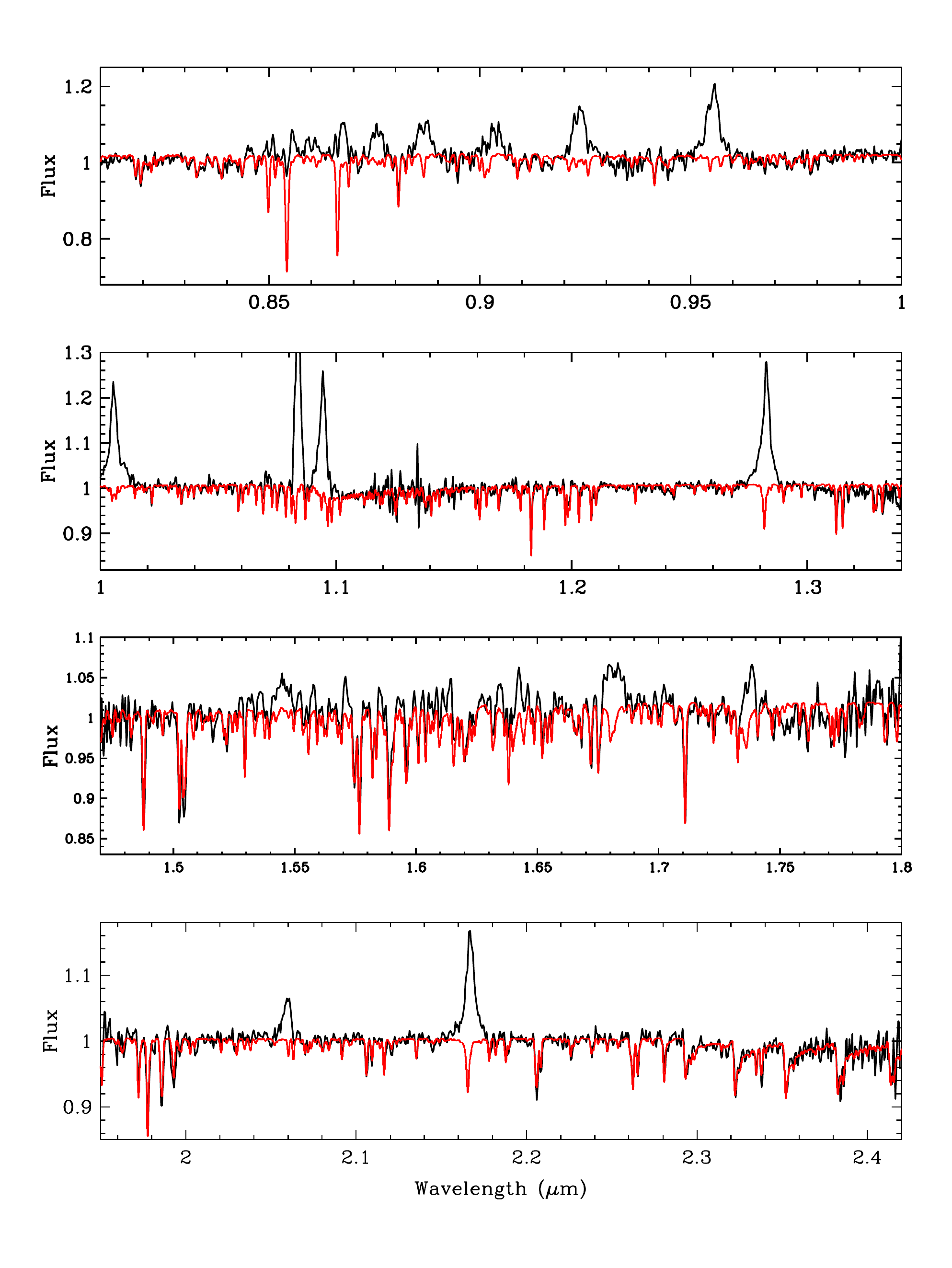}}}
\caption{The spectrum of RU Peg (black) and the best fitting model
spectrum (red).}
\label{rupeg}
\end{figure}

\renewcommand{\thefigure}{7}
\begin{figure}[htb]
\centerline{{\includegraphics[width=12cm]{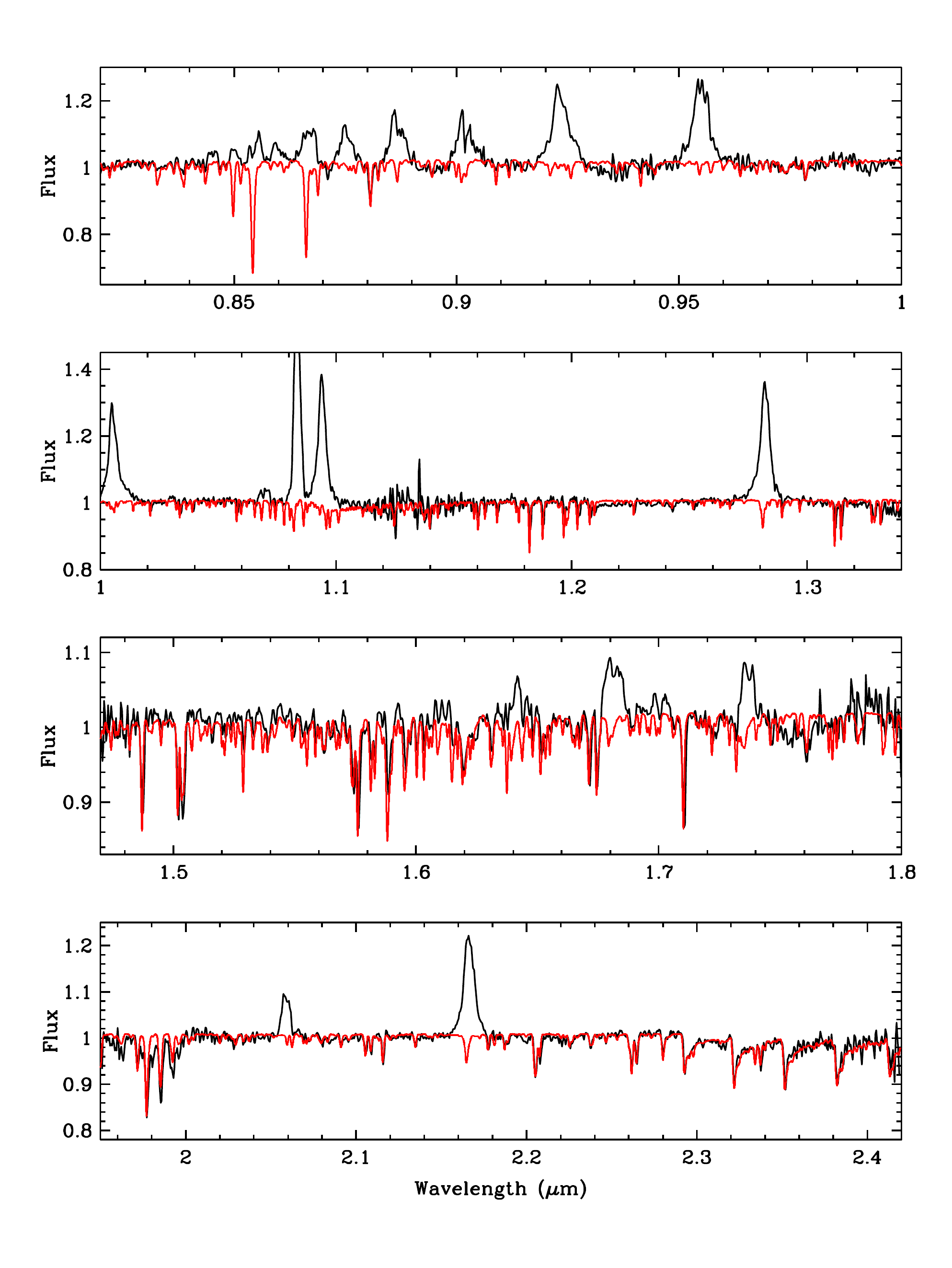}}}
\caption{The spectrum of SS Cyg (black) and the best fitting model
spectrum (red).}
\label{sscyg}
\end{figure}

\renewcommand{\thefigure}{8a}
\begin{figure}[htb]
\centerline{{\includegraphics[width=12cm]{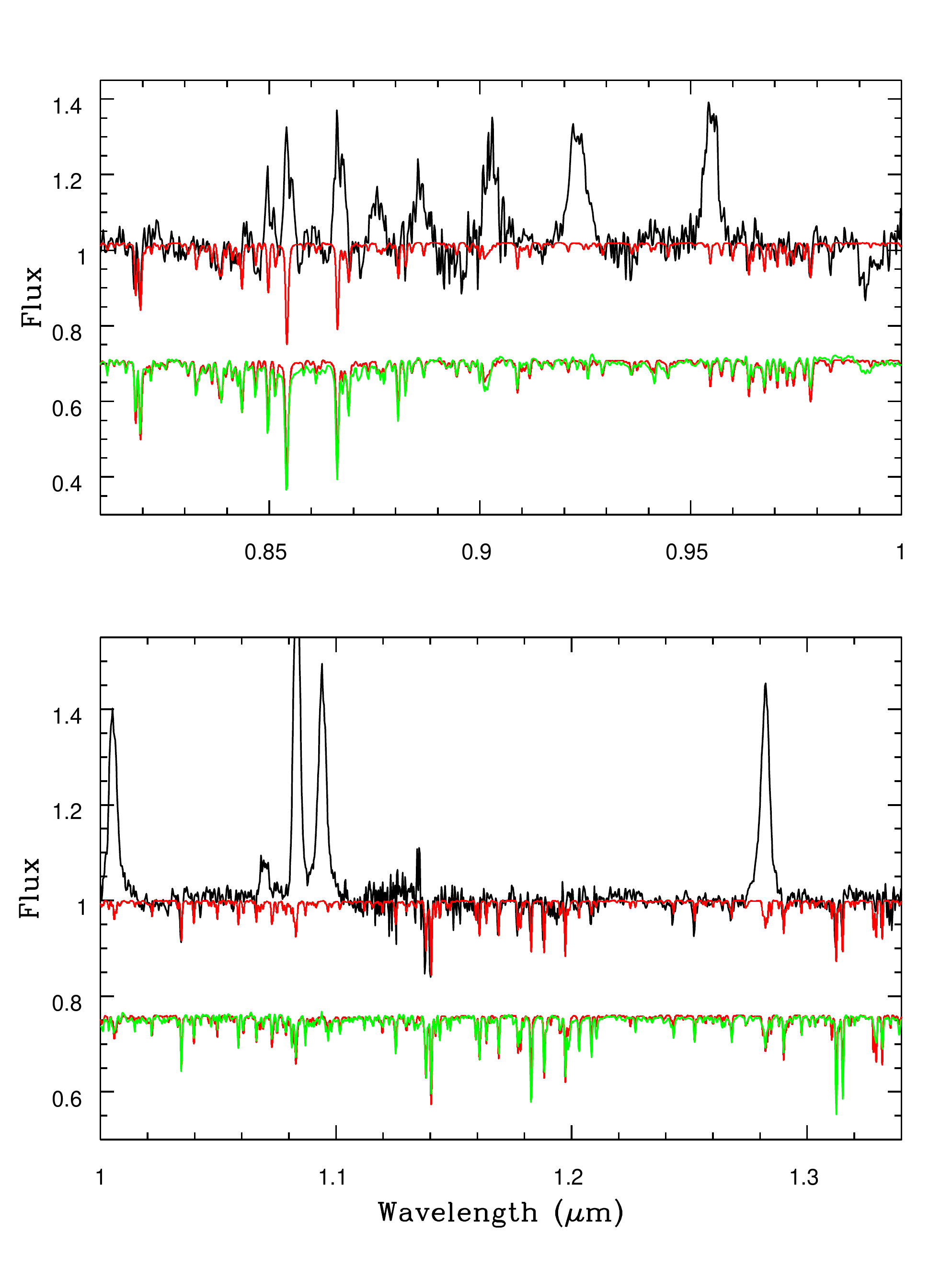}}}
\caption{The spectrum of RX And (black) and HD 19305 (green) in the $I$-band
(top panel) and in the $J$-band (bottom panel). For HD 19305 a model spectrum 
with T$_{\rm eff}$ = 3750 K and [Fe/H] = 0.0 is plotted in red. For RX And, 
the synthetic spectrum also has T$_{\rm eff}$ = 3750 K, but [Fe/H] = $-$0.3.}
\label{rxand}
\end{figure}
\renewcommand{\thefigure}{8b}
\begin{figure}[htb]
\centerline{{\includegraphics[width=12cm]{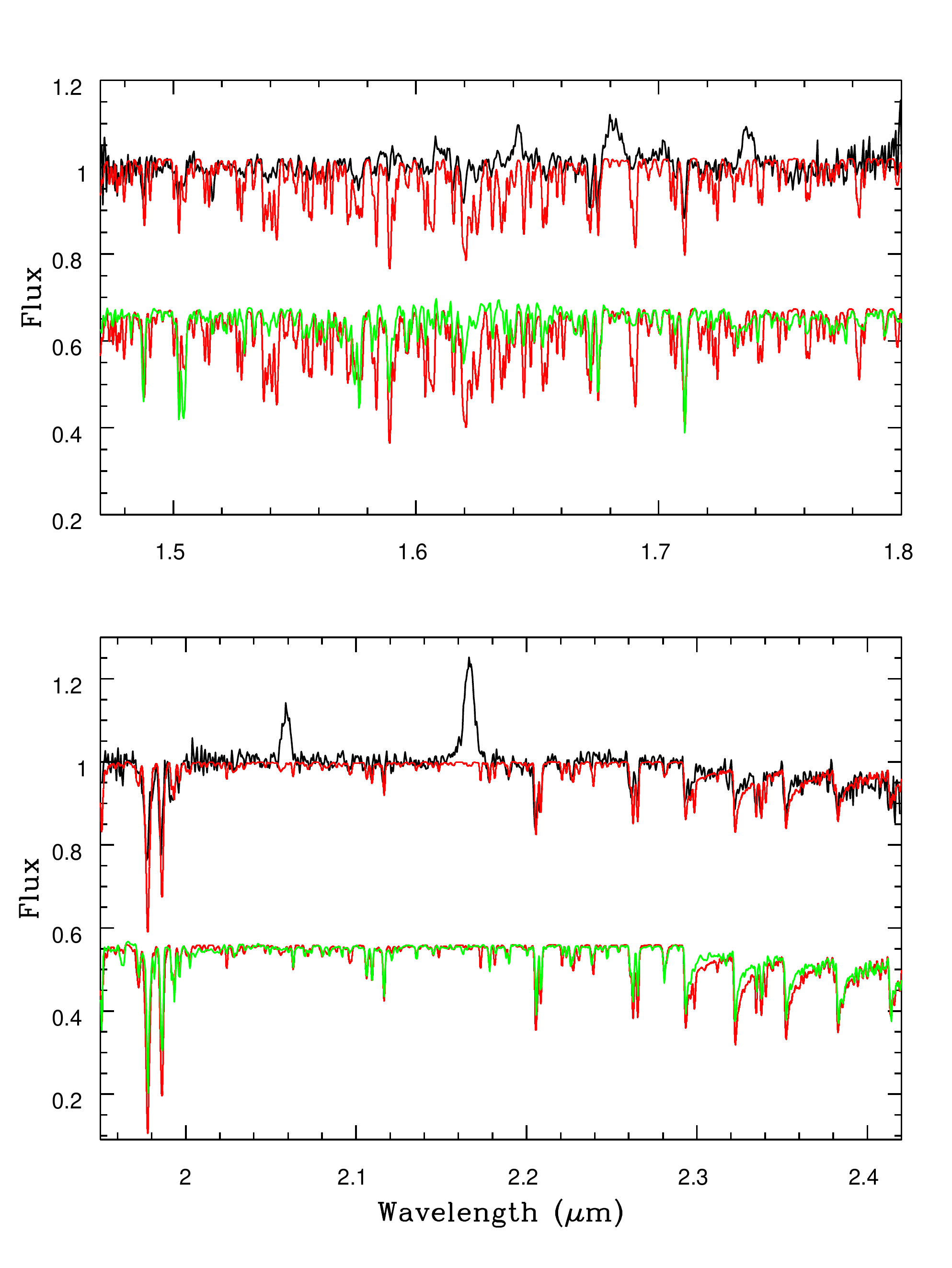}}}
\caption{The same as 7a, but for the $H$-band (top), and $K$-band (bottom).}
\label{rxand2}
\end{figure}
\renewcommand{\thefigure}{9}
\begin{figure}[htb]
\centerline{{\includegraphics[width=12cm]{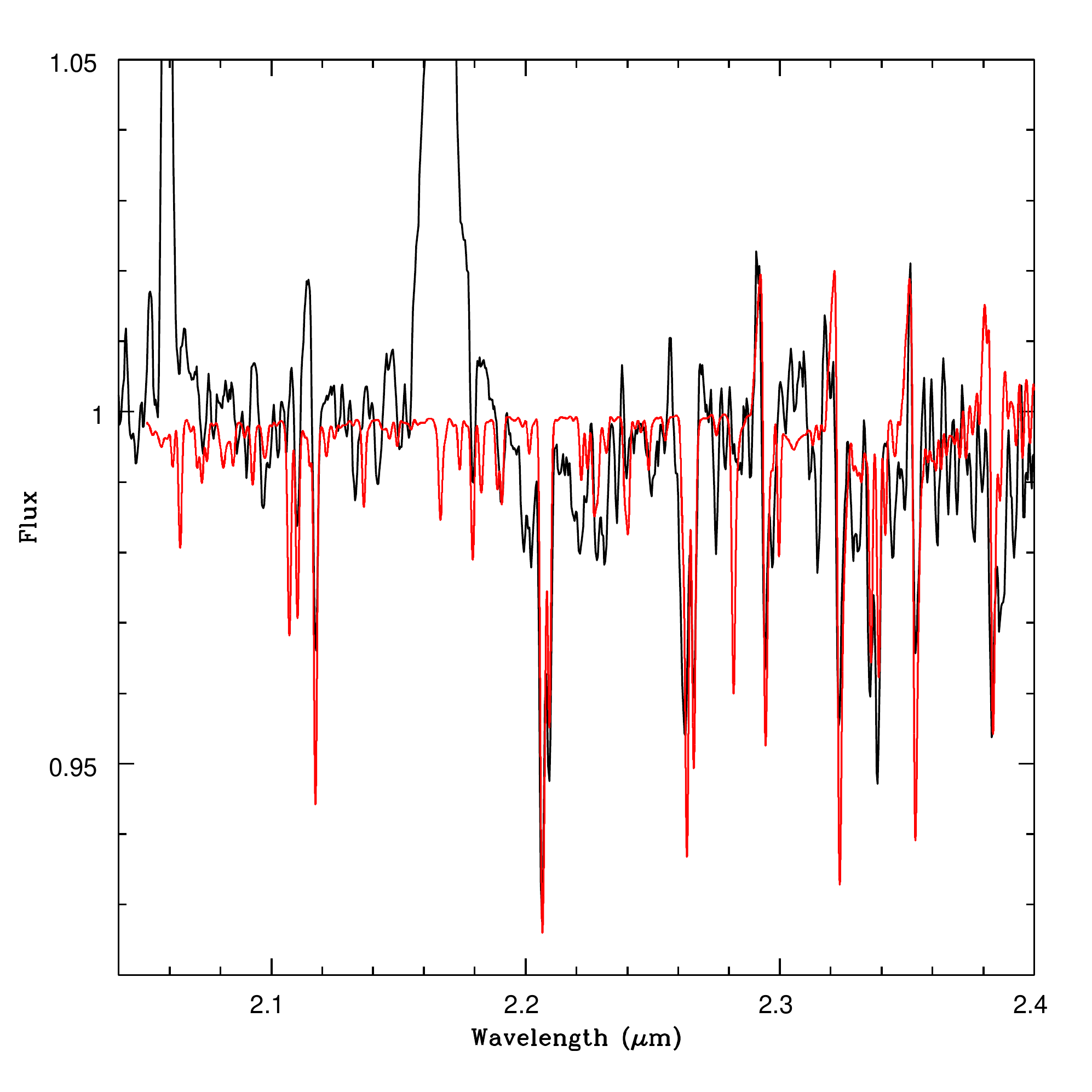}}}
\caption{The NIRSPEC $K$-band spectrum of CZ Ori (black) with a CO emission 
spectrum with ($v$sin$i$ = 800 km s$^{\rm -1}$) that has been added to a 
synthetic spectrum (red) with T$_{\rm eff}$ = 4250 K, and [Fe/H] = $-$1.0}
\label{czori}
\end{figure}

\renewcommand{\thefigure}{10}
\begin{figure}[htb]
\centerline{{\includegraphics[width=12cm]{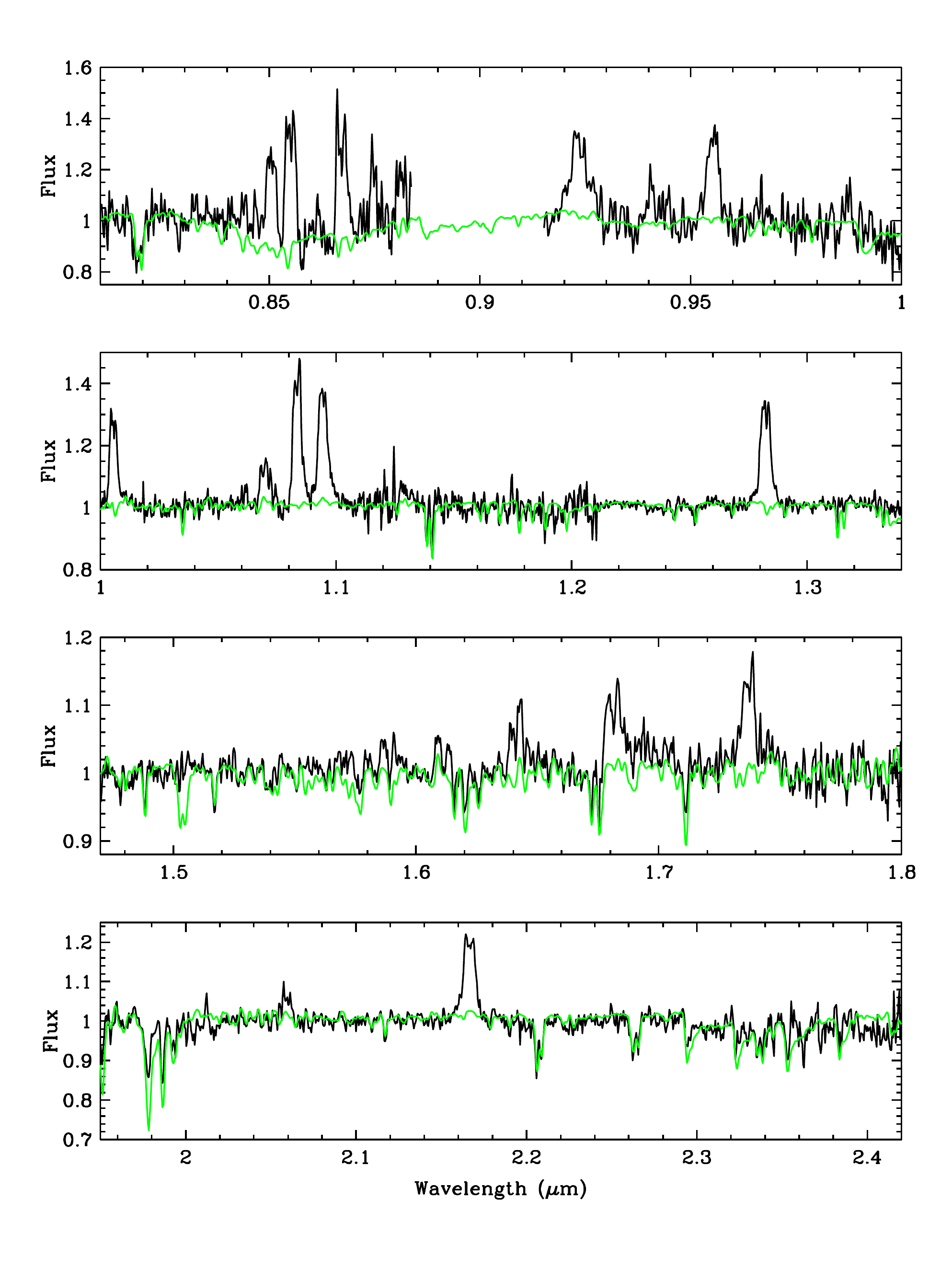}}}
\caption{The SPEX data for SS Aur (black) compared to Gl 388 (green), an M3V 
with [Fe/H] = +0.11. We have rotationally broadened the template spectrum to a 
velocity of $v$sin$i$ = 125 km s$^{\rm -1}$ to better match SS Aur.}
\label{ssaur}
\end{figure}
\renewcommand{\thefigure}{11}
\begin{figure}[htb]
\centerline{{\includegraphics[width=12cm]{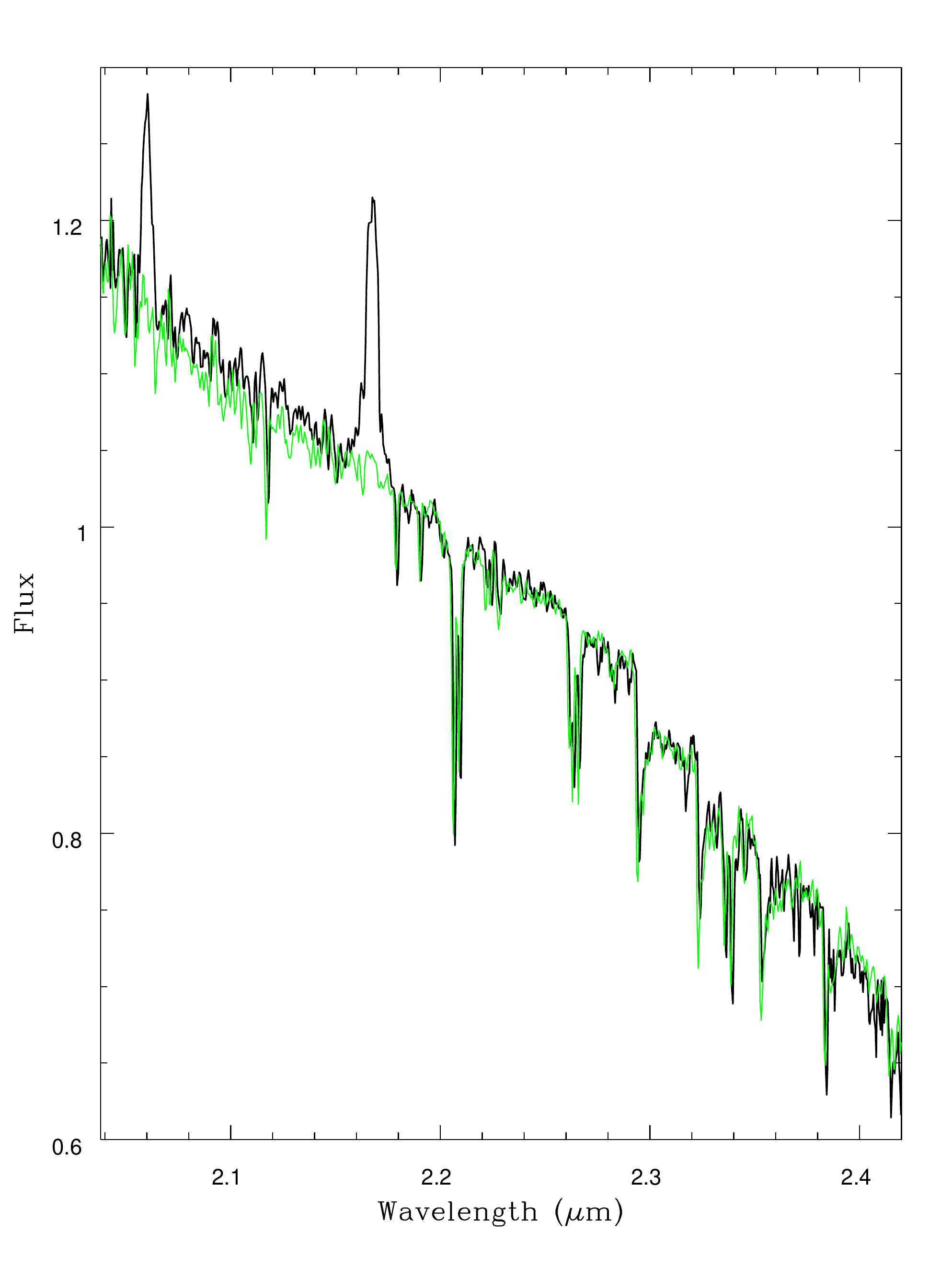}}}
\caption{The NIRSPEC $K$-band data for SS Aur (black) compared to Gl 388 
(green). }
\label{ssaurkeck}
\end{figure}
\clearpage
\renewcommand{\thefigure}{12}
\begin{figure}[htb]
\centerline{{\includegraphics[width=12cm]{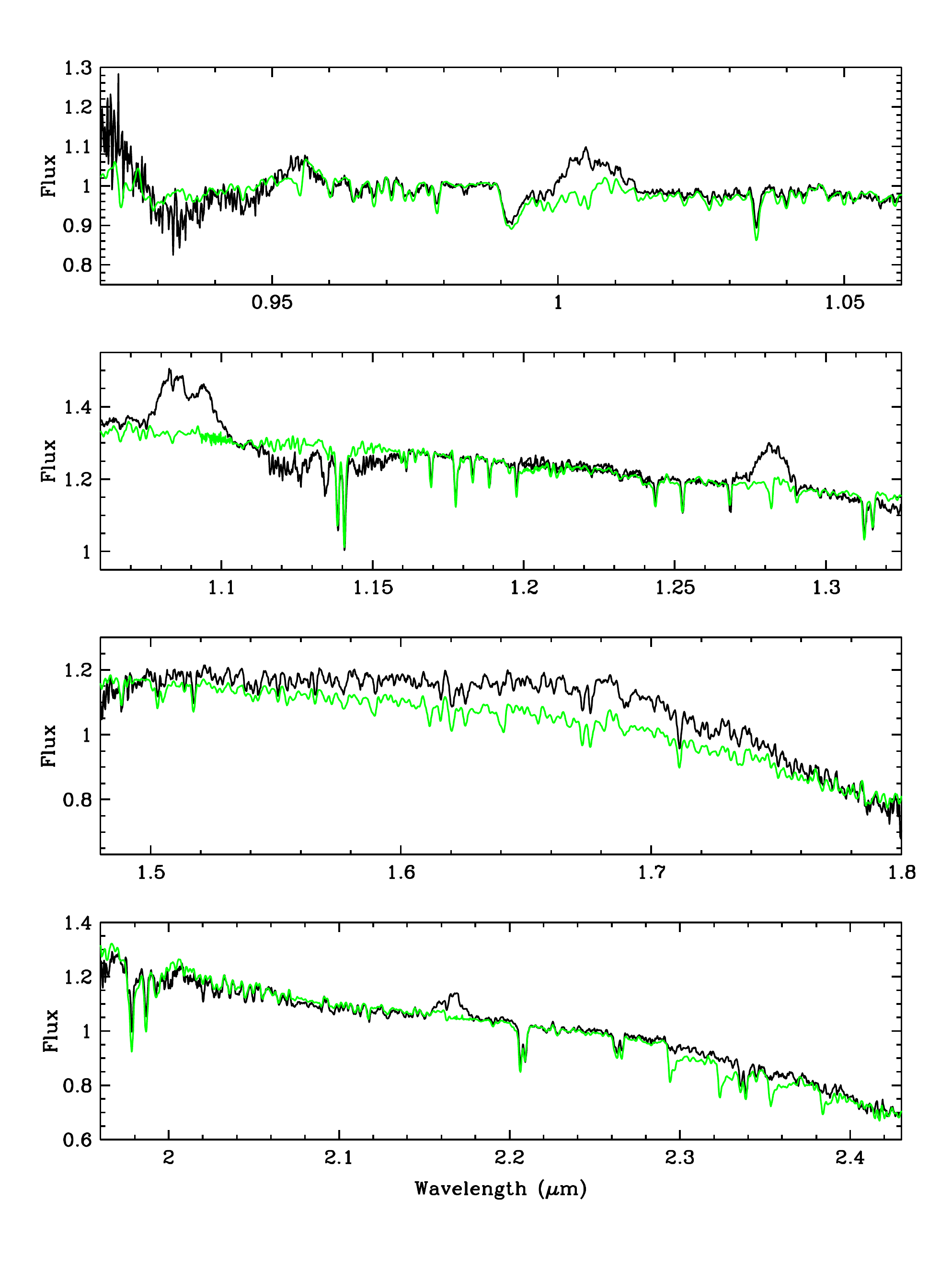}}}
\caption{The TripleSpec spectra of U Gem (black), and GJ 402 (green). Both
objects appear to have super-solar metallicities, however, U Gem has
extremely weak CO features. The deviation of the continuum in the $H$-band
corresponds with the shape of the H$^{-}$ opacity minimum, suggesting
that the secondary star of U Gem has a lower gravity than GJ 402.}
\label{ugemspec}
\end{figure}

\clearpage
\renewcommand{\thefigure}{13}
\begin{figure}[htb]
\centerline{{\includegraphics[width=12cm]{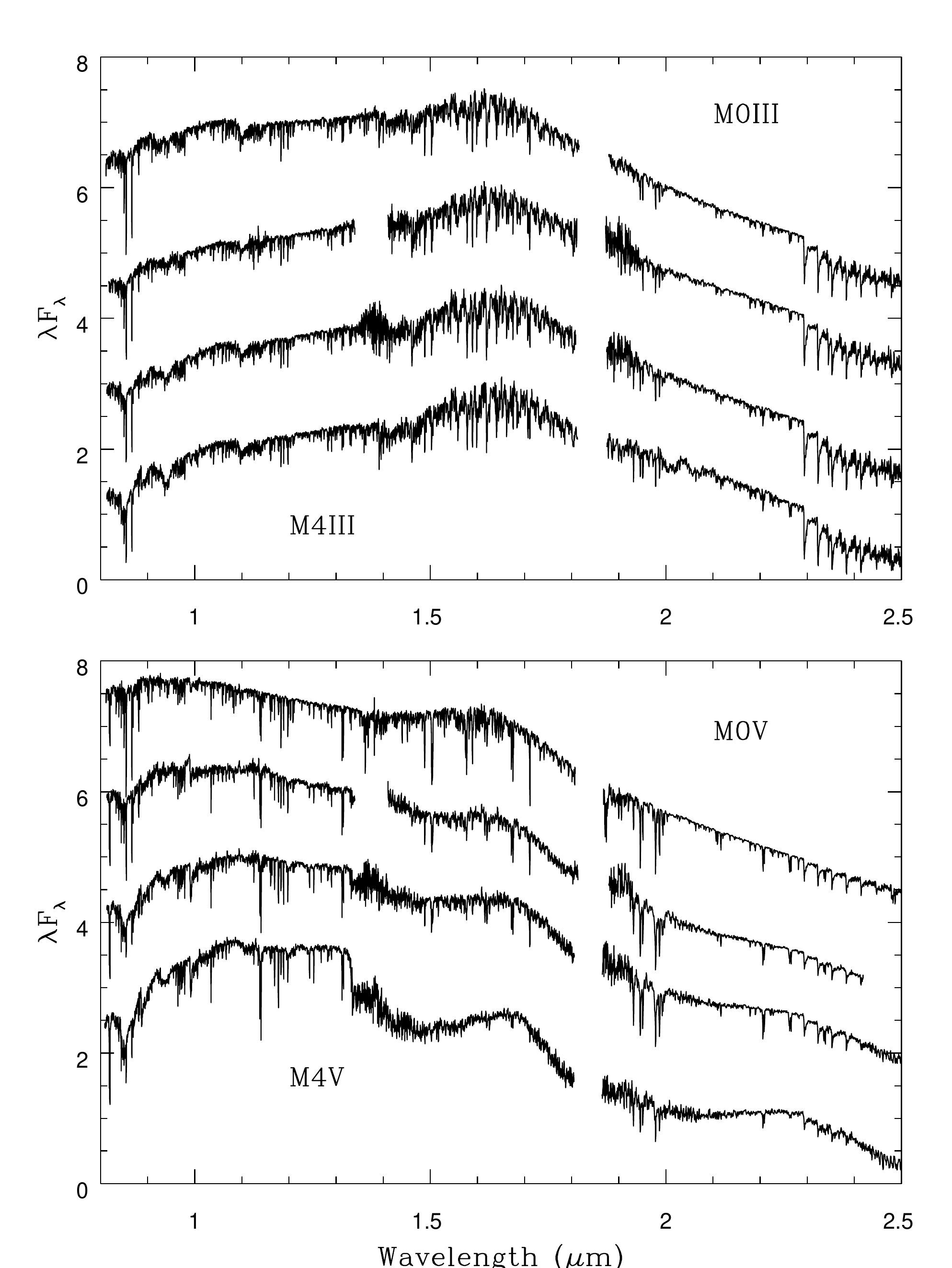}}}
\caption{The NIR spectra of M giants (top panel) and M dwarfs (bottom
panel), from M0 to M4 (from the IRTF Spectral Libary).}
\label{giantdwarf}
\end{figure}
\clearpage
\renewcommand{\thefigure}{14}
\begin{figure}[htb]
\centerline{{\includegraphics[width=12cm]{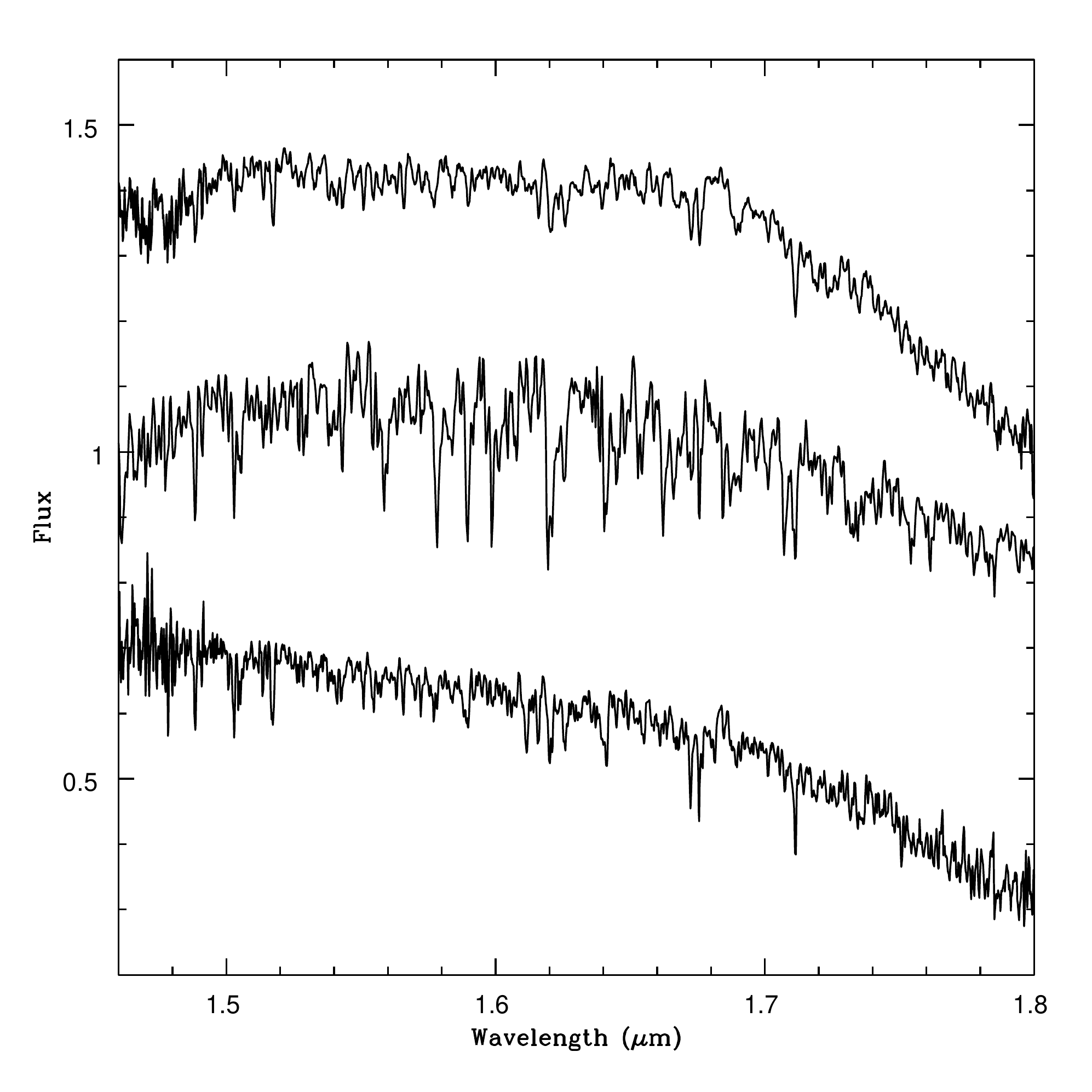}}}
\caption{The $H$-band spectra of U Gem (top) and GJ 402 (bottom),
compared to that of HD4408 (middle), an M4III (from the IRTF Spectral Libary).}
\label{ugemHcomp}
\end{figure}
\clearpage
\renewcommand{\thefigure}{15}
\begin{figure}[htb]
\centerline{{\includegraphics[width=12cm]{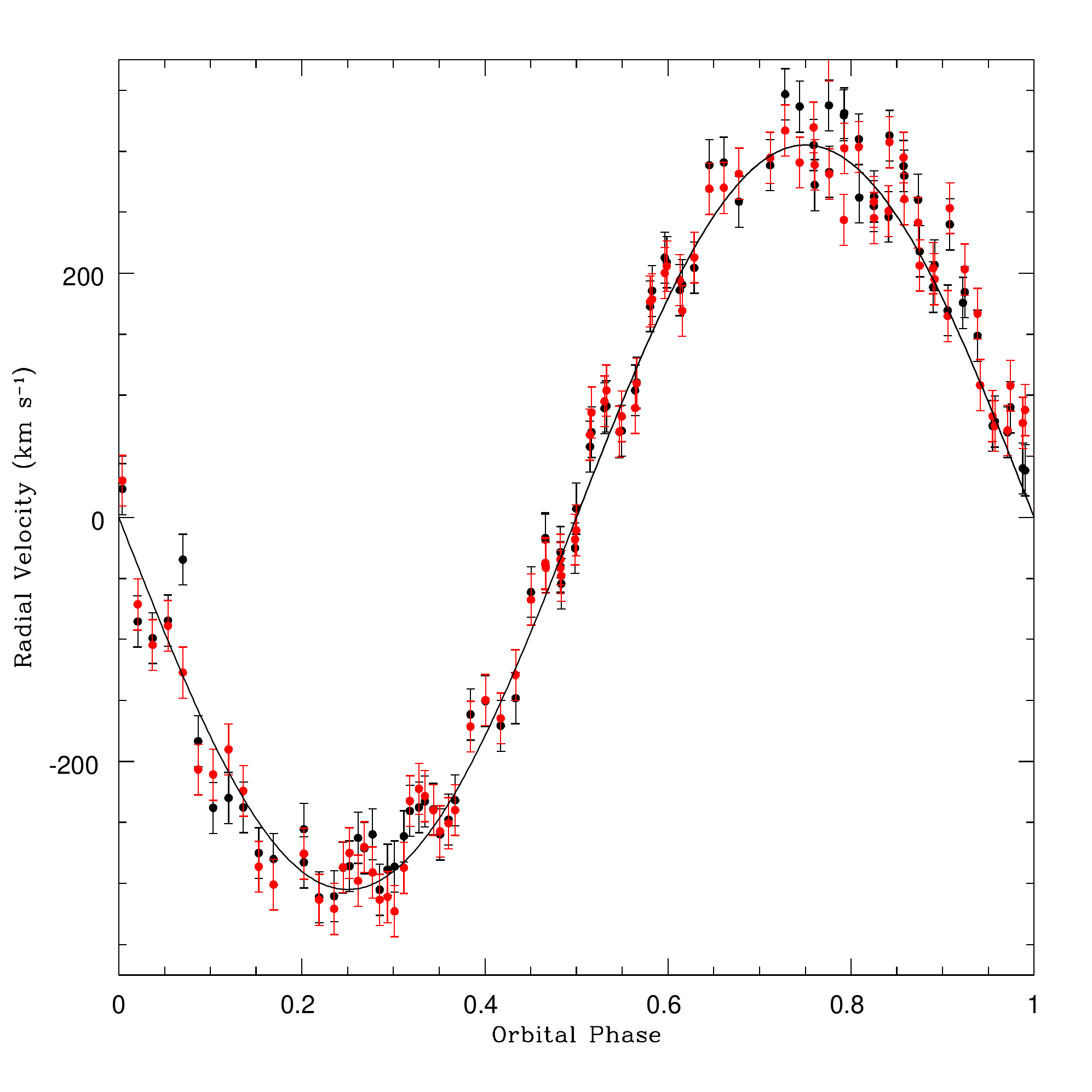}}}
\caption{The TripleSpec radial velocity curve for U Gem. The black dots
are the values for the bluer component of the Na I doublet in the $K$-band,
and the red dots are for the red component. A best-fit sinewave with an
amplitude of 310 km s$^{\rm -1}$ is superposed.  }
\label{ugemrv}
\end{figure}
\clearpage
\renewcommand{\thefigure}{16}
\begin{figure}[htb]
\centerline{{\includegraphics[width=12cm]{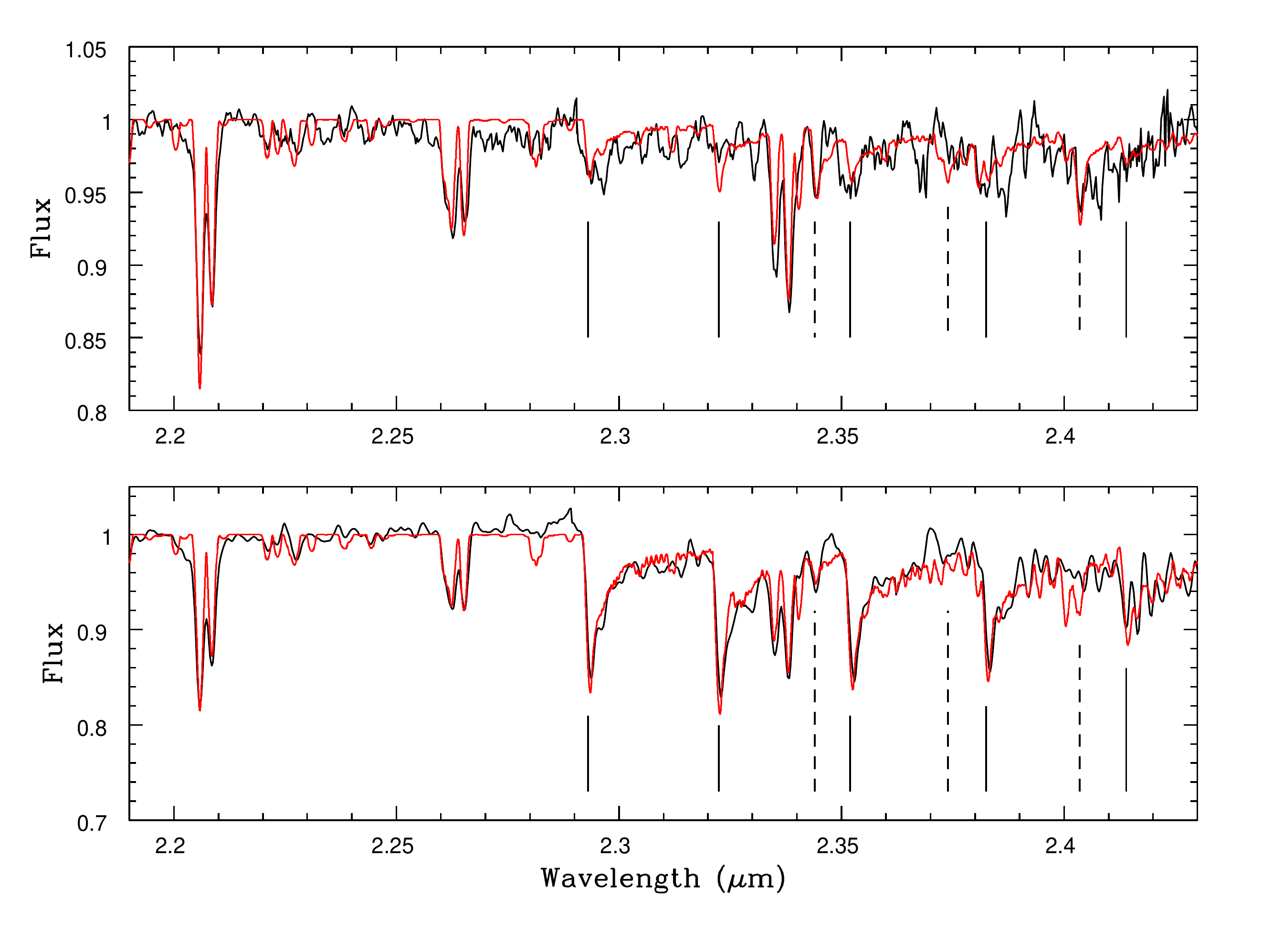}}}
\caption{The red end of the $K$-band spectrum for U Gem (black, top panel), and 
GJ 402 (black, bottom panel). We plot a synthetic spectra (red) with 
T$_{\rm eff}$
= 3200 K and [Fe/H] = 0.0, in black in both panels. The vertical solid
lines are the location of the $^{\rm 12}$CO bandheads, while the vertical
dashed lines are for the $^{\rm 13}$CO features.}
\label{ugemKcomp}
\end{figure}
\clearpage
\renewcommand{\thefigure}{17}
\begin{figure}[htb]
\centerline{{\includegraphics[width=12cm]{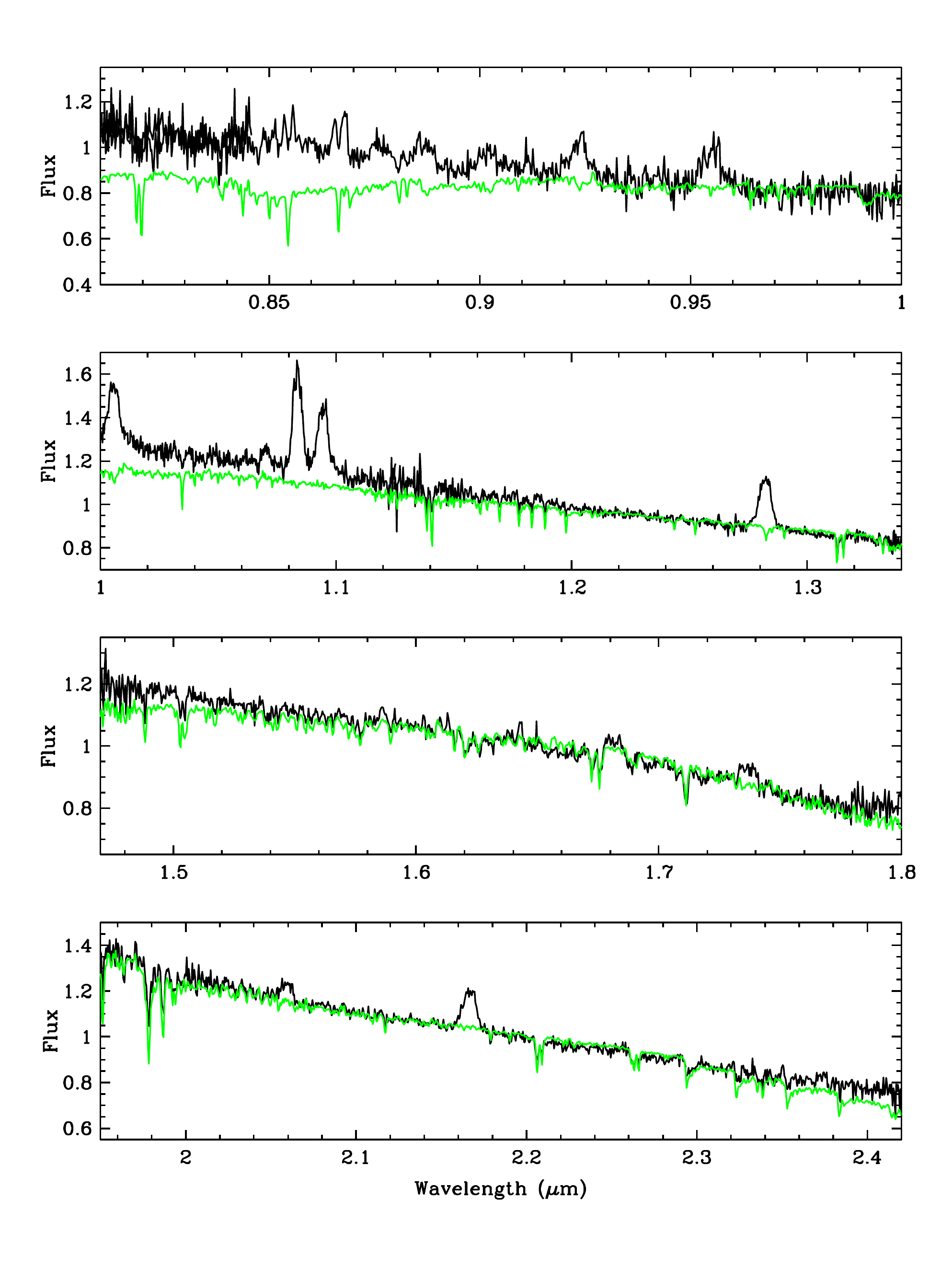}}}
\caption{The SPEX data for WW Cet (black) compared to the IRTF template Gl 
381 (M2.5V).
There is a small amount of contamination from the white dwarf $+$ accretion
disk that leads to a blue excess in most of the bands. Note the influence
of the Ca II triplet absorption on the H I Paschen continuum.
}
\label{wwcet}
\end{figure}
\clearpage
\renewcommand{\thefigure}{18}
\begin{figure}[htb]
\centerline{{\includegraphics[width=12cm]{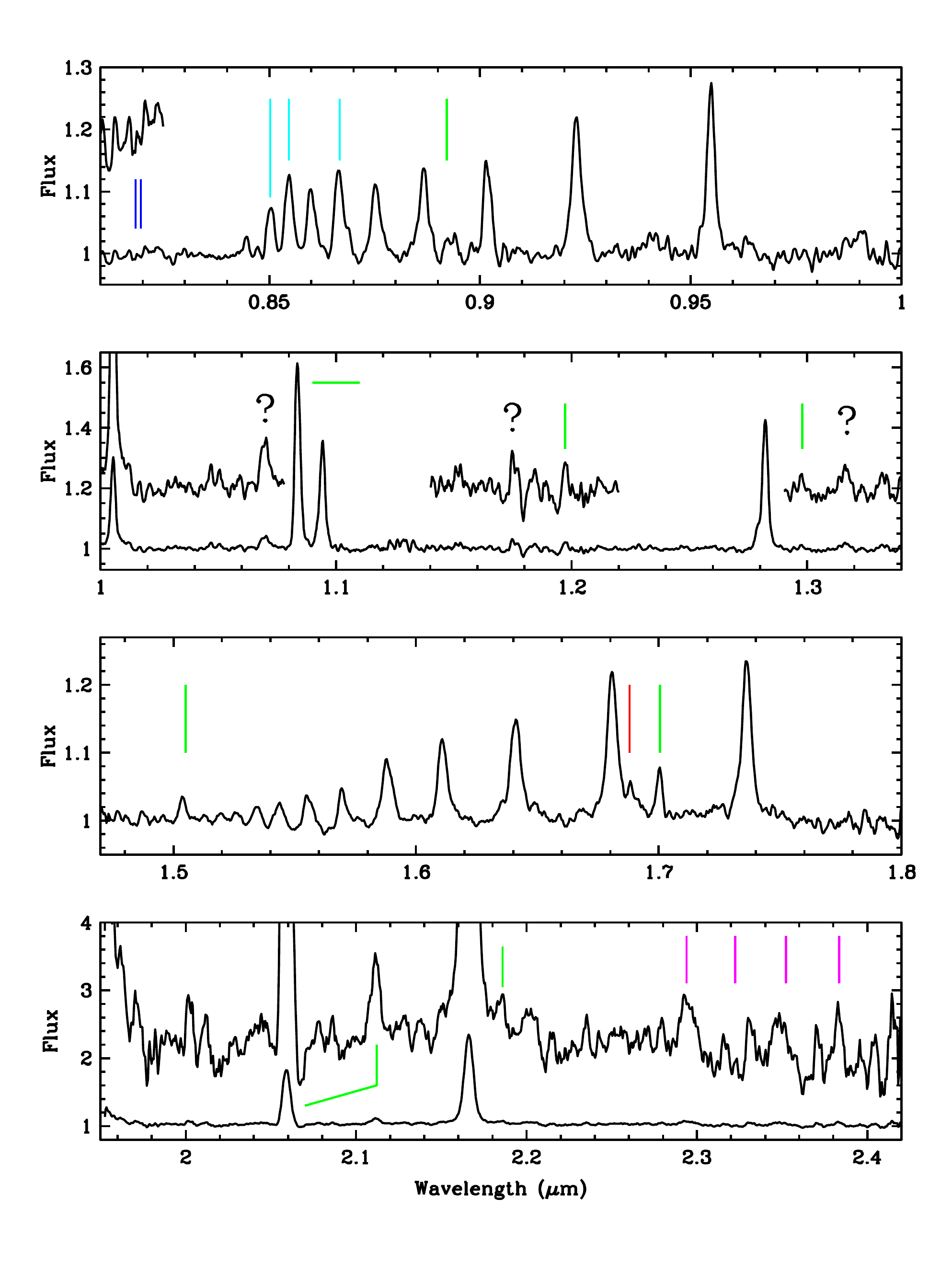}}}
\caption{The spectrum of LS Peg. The unmarked emission lines are due to H I,
while those indicated by green line segments are from helium. The main
$^{\rm 12}$CO bandheads are indicated by vertical magenta lines. [Fe II]
is in red, the Ca II triplet in cyan, and the Na I doublet at 0.82 $\mu$m in
blue. Unidentified emission lines are indicated with a ``?''.
}
\label{lspeg}
\end{figure}
\clearpage
\renewcommand{\thefigure}{19}
\begin{figure}[htb]
\centerline{{\includegraphics[width=12cm]{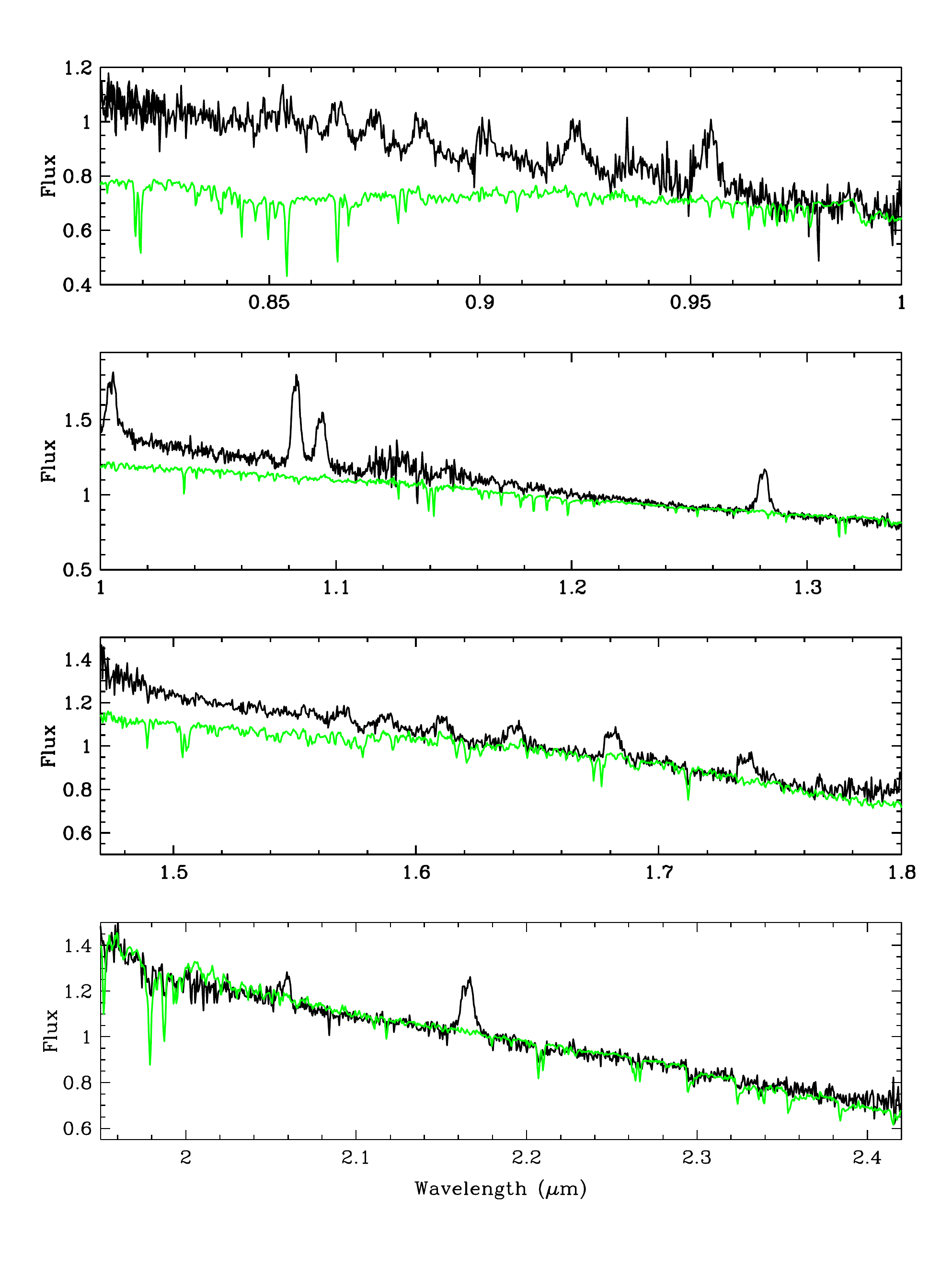}}}
\caption{The IRTF SPEX data for KT Per (black), and Gl 806 (green).}
\label{ktper}
\end{figure}

\clearpage
\renewcommand{\thefigure}{20}
\begin{figure}[htb]
\centerline{{\includegraphics[width=12cm]{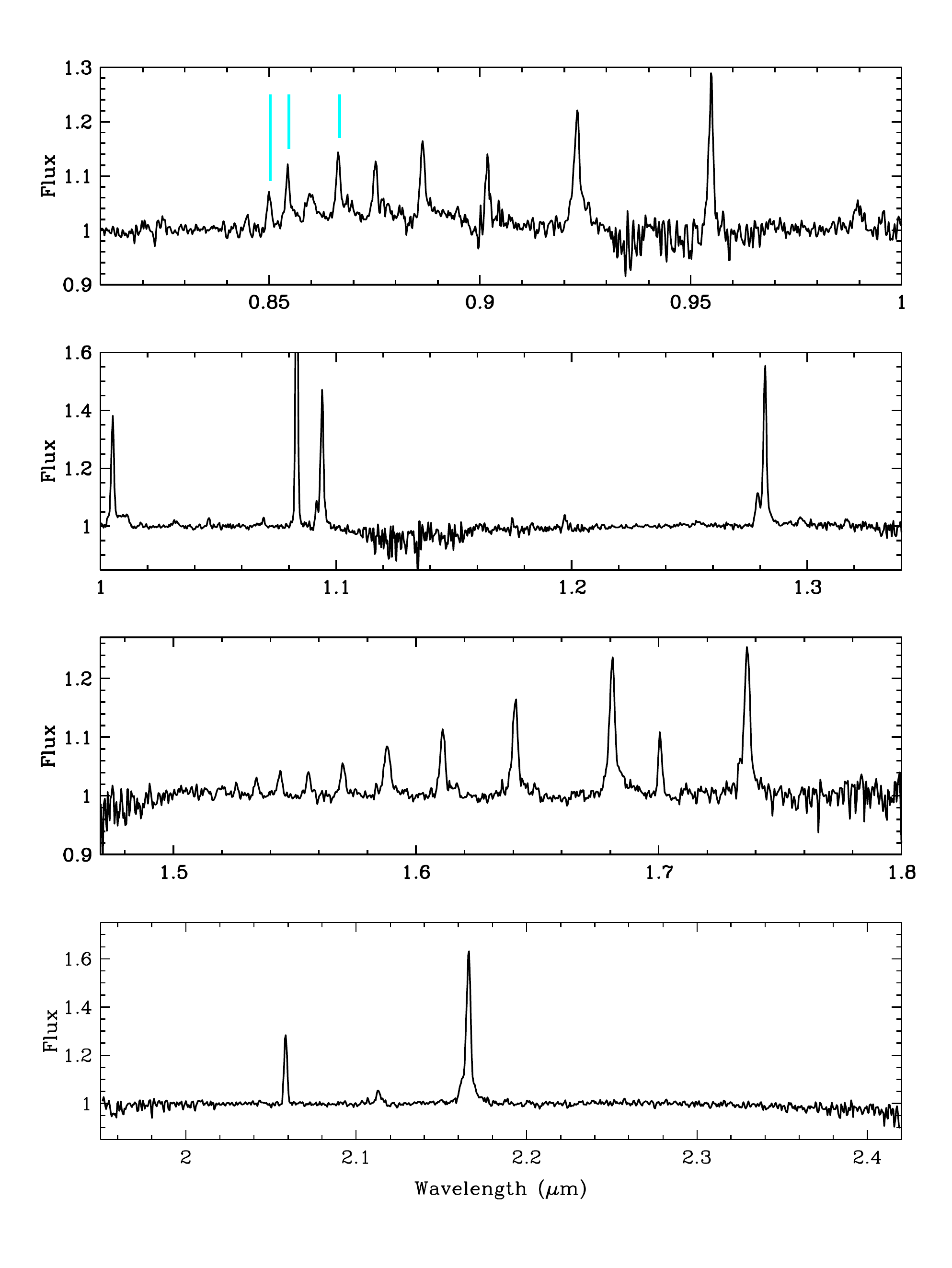}}}
\caption{The IRTF SPEX data for TT Ari, the location of the Ca II emission
lines have been indicated with vertical cyan lines.}
\label{ttari}
\end{figure}

\clearpage
\renewcommand{\thefigure}{21}
\begin{figure}[htb]
\centerline{{\includegraphics[width=12cm]{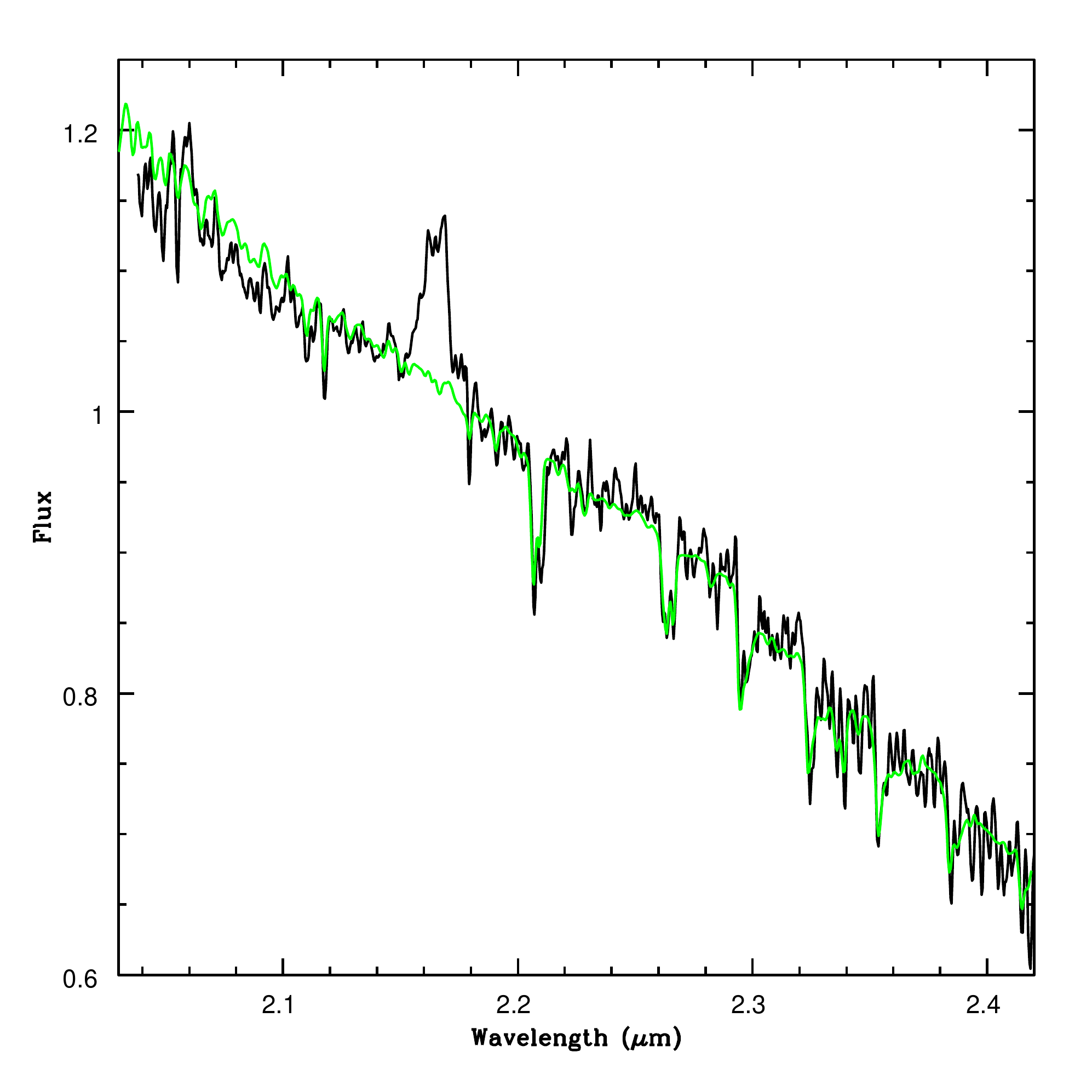}}}
\caption{The NIRSPEC data for CW Mon (black), compared to the M2V template
Gl 806 (green).}
\label{cwmon}
\end{figure}

\clearpage
\renewcommand{\thefigure}{22}
\begin{figure}[htb]
\centerline{{\includegraphics[width=12cm]{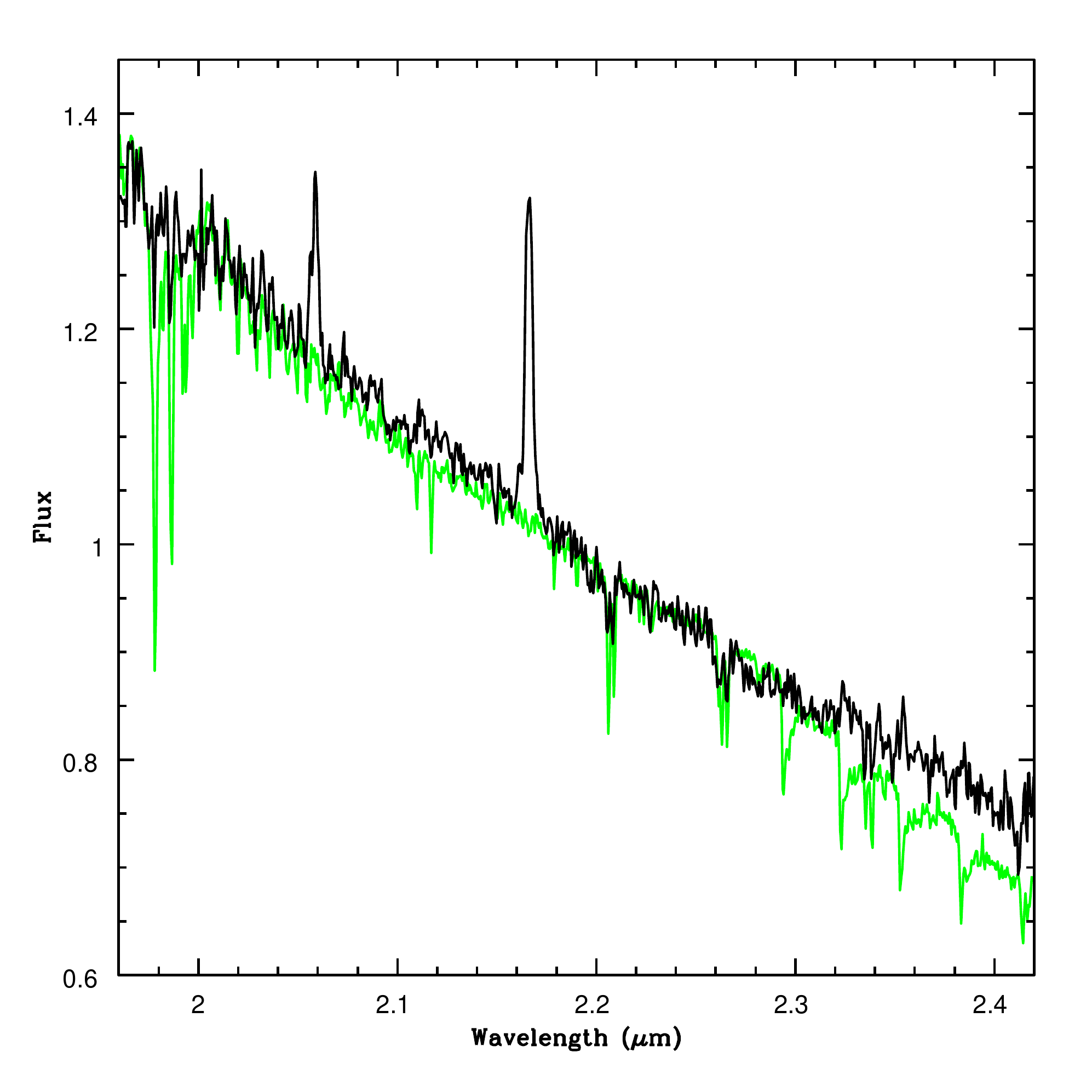}}}
\caption{The SPEX data for UU Aql (black), compared to the M2V template
Gl 806 (green).}
\label{uuaql}
\end{figure}

\clearpage
\renewcommand{\thefigure}{23}
\begin{figure}[htb]
\centerline{{\includegraphics[width=12cm]{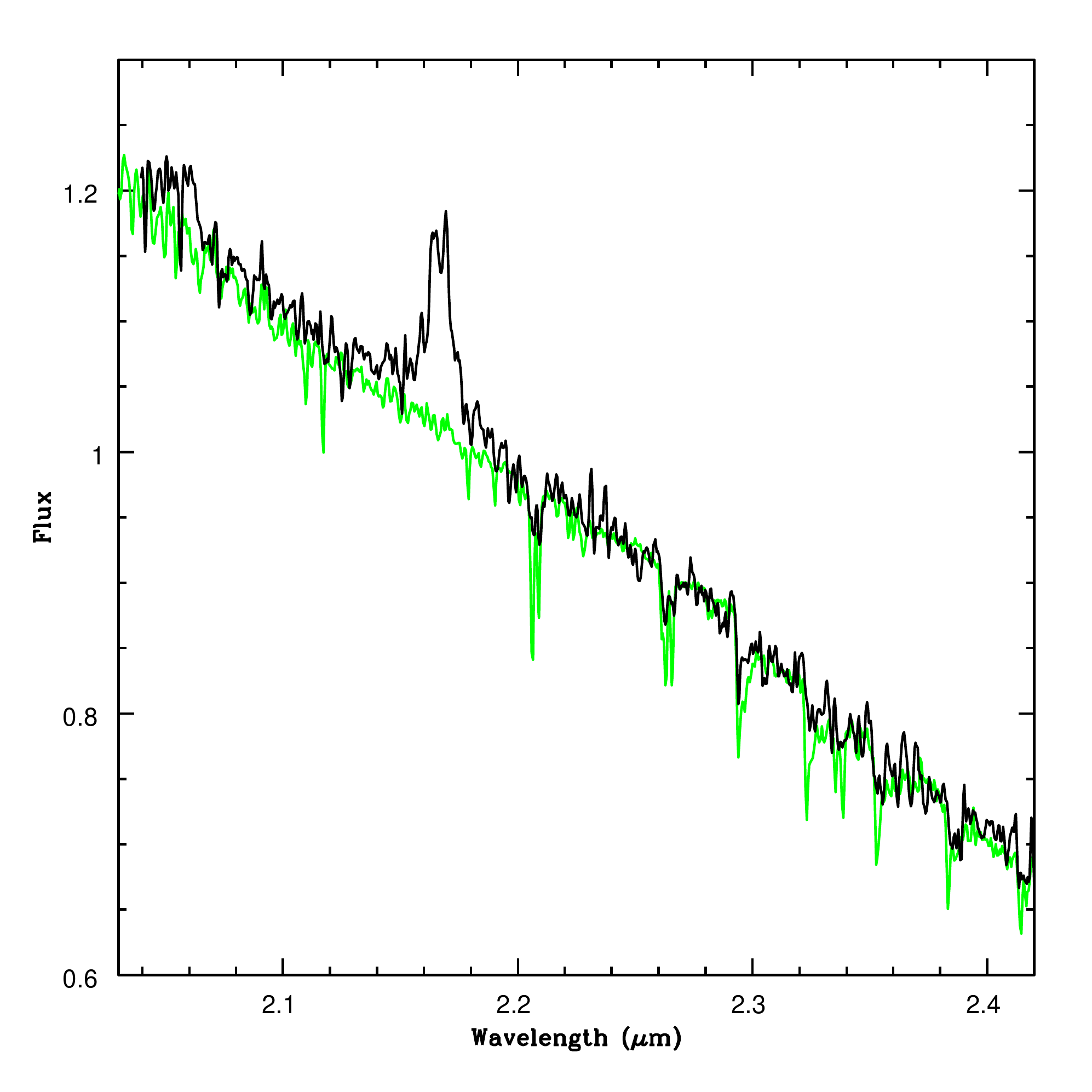}}}
\caption{The NIRSPEC data for CN Ori (black), compared to the M2V template
Gl 806 (green).}
\label{cnori}
\end{figure}
\clearpage
\renewcommand{\thefigure}{24}
\begin{figure}[htb]
\centerline{{\includegraphics[width=12cm]{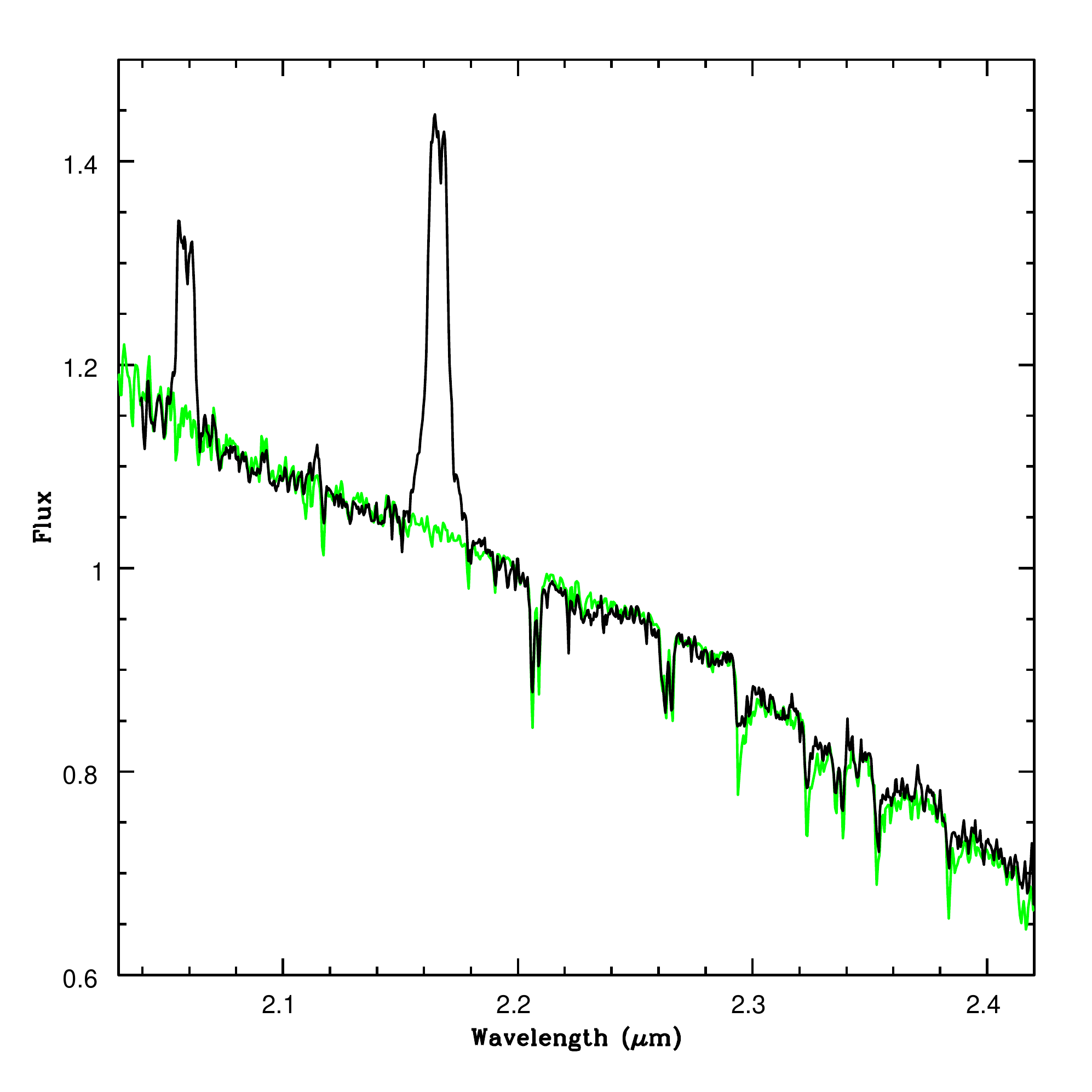}}}
\caption{The NIRSPEC data for YY/DO Dra (black), compared to the M2.5V template
Gl 381 (green).}
\label{yydra}
\end{figure}
\clearpage
\renewcommand{\thefigure}{25}
\begin{figure}[htb]
\centerline{{\includegraphics[width=12cm]{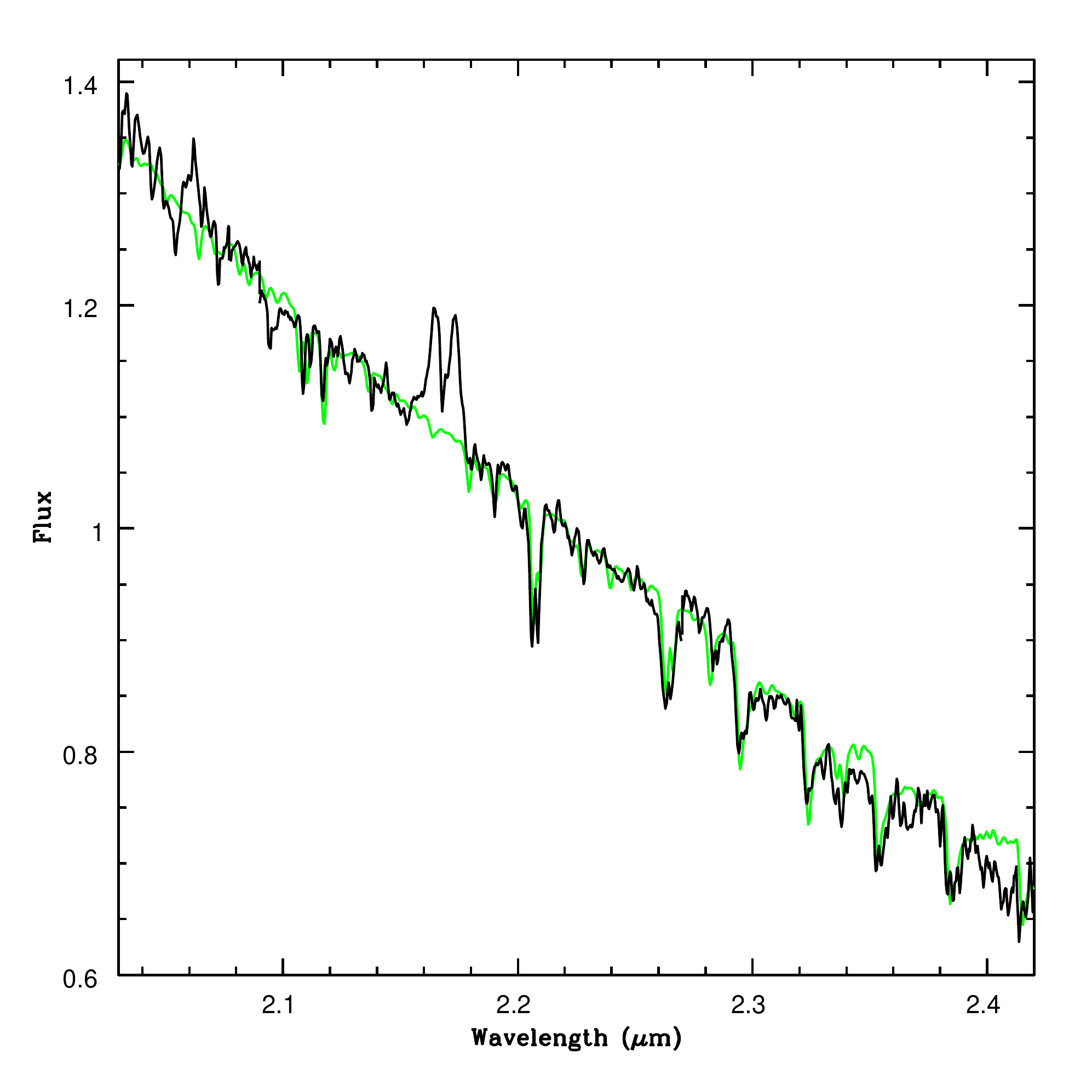}}}
\caption{The NIRSPEC data for IP Peg (black), compared to the M0V template
HD 13905 (green).}
\label{ippeg}
\end{figure}
\renewcommand{\thefigure}{26}
\begin{figure}[htb]
\centerline{{\includegraphics[width=12cm]{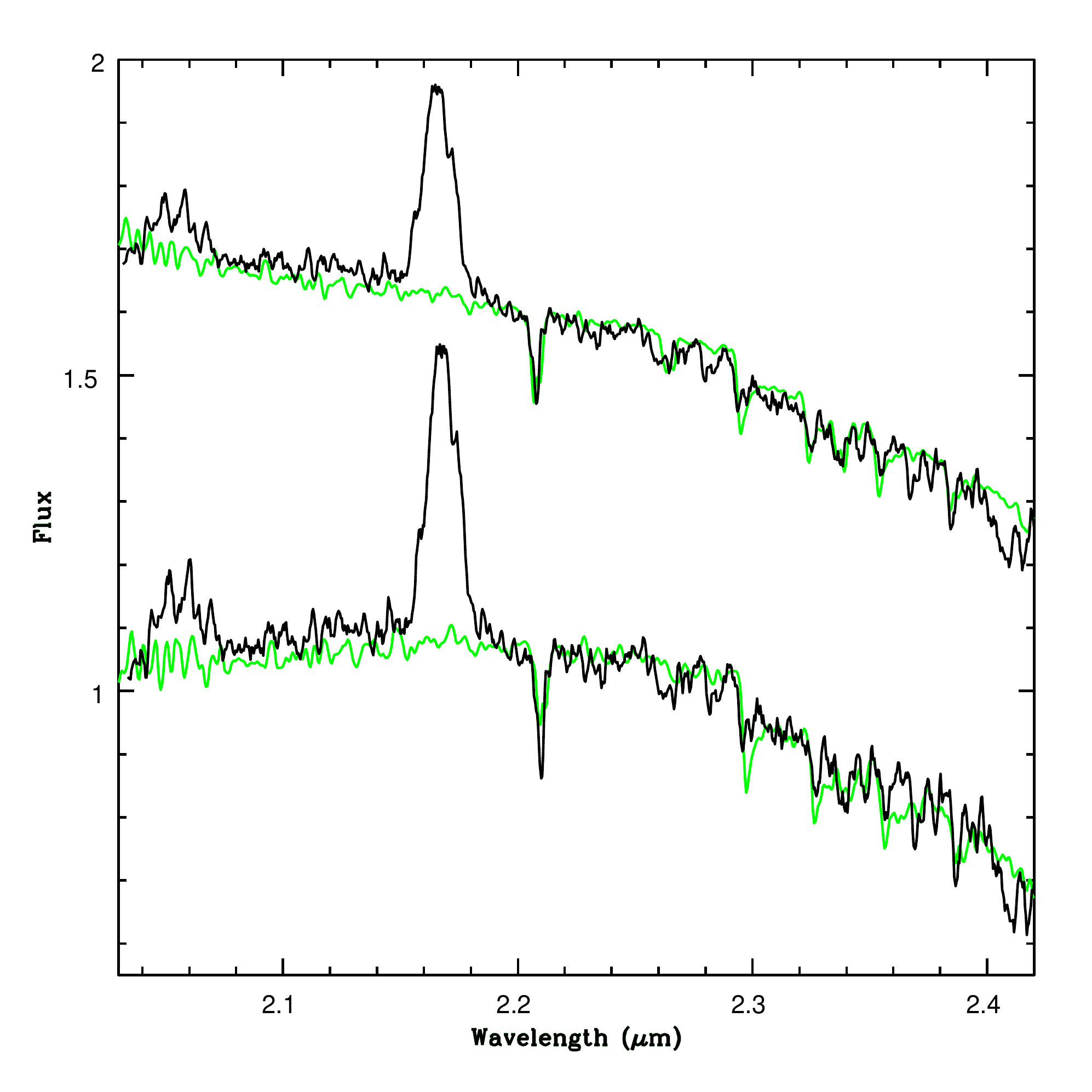}}}
\caption{The Doppler-corrected NIRSPEC data for RZ Leo (top, black, boxcar
smoothed by 5 pixels), compared to (in green) the M4.5V template Gl 268AB.
We have subtracted-off a hot blackbody that supplies 30\% of the $K$-band
flux from the spectrum of RZ Leo (bottom, black), and compare it to the M6.5V 
template GJ 1111 (green).}
\label{rzleo}
\end{figure}

\renewcommand{\thefigure}{27}
\begin{figure}[htb]
\centerline{{\includegraphics[width=12cm]{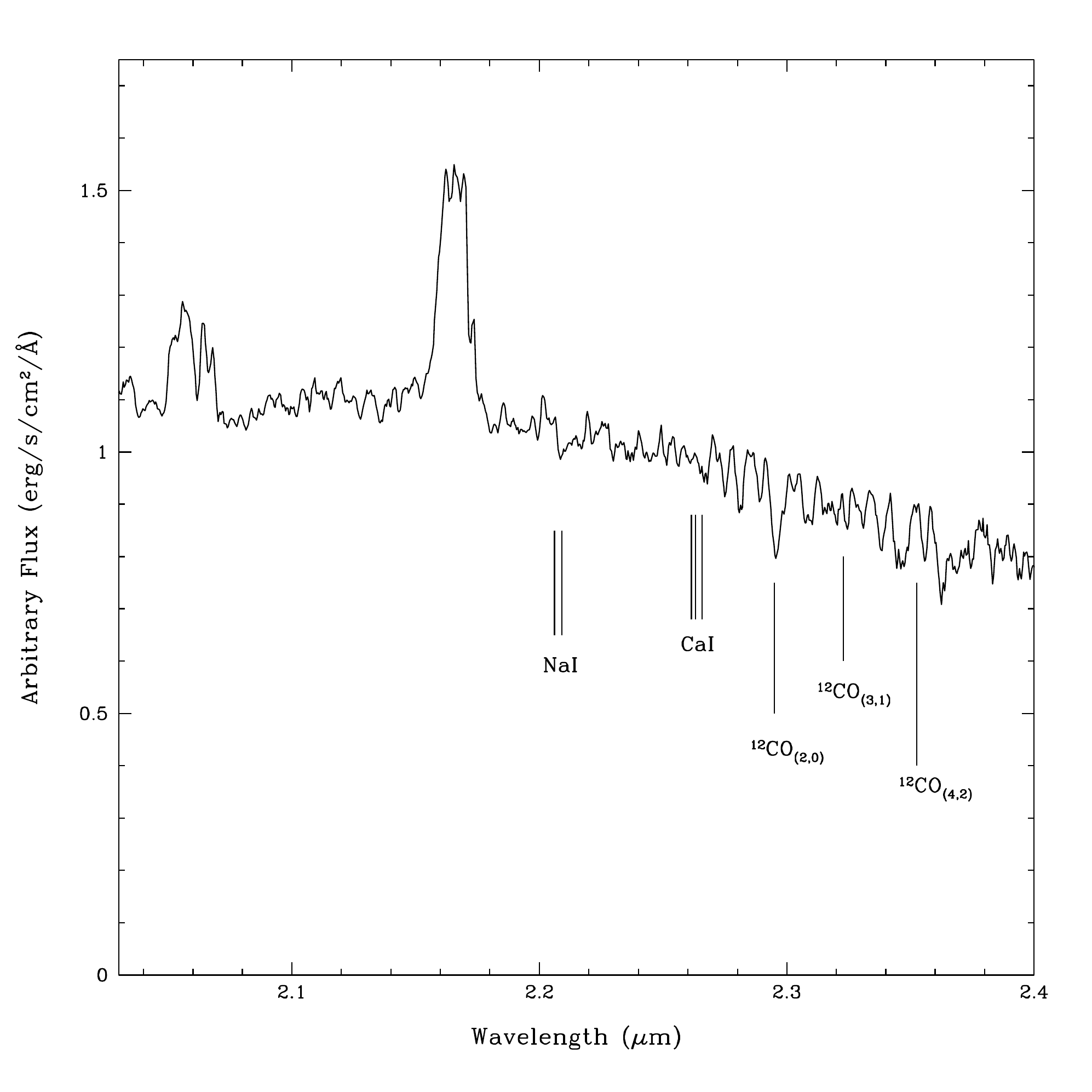}}}
\caption{The NIRSPEC data for WX Cet (smoothed by 30\AA). The locations of
the most prominent absorption features in late-type dwarfs have been
indicated.} 
\label{wxcet}
\end{figure}
\clearpage
\renewcommand{\thefigure}{28}
\begin{figure}[htb]
\centerline{{\includegraphics[width=12cm]{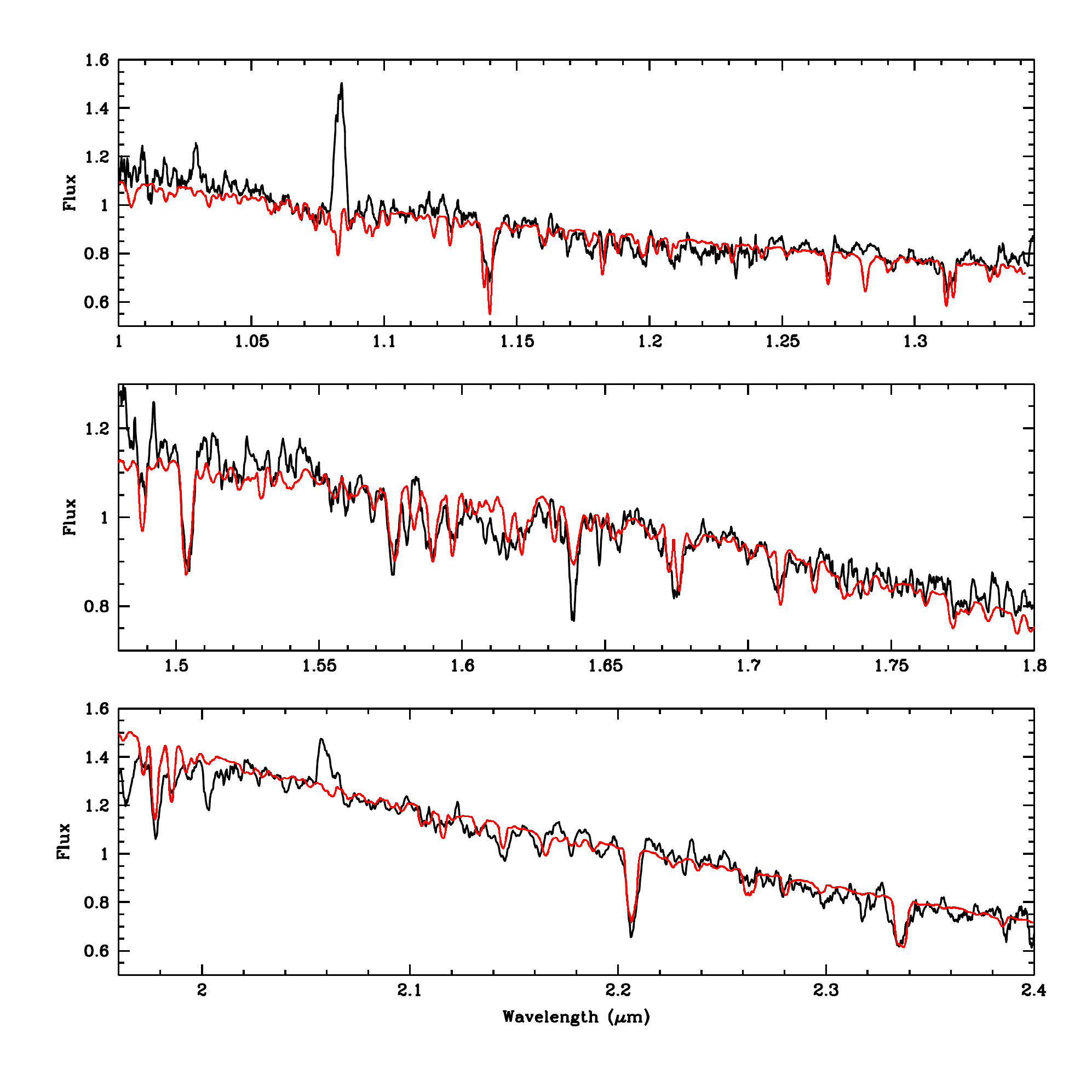}}}
\caption{The TripleSpec spectrum for QZ Ser (black, smoothed by 20\AA),
and a synthetic spectrum with T = 4500 K (red) having the abundance 
patterns noted in the text. Do to the peculiar nature of the spectrum
of QZ Ser, we have not continuum-divided it like the others. Instead,
we have multiplied the synthetic spectrum by a 4500 K blackbody.}
\label{qzser}
\end{figure}
\clearpage
\renewcommand{\thefigure}{29}
\begin{figure}[htb]
\centerline{{\includegraphics[width=12cm]{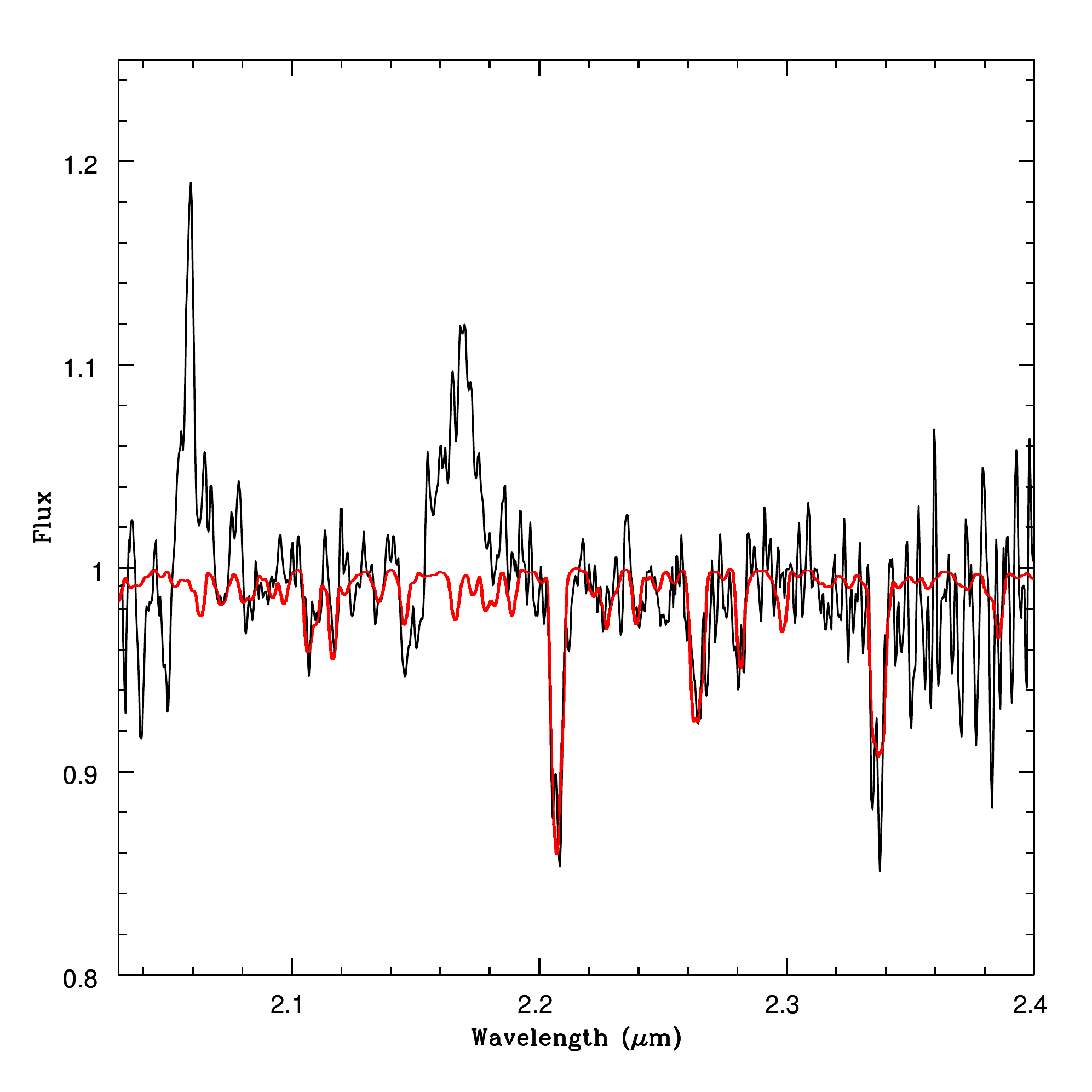}}}
\caption{The spectrum of EI Psc (black), boxcar smoothed by 3 pixels,
and a similarly smoothed synthetic spectrum (red) with T = 4500 K, log$g$
= 4.5, and [Fe/H] = $-$0.2. In this model, sodium is enhanced and carbon is
highly deficient.}
\label{eipsc}
\end{figure}
\renewcommand{\thefigure}{30}
\begin{figure}[htb]
\centerline{{\includegraphics[width=12cm]{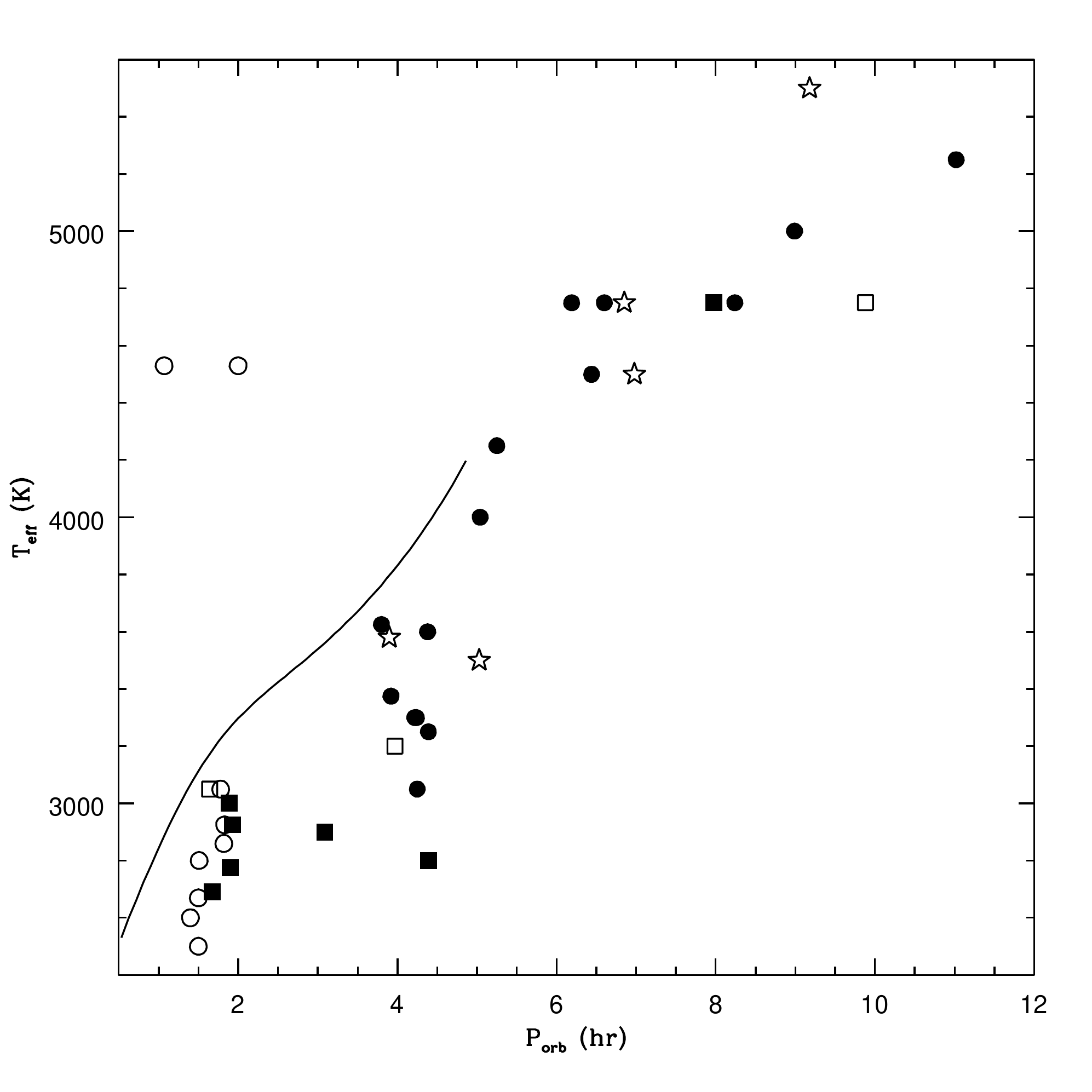}}}
\caption{The T$_{\rm eff}$ vs. P$_{\rm orb}$ relationship for CVs using the 
results in Table \ref{results}. U Gem-type dwarf novae are plotted as filled 
circles, while SU UMa type dwarf-novae are plotted as open circles. Z Cam-type
dwarf novae are plotted with star symbols. Polars are plotted as filled
squares, and IPs as open squares. The solid line was constructed using the 
mass-radius relationship for late-type main sequence stars from Mann et al. 
(2013).}
\label{teffvsporb}
\end{figure}

\end{document}